\documentclass[a4paper]{aa}  
\usepackage{graphicx}
\usepackage{natbib}
\usepackage{latexsym,amsfonts,amssymb}
\usepackage{hyperref}
\usepackage{breakurl}
\usepackage{aalongtable}
\usepackage{rotating}
\usepackage{xspace}
\usepackage{txfonts}

\newcommand{\BPZ}{\texttt{BPZ}\xspace}
\newcommand{\CARS}{\texttt{CARS}\xspace}
\newcommand{\dsnine}{\texttt{ds9}\xspace}
\newcommand{\Elixir}{\texttt{Elixir}\xspace}
\newcommand{\GaBoDS}{\texttt{GaBoDS}\xspace}
\newcommand{\Astrometrix}{\texttt{Astrometrix}\xspace}
\newcommand{\MegaPipe}{\texttt{MegaPipe}\xspace}
\newcommand{\MegaPrime}{\texttt{MegaPrime}\xspace}
\newcommand{\Photometrix}{\texttt{Photometrix}\xspace}
\newcommand{\saoimage}{\texttt{saoimage}\xspace}
\newcommand{\SExtractor}{\texttt{SExtractor}\xspace}
\newcommand{\Swarp}{\texttt{Swarp}\xspace}
\newcommand{\THELI}{\texttt{THELI}\xspace}
\newcommand{\ww}{\texttt{WeightWatcher}\xspace}
\newcommand{\Wone}{\texttt{W1}\xspace}
\newcommand{\Wthree}{\texttt{W3}\xspace}
\newcommand{\Wfour}{\texttt{W4}\xspace}
\newcommand{\Tthree}{\texttt{T0003}\xspace}
\newcommand{\myarcsec}{\hbox{$.\!\!^{\prime\prime}$}}
\newcommand{\myarcmin}{\hbox{$.\!\!^{\prime}$}}
\newcommand{\sectionref}[1]{Sect.~\ref{#1}}
\newcommand{\appendixref}[1]{App.~\ref{#1}}
\newcommand{\figref}[1]{Fig.~\ref{#1}}
\newcommand{\eqref}[1]{eq.~(\ref{#1})}
\newcommand{\tabref}[1]{Table~\ref{#1}}%
\newcommand{\refcomm}[1]{#1}
\newcommand{\NiT}{\ensuremath{N_{1}^{\rm T}}}
\newcommand{\NjT}{\ensuremath{N_{2}^{\rm T}}}
\newcommand{\Nio}{\ensuremath{N_{1}^{\rm o}}}
\newcommand{\Njo}{\ensuremath{N_{2}^{\rm o}}}
\newcommand{\fij}{\ensuremath{f_{12}}}
\newcommand{\fji}{\ensuremath{f_{21}}}
\newcommand{\wiio}{\ensuremath{\omega_{11}^{\rm o}}}

\newcommand{\wijo}{\ensuremath{\omega_{12}^{\rm o}}}
\newcommand{\wijT}{\ensuremath{\omega_{12}^{\rm T}}}
\newcommand{\wiiT}{\ensuremath{\omega_{11}^{\rm T}}}
\newcommand{\wjjT}{\ensuremath{\omega_{22}^{\rm T}}}
\newcommand{\Dio}{\ensuremath{(D_{1}D_{1})_\theta^{\rm{o}}}}

\newcommand{\Dijo}{\ensuremath{(D_{1}D_{2})_\theta^{\rm{o}}}}
\newcommand{\DijT}{\ensuremath{(D_{1}D_{2})_\theta^{\rm{T}}}}
\newcommand{\DiT}{\ensuremath{(D_{1}D_{1})_\theta^{\rm{T}}}}
\newcommand{\DjT}{\ensuremath{(D_{2}D_{2})_\theta^{\rm{T}}}}
\newcommand{\RR}{\ensuremath{(R_1R_2)_\theta}}

\def\fow{f_{\rm 12}}
\def\fwo{f_{\rm 21}}
\def\wooT{\omega_{\rm 11}^{\rm T}}
\def\wooo{\omega_{\rm 11}^{\rm o}}
\def\wwwo{\omega_{\rm 22}^{\rm o}}
\def\wowT{\omega_{\rm 12}^{\rm T}}
\def\Noo{N_{\rm 1}^{\rm o}}
\def\Nwo{N_{\rm 2}^{\rm o}}
\def\NoT{N_{\rm 1}^{\rm T}}
\def\NwT{N_{\rm 2}^{\rm T}}
\begin{document}
    \title{CARS: The CFHTLS-Archive-Research Survey}
      \titlerunning{CARS - Five-band multi-colour data from 37
        sq. deg. archival CFHTLS observations}

      \subtitle{I. Five-band multi-colour data from 37 sq. deg.
        CFHTLS-Wide observations\thanks{Based on observations obtained
          with MegaPrime/MegaCam, a joint project of CFHT and
          CEA/DAPNIA, at the Canada-France-Hawaii Telescope (CFHT)
          which is operated by the National Research Council (NRC) of
          Canada, the Institut National des Sciences de l'Univers of
          the Centre National de la Recherche Scientifique (CNRS) of
          France, and the University of Hawaii. This work is based in
          part on data products produced at TERAPIX and the Canadian
          Astronomy Data Centre (CADC) as part of the
          Canada-France-Hawaii Telescope Legacy Survey, a
          collaborative project of NRC and CNRS.}}

    \author{T.~Erben\inst{1}
            \and 
            H.~Hildebrandt\inst{2,1}
            \and
            M.~Lerchster\inst{3,4}
            \and
            P.~Hudelot\inst{1,6}
            \and
            J.~Benjamin\inst{5}
            \and 
            L.~van~Waerbeke\inst{5}
            \and 
            T.~Schrabback\inst{2,1}
            \and 
            F.~Brimioulle\inst{3}
            \and 
            O.~Cordes\inst{1}
            \and
            J.~P.~Dietrich\inst{7}
            \and 
            K.~Holhjem\inst{1}
            \and 
            M.~Schirmer\inst{1}
            \and 
            P.~Schneider\inst{1}
           }
      \authorrunning{T. Erben et al.}        
          
      \offprints{T. Erben \email{terben@astro.uni-bonn.de}}

      \institute{Argelander-Institut f\"ur Astronomie, University of Bonn, 
                 Auf dem H\"ugel 71, 53121 Bonn, 
                 Germany
                 \and
                 Sterrewacht Leiden, Leiden University, Niels Bohrweg 2, 
                 2333 CA Leiden, The Netherlands
                 \and
                 University Observatory Munich, Department of Physics, 
                 Ludwigs-Maximillians University Munich, 
                 Scheinerstr. 1, 81679 Munich, Germany
                 \and
                 Max-Planck-Institut f\"ur extraterrestrische Physik, 
                 Giessenbachstra{\ss}e, 
                 85748 Garching, Germany
                 \and       
                 Department of Physics and Astronomy, 
                 University of British Columbia,
                 Vancouver, BC V6T 1Z1, Canada
                 \and
                 Institut d'Astrophysique de Paris, 
                 98bis. bd Arago, 75014 Paris, France
                 \and
                 ESO, Karl-Schwarzschild-Strasse
                 2, 85748 Garching, Germany
                }

   \date{Received ....; accepted ...}

 
  \abstract
  {We present the CFHTLS-Archive-Research Survey (\CARS). It is a
    virtual multi-colour survey which is based on public
    archive images from the Deep and Wide components of the
    CFHT-Legacy-Survey (CFHTLS).  Our main scientific interests in the
    CFHTLS Wide-part of \CARS are optical searches for galaxy clusters from
    low to high redshift and their subsequent study with photometric
    and weak-gravitational lensing techniques.}
{As a first step of the \CARS project we
  present multi-colour catalogues from 37 sq. degrees
  of the CFHTLS-Wide component. Our aims are first to create
  astrometrically and photometrically well calibrated co-added images
  from publicly available CFHTLS data. Second goal are five-band
  ($u^*g'r'i'z'$) multi-band catalogues with an emphasis on reliable
  estimates for object colours. These are subsequently used for
  photometric redshift estimates.}
{We consider all CFHTLS-Wide survey
      pointings which were publicly available on January 2008 and which
      have five-band coverage in $u^*g'r'i'z'$. The data are 
      calibrated and processed with our
      \GaBoDS/\THELI image processing pipeline.  The quality of the
      resulting images is thoroughly checked against the
      Sloan-Digital-Sky Survey (SDSS) and already public high-end
      CFHTLS data products. From the co-added images we extract source
      catalogues and determine photometric redshifts using the public
      code \texttt{Bayesian Photometric Redshifts (\BPZ)}. Fifteen of
      our survey fields have direct overlap with public spectra from
      the VIMOS VLT deep (VVDS), DEEP2 and SDSS redshift surveys which we
      use for calibration and verification of our redshift estimates.
      Furthermore we apply a novel technique, based on studies of
      the angular galaxy cross-correlation function, to quantify the
      reliability of photo-$z$'s.}
{With this paper we present 37
      sq. degrees of homogeneous and high quality five-colour
      photometric data from the CFHTLS-Wide survey. The median seeing
      of our data is better than $0\myarcsec 9$ in all bands and our
      catalogues reach a $5\sigma$ limiting magnitude of about
      $i'_{\rm AB}\approx 24.5$. Comparisons with the SDSS indicate
      that most of our survey fields are photometrically 
      calibrated to an accuracy of $0.04$mag or better. This
      allows us to
      derive photometric redshifts of homogeneous quality over the
      whole survey area. The accuracy of our high-confidence photo-$z$
      sample (10-15 galaxies per sq. arcmin) is estimated \refcomm{with
      external spectroscopic data} to
      $\sigma_{\Delta_z/(1+z)}\approx 0.04-0.05$ up to $i'_{\rm
        AB}<24$ with typically only 1-3\% outliers.  In the spirit of
      the Legacy Survey we make our catalogues available to
      the astronomical community. Our products consist of multi-colour
      catalogues and supplementary information such as image masks and
      JPEG files to visually inspect our catalogues.  Interested users
      can obtain the data by request to the authors.}
   {}


   \maketitle
%

\section{Introduction}
Being the signposts of the largest density peaks of the cosmic matter
distribution, clusters of galaxies are of particular interest for
cosmology. The statistical distribution of clusters as a function of
mass and redshift forms one of the key cosmological probes.  Since
their dynamical or evolutionary timescale is not much shorter than the
Hubble time, they contain a `memory' of the initial conditions for
structure formation \citep[e.g.][]{bog01}.  The population of clusters
evolves with redshift, and this evolution depends on the cosmological
model \citep[e.g.][]{ecf96}.  Therefore, the redshift dependence of
the cluster abundance has been used as a cosmological test
\citep[e.g.][]{bah98,brt99,sbc03,scb03}. A prerequisite for these
studies are large and homogeneous cluster samples with well-understood
selection functions. Consequently, a large variety of systematic
searches has been performed in various parts of the electromagnetic
spectrum.  The most extensive cluster searches and cosmological
studies were performed in X-Rays \citep[see
e.g.][]{bvh00,reb02,bsg04,mae07} and in the optical \citep[see
e.g.][]{plg96,ocb99,gly00,gsn02,bma03,gym07,kma07}; see also
\citet{gal08} for a concise review of various cluster detection
algorithms in the optical. Each of the cluster searches relies on
certain cluster properties such as X-Ray emission of the hot
intra-cluster gas or an optical overdensity of red galaxies and may
introduce systematic biases in the candidate list creation.  Hence, a
careful comparison and selection with different methods on the same
area of the sky is essential to obtain a comprehensive understanding
of galaxy clusters and their mass properties.

The Wide part of the Canada-France-Hawaii-Telescope Legacy Survey
(CFHTLS-Wide) \refcomm{is an optical Wide-Field-Imaging-Survey particularly
well suited for such studies.} When completed it will cover 170 sq.
deg.  in the five optical Sloan filters $u^*g'r'i'z'$ to a limiting
magnitude of $i'_{\rm AB}\approx 24.5$. The unique combination of
area, depth and wavelength coverage allows the application of a
variety of currently available optical search algorithms. For
instance, the Postman matched filter technique \citep[see][]{plg96}
applies an overdensity and luminosity function filter to photometric
data of a single band survey. It can provide high-confidence samples
in the low- and medium redshift range \citep[see e.g.][]{ocb99,obd01}.
The Red-Cluster-Sequence algorithm scans a two-filter survey for the
Red Sequence of elliptical galaxies and is mainly used for the medium
to high redshift regime with the $r$ and $z$ filters \citep{gly00}.
The existence of five bands in the CFHTLS-Wide allows us to estimate
photometric redshifts and the application of techniques using distance
information \citep[e.g.][]{mnr05}.  Furthermore, one of the main goals
of the CFHTLS-Wide are weak gravitational lensing studies of the
large-scale structure distribution \citep[see e.g.][for recent
results]{hmv06,fsh08}. This will allow us to complement and to
directly compare optical cluster searches with candidates from weak
lensing mass reconstructions and shear peak detections
(see e.g. \citeauthor{sch96b} \citeyear{sch96b}; \citeauthor{ewm00} 
\citeyear{ewm00}; \citeauthor{bas01} \citeyear{bas01};
\citeauthor{wtm01} (\citeyear{wtm01,wmt03}); \citeauthor{dpl03}
\citeyear{dpl03}; \citeauthor{hes05} \citeyear{hes05};
\citeauthor{wdh06} \citeyear{wdh06}; \citeauthor{seh07}
\citeyear{seh07}; \citeauthor{del07} \citeyear{del07}).
To perform these galaxy cluster studies, we perform an extensive
Archive-Research programme on publicly available data from the
CFHTLS-Wide.  We baptise our survey the CFHTLS-Archive-Research Survey
(CARS in the following).

This paper marks the first step of our science programme on a
significant area of CARS. We describe our data handling and the
creation of multi-colour catalogues, including a first set of
photometric redshifts, on 37 sq. degrees of five-colour CARS data.
 
The article is organised as follows: \sectionref{sec:CARSdata} gives a
short overview on our current data set; a detailed description of our
complete image data handling is given in \appendixref{app:processing}.
\sectionref{sec:catalogues} and \sectionref{sec:photoz} summarise the
multi-colour catalogue creation and the photometric redshift
estimation together with a thorough quantification of their quality.
We continue to describe our data products (\sectionref{sec:releaseproducts})  
and finish with our conclusions in \sectionref{sec:summary}.


%

\section{The data}
\label{sec:CARSdata}
The current set of \CARS data consists of a subset of the synoptic
CFHTLS-Wide observations which is one of three independent parts of
the Canada-French-Hawaii-Telescope Legacy Survey (CFHTLS). It is a
very large, 5-year project designed and executed jointly by the
Canadian and French communities. The survey started in spring 2003 and
is planned to finish during 2008.  All observations are carried out
with the \MegaPrime instrument mounted at the Canada-France-Hawaii
Telescope (CFHT).  \MegaPrime \citep[see e.~g.][]{bca03} is an optical
multi-chip instrument with a $9\times 4$ CCD array ($2048\times 4096$
pixel in each CCD; $0\myarcsec 186$ pixel scale; $\approx
1^{\circ}\times 1^{\circ}$ total field-of-view).  When completed, the
CFHTLS-Wide will cover 170 sq.  deg. in four high-galactic-latitude
patches \texttt{W1-W4} of 25 to 72 square degrees through the five
optical filters $u^*g'r'i'z'$ down to a magnitude of $i'_{\rm
  AB}\approx 24.5$. See
\burl{http://www.cfht.hawaii.edu/Science/CFHLS/} and
\burl{http://terapix.iap.fr/cplt/oldSite/Descart/summarycfhtlswide.html}
for further information on survey goals and survey implementation.

Since June 2006 CFHTLS observations are publicly released to the
astronomical community via the Canadian Astronomy Data Centre
(CADC)\footnote{see \burl{http://www1.cadc-ccda.hia-iha.nrc-cnrc.gc.ca/cadc/}}.
At the time of writing raw and \Elixir preprocessed images (see below),
together with auxiliary meta-data, can be obtained 13 months after
observations.

\begin{figure*}[ht]
  \centering
  \includegraphics[width=0.95\columnwidth]{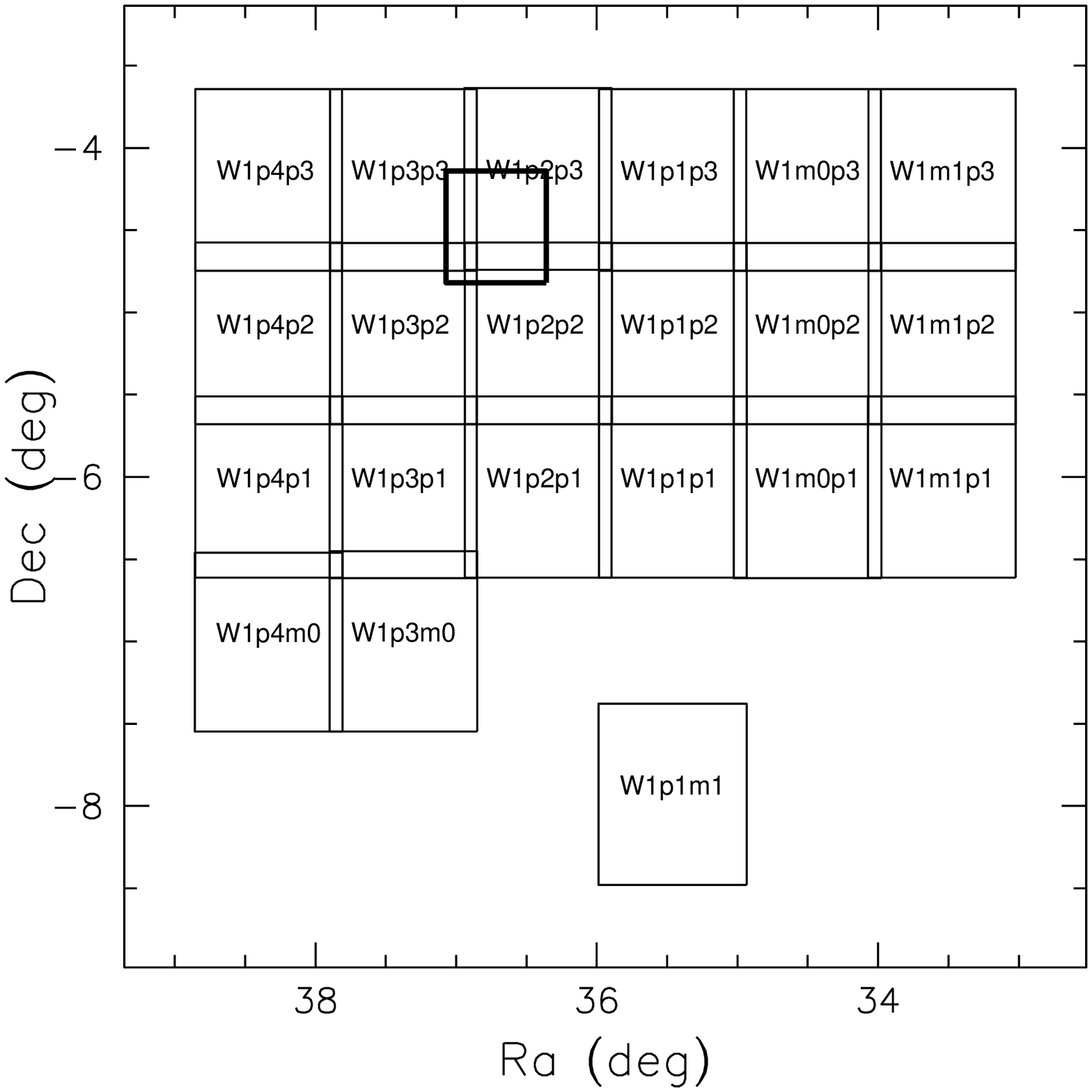}
  \includegraphics[width=0.95\columnwidth]{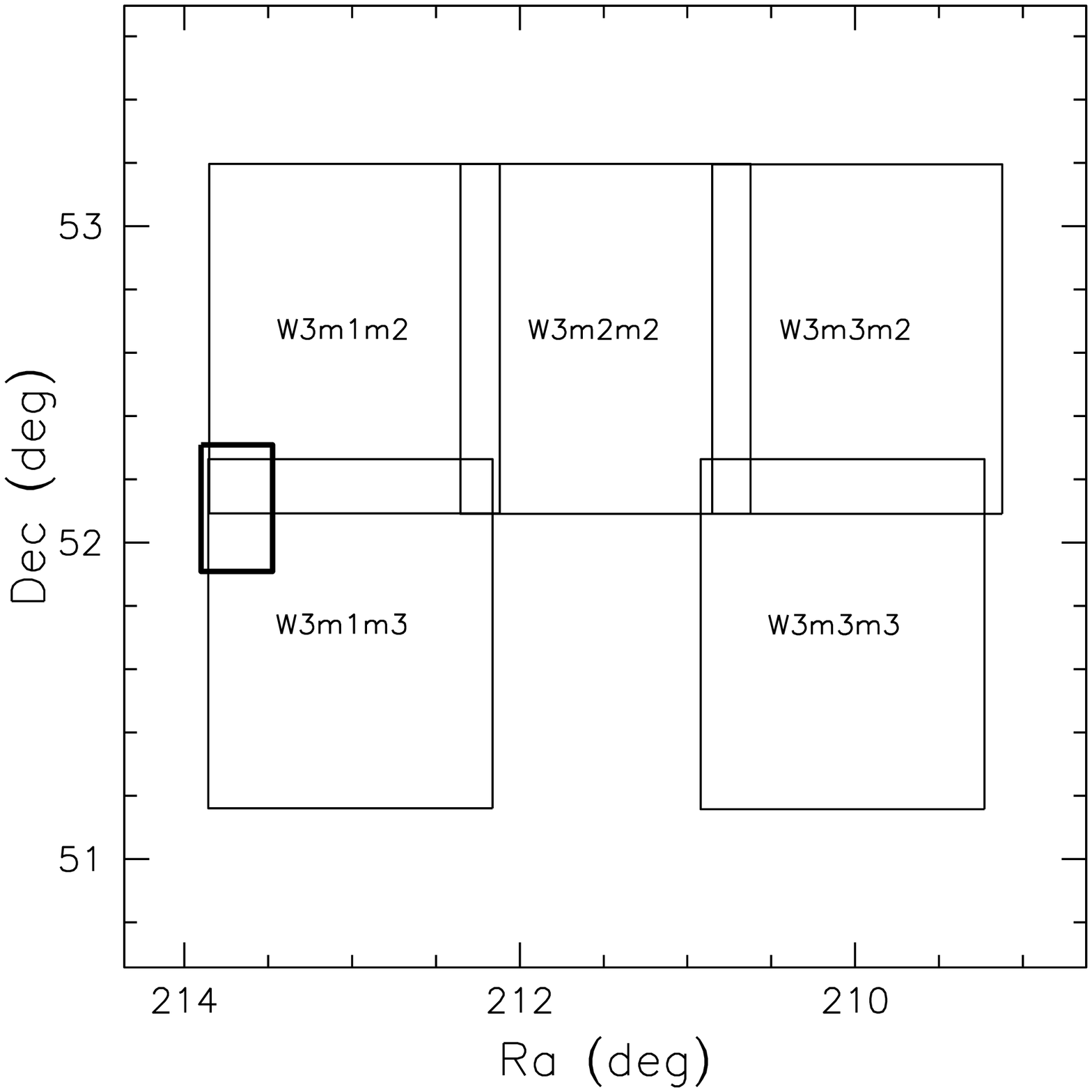}
  \includegraphics[width=0.95\columnwidth]{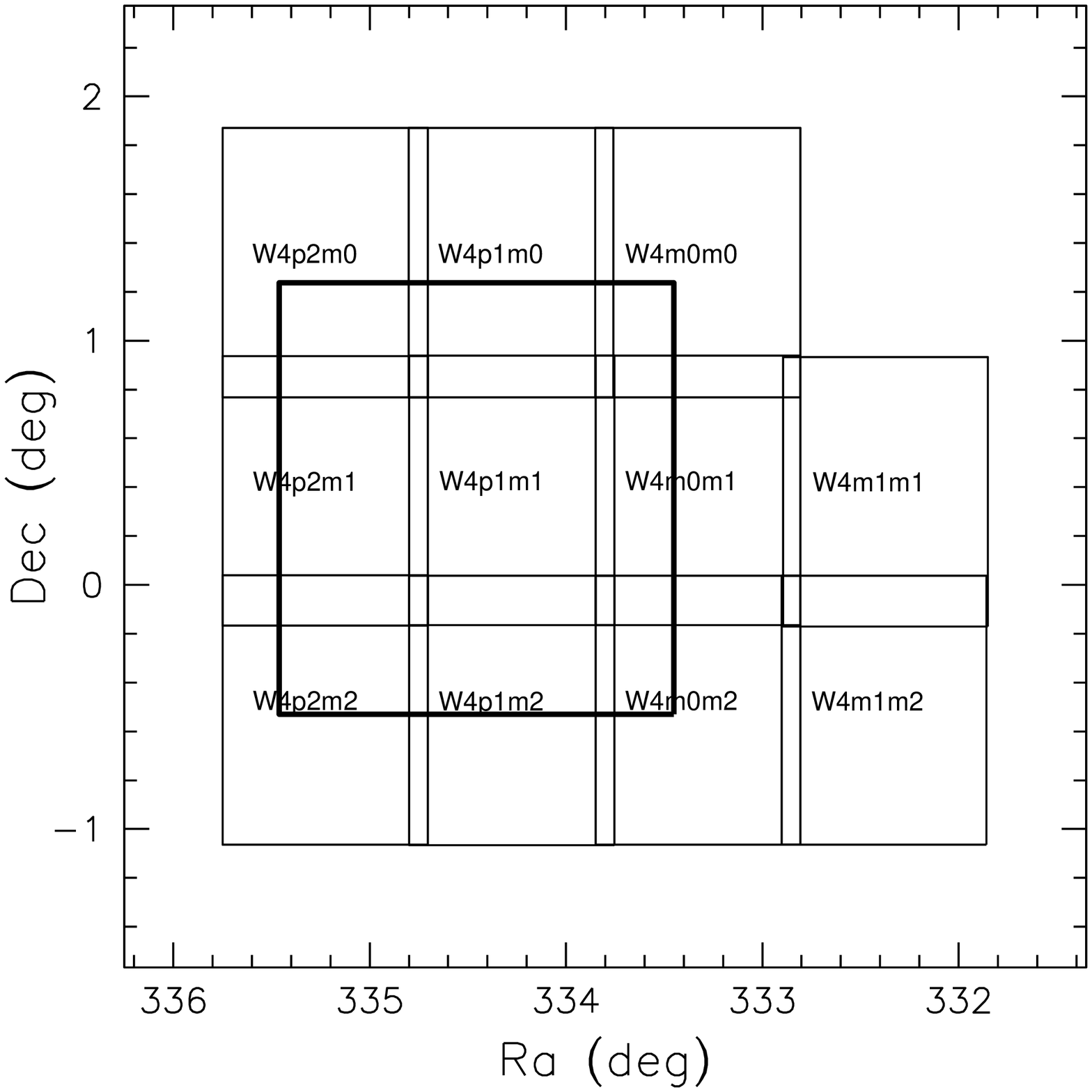}
  \caption{\label{fig:fieldlayout} Layouts of the current three \CARS:
    The \CARS data of this work are split
    up in the three CFHTLS-Wide patches \texttt{W1} (21 sq. degrees;
    patch centre: Ra=02:18:00, Dec=$-$07:00:00), \texttt{W3} (five sq.
    degrees; patch centre: Ra=14:17:54, Dec=+54:30:31) and \texttt{W4}
    (11 sq. degrees; patch centre: Ra=22:13:18, Dec=+01:19:00). In 
    areas covered by thick lines spectra from various surveys are
    publicly available for photo-$z$ calibration and verification 
    (see text for details).}
\end{figure*}

For the current work we consider all \Elixir processed CFHTLS-Wide
fields with observations in all five optical colours $u^*g'r'i'z'$
which were publicly available on 18/01/2008, i.~e. observed until
18/12/2006. In total, the current \CARS set consists of 37 sq. degrees
split between the three CFHTLS-Wide patches \texttt{W1} (21 sq.
degrees), \texttt{W3} (five sq. degrees) and \texttt{W4} (eleven sq.
degrees).  \refcomm{The area around the defined patch centres
(\texttt{W1}: Ra=02:18:00, Dec=$-$07:00:00, \texttt{W3}:
Ra=14:17:54, Dec=+54:30:31 and \texttt{W4}: Ra=22:13:18,
Dec=+01:19:00) are covered on a regular grid with pointed
observations. Names of individual \CARS pointings are constructed
like \texttt{W1m1p2} (read ``\texttt{W1} minus 1 plus 2''; see also
\figref{fig:fieldlayout}). They indicate the patch and the
separation (approximately in degrees) from the patch centre, e.~g.
pointing \texttt{W1m1p2} is about one degree west and two degrees
north from the \texttt{W1} centre.  The overlap of adjacent
pointings is about $3\myarcmin 0$ in Ra and $6\myarcmin 0$ in Dec.
The exact layout of the \CARS survey fields is shown in
\figref{fig:fieldlayout}. All three patches are covered by
spectroscopic surveys which allow us to calibrate and to verify
photometric redshift estimates later on; \texttt{W1} and \texttt{W4}
overlap with the VIMOS VLT Deep Survey [VVDS; see
\citet{lvg05,glg08}],
\texttt{W3} with the DEEP2 galaxy redshift survey
\citep[see][]{dgk07}. Moreover, \CARS data from patches \texttt{W3}
and \texttt{W4} have complete overlap with the
Sloan-Digital-Sky-Survey (SDSS; see e.~g.  \citealt{aaa07}). From
patch \texttt{W1} only the southern pointings \texttt{W1p3m0},
\texttt{W1p4m0} and \texttt{W1p1m1} have SDSS overlap.}

The \Elixir data preprocessing performed at CFHT \citep[see][]{mac04} 
includes removal of the 
instrumental signature from raw data [bias/dark subtraction;
flat-fielding; fringe correction in $i'$ and $z'$ data] and
absolute photometric calibration [determination of zeropoints, 
colour terms and extinction coefficients, corrections for scattered 
light effects which lead to significant inhomogeneous photometric 
zeropoints across the CCD mosaic \citep[see
e.g.][]{msj01,kgo04a,reg07}]. 
The data is accompanied with comprehensive information on the observing conditions 
(seeing, sky-transparency, sky-background level) for each exposure\footnote{see
\burl{http://www.cfht.hawaii.edu/Science/CFHTLS-DATA/exposurescatalogs.html}}. 

After downloading all data from CADC and rejecting exposures with a
problematic CFHT quality assessment we further process the data on a
pointing/colour basis with our \GaBoDS/\THELI pipeline to produce deep
co-added images for scientific exploitation.  Our algorithms and
software modules to process multi-chip cameras are described in
\citet{esd05} and most of the details do not need to be repeated here.
For the interested reader we give in
\appendixref{app:processing} a thorough description of the \CARS data
handling, data peculiarities and the pipeline upgrades/extensions
necessary to smoothly and automatically process \MegaPrime data. In
addition, a comprehensive assessment of the astrometric and
photometric quality of our data, together with a comparison to previous
releases of CFHTLS data can be found there. We conclude that the \CARS
data set is accurately astrometrically and photometrically calibrated
for multi-colour photometric and lensing studies. 

\begin{figure*}[ht]
  \centering
  \fbox{\includegraphics[width=0.45\textwidth]{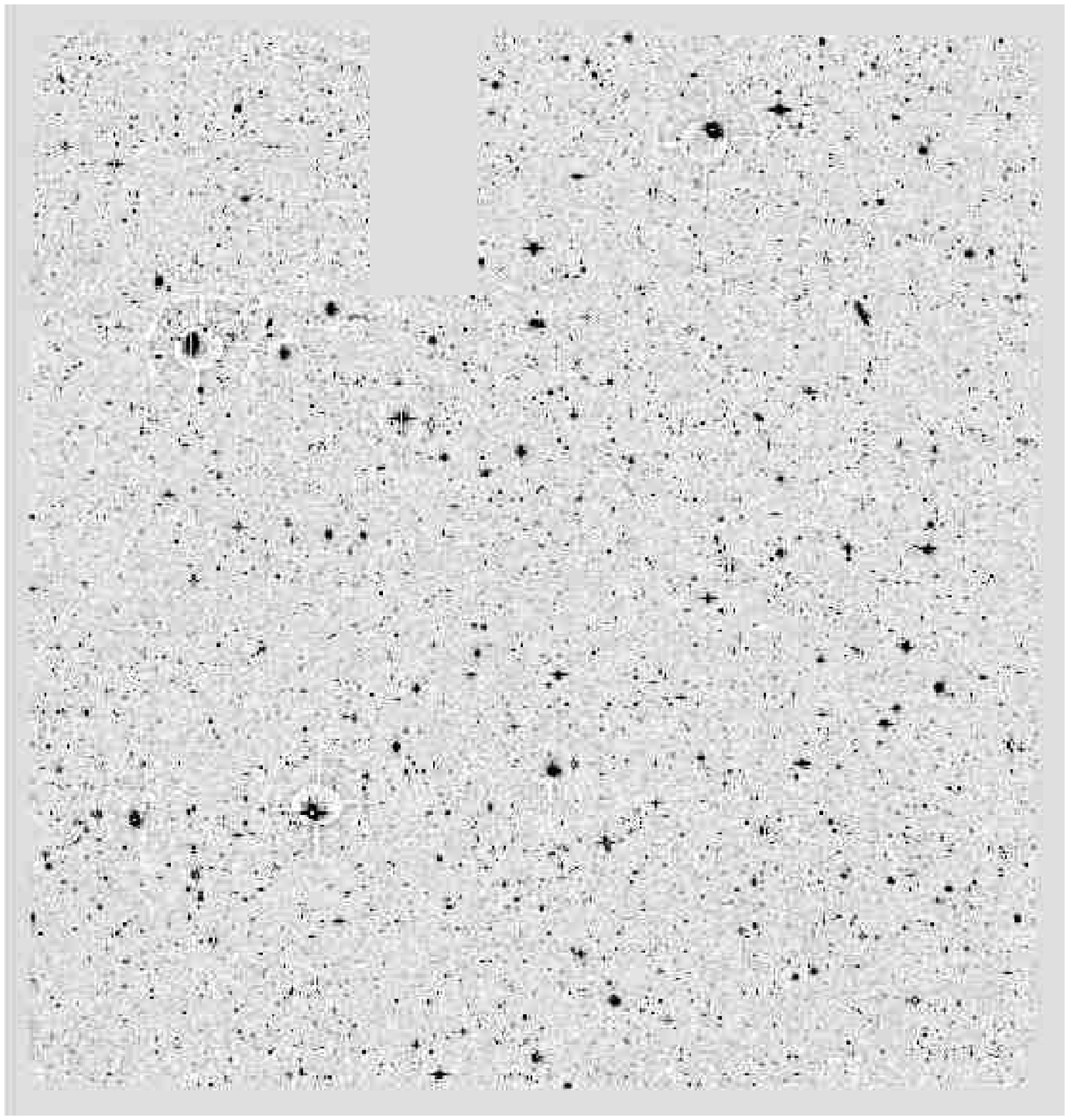}}
  \fbox{\includegraphics[width=0.45\textwidth]{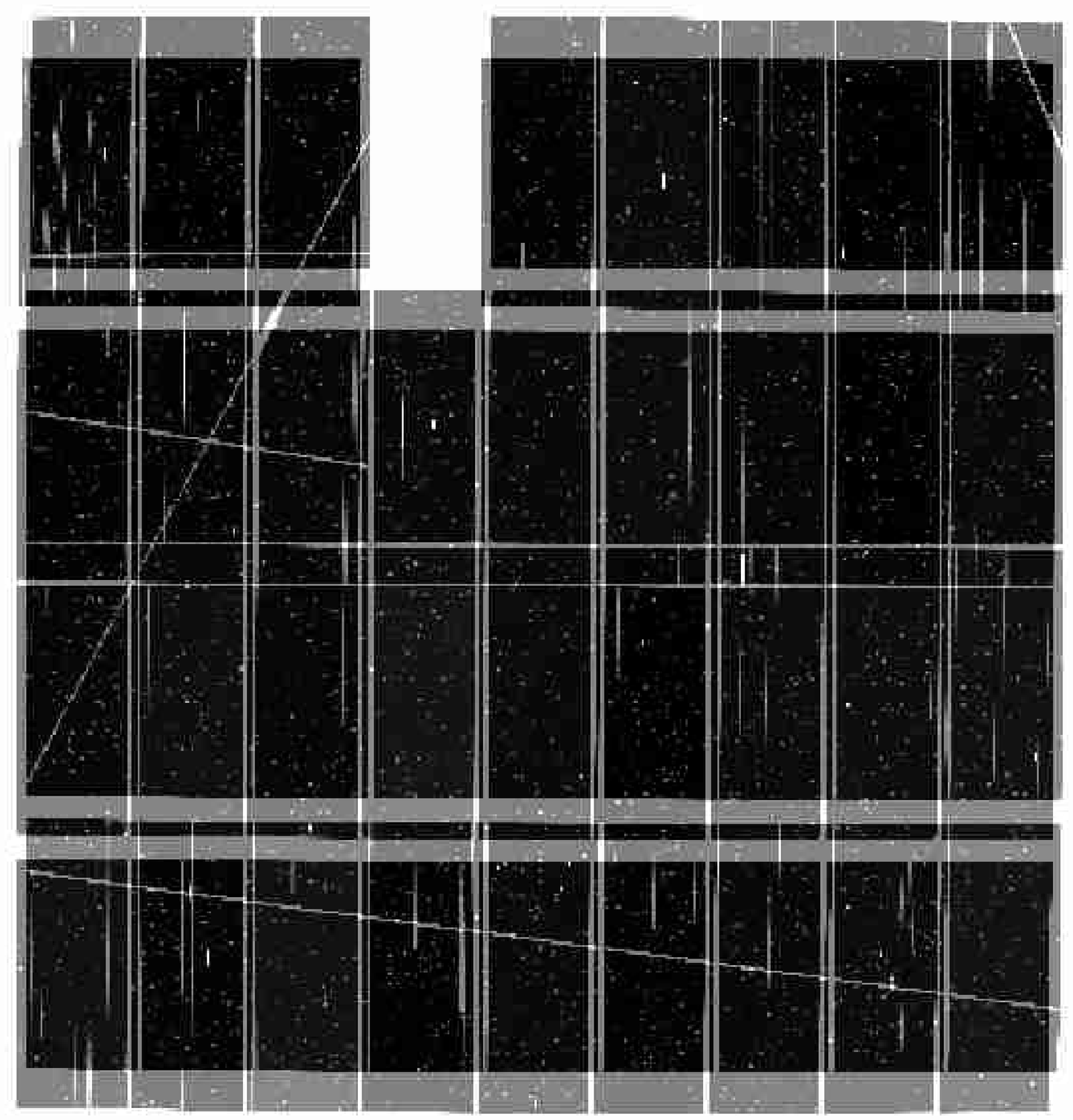}}
  \caption{\label{fig:finalimages} Co-added \CARS data: We show the
    final science image of the $r'-$ band observations from field
    \texttt{W1p3p1} (left panel) and the accompanying weight map (right
    panel). The weight
    map shows five extended satellite tracks which were automatically
    identified and masked before image co-addition (see
    \appendixref{sec:data-preprocessing}
    for details). In several \CARS pointings/colours
    individual chips did not contain useful data and were hence
    excluded from the analysis. The pointing shown suffered from this
    problem in the uppermost row.}
\end{figure*}

\refcomm{In the following we give a very brief summary of the most 
important \CARS data characteristics:
The first products of the \THELI processing are 185 co-added science
images accompanied by weight maps which characterise their noise
properties \citep[see e.g. Sect. 6 of][ for a discussion on the role of weight
images in the object detection process]{esd05}. 
Our image stacking procedure first automatically
identifies image defects (hot and cold pixel, cosmic ray hits and
satellite tracks) in the individual frames and assigns them zero
weight in the co-addition process. The stacking itself is a
statistically optimal, weighted mean co-addition taking into account 
sky background variations and
photometric zeropoint variations in the individual frames (see
Sects. 6 and 7 from \citealt{esd05} and \appendixref{app:processing}
for further details). Identifying and masking image defects
in the individual images before co-addition allows us to obtain clean 
stacked images
also if only very few input images are contributing. This is essential
for the processing of \CARS $r'$-band data where most pointings are
covered only by two individual exposures.}  
As an example, we show in \figref{fig:finalimages} the
final co-added $r'-$band image of the field \texttt{W1p3p1}.
\refcomm{We perform numerous internal and external tests to quantify the
astrometric and photometric properties of our data. In 
\appendixref{sec:astr-calibr} we conclude that the internal
astrometric accuracy of our data, i.~e. the accuracy with which we can
align individual exposures of a colour and pointing, is 
$0\myarcsec 03-0\myarcsec 04$ (1/5th of a \MegaPrime pixel) over the
whole
field-of-view of \MegaPrime; our absolute astrometric frame is
given by the \texttt{USNO-B1} catalogue \citep[see][]{mlc03}. The
co-added images of the different colours from each pointing are
aligned to sub-pixel precision in all cases.}

\refcomm{We quantify the quality of the photometric calibration of our data in
\appendixref{sec:phot-calibr}, \appendixref{sec:sloancomp} and 
\appendixref{sec:comparison-cars-data}. First, we investigate
photometric flatness over the \MegaPrime field-of-view. We use data
from the CFHTLS-Deep survey which keeps observing four sq. degrees over the
whole five-year period of the CFHT-Legacy Survey. This allows us to
create image stacks from different periods and to investigate
photometric consistency.  Magnitude comparisons of co-additions
obtained from the three years 2003, 2004 and 2005 indicate uniform 
photometric properties with a
dispersion of $\sigma_{int, u^*g'r'i'} \approx 0.01-0.02$ in
$u^*g'r'i'$ and about $\sigma_{int, z'} \approx 0.03-0.04$ in $z'$. We
attribute higher residuals in $z'$ to fringe residuals in this band.}

\refcomm{Our absolute magnitude zeropoints are tested against
photometry in the SDSS and against previous data releases of the
CFHTLS. Our comparison with the SDSS shows that our absolute
photometric calibration agrees with Sloan to $\sigma_{abs, g'r'i'}
\approx 0.01-0.04$ mag in $g'r'i'$ and $\sigma_{abs, z'} \approx
0.03-0.05$ mag for $z'$. While the calibration in these four bands
seems to be unbiased, we observe, at the current stage, a systematic
magnitude offset in $u^*$ of about 0.1mag with respect to Sloan
(\CARS magnitudes appear fainter than Sloan). For $u^*$-band data
from spring to fall 2006 our analysis suggests an \Elixir
calibration problem leading to offsets of 0.2-0.3 mag in
$u^*$.}

\refcomm{Finally, we directly compared our flux measurements with those of
the previous
CFHTLS Terapix \Tthree release and Stephen Gwyn's \MegaPipe project 
\citep[see][]{gwy08}. The measurements to \Tthree are in very good
agreement with typical dispersions of 0.02mag; in many cases larger 
scatters are observed with respect to the \MegaPipe data. Private
communication with S. Gwyn suggests that several \MegaPipe stacks 
suffer from the accidental inclusion of images obtained under
unfavourable photometric conditions; see 
\appendixref{sec:comparison-cars-data} and
\appendixref{sec:phot-calibr-summ} for further details.}
  
\refcomm{\tabref{tab:characteristics}  lists  average  properties  for
seeing and limiting magnitude values  in our survey data. The quoted
values  for exposure  time (we  list the  typical exposure  time per
dither, the number of dithered observations per colour and the total
exposure  time  in  parentheses),  limiting  magnitudes  and  seeing
correspond to  a typical field and  hence give a  good indication of
what can be  expected from the data. The  seeing values (\SExtractor
parameter  FWHM\_IMAGE  for  stellar  sources)  are  the  median  of
measured  seeing values  from  all co-added  science  images in  the
corresponding  filters.   We  note  that  we  measure  a  seeing  of
$1\myarcsec 0$ or below for all co-added \CARS stacks except for the
$u^*$-band image of \texttt{W1p3p3}  for which we obtain $1\myarcsec
1$.  The  limiting magnitude is defined as  the 5$-\sigma$ detection
limit    in    a     $2\myarcsec    0$    aperture    via    $m_{\rm
  lim}=ZP-2.5\log(5\sqrt{N_{\rm pix}}\sigma_{\rm  sky})$, where $ZP$
is the magnitude zeropoint, $N_{\rm pix}$ is the number of pixels in
a circle with  radius $2\myarcsec 0$ and $\sigma_{\rm  sky}$ the sky
background  noise variation.  The actual  numbers for  $m_{\rm lim}$
in \tabref{tab:characteristics}
were obtained from the field \texttt{W4p2m0}. It represents a \CARS
pointing with typical properties concerning exposure times and image 
seeing. A more detailed table
listing these quantities for each individual field can be found in
\appendixref{sec:CARSquality}.}

The described imaging data form the basis for the subsequent multi-colour
catalogue creation and photo-$z$ estimation.
\begin{table}
  \caption{Characteristics of the \CARS co-added science data: The table
    shows basic average properties of our final science data (see text
    for an explanation of the columns).}           
\label{tab:characteristics}      
\centering          
\begin{tabular}{llll}    
\hline\hline       
\multicolumn{1}{c}{Filter} & \multicolumn{1}{c}{expos. time [s]} &
\multicolumn{1}{c}{$m_{\rm lim}$ [AB mag]} & \multicolumn{1}{c}{seeing [$''$]}\\ 
\hline
$u^* (u.MP9301) $ & $5\times 600$ (3000) & 25.24 & 0.87\\ 
$g' (g.MP9401) $ & $5\times 500$ (2500) & 25.30 & 0.85\\ 
$r' (r.MP9601) $ & $2\times 500$ (1000) & 24.36 & 0.79\\ 
$i' (i.MP9701) $ & $7\times 615$ (4305) & 24.68 & 0.71\\ 
$z' (z.MP9801) $ & $6\times 600$ (3600) & 23.20 & 0.66\\ 
\hline                  
\end{tabular}
\end{table}
%

%
\section{The multi-colour catalogues}
\label{sec:catalogues}
Our procedures to create multi-colour catalogues for the five-band
\CARS data are similar to the ones presented in \cite{hed06} where we
studied Lyman-break galaxies in the ESO Deep Public Survey (DPS).

\subsection{Preparation and PSF equalisation}
In order to estimate unbiased colours it is necessary to measure object
fluxes in the same physical apertures in each band, i.e. for a given
object the same physical parts of the object should be measured in the
different bands. Since the PSF usually varies from band to band we
apply a convolution to degrade the seeing of all images of one field
to the PSF size of the image with the worst seeing. Assuming a
Gaussian PSF we first measure the seeing and then calculate
appropriate filter functions by the following formula:

\begin{equation}
\sigma_{\mathrm{filter, k}}=\sqrt{\sigma_{\mathrm{worst}}^2-\sigma_{\mathrm{k}}^2}\:,
\end{equation}
with $\sigma_{\mathrm{filter, k}}$ being the width of the Gaussian
filter for convolution of the k'th image, $\sigma_{\mathrm{worst}}$
being the PSF size of the image with the worst seeing, and
$\sigma_{\mathrm k}$ being the PSF size of the k'th image. By doing so
we neglect the non-Gaussianity of a typical ground-based PSF.
Nevertheless, experience with the DPS shows that our procedure is
sufficient to estimate reliable colours if the seeing values in the
individual colours are subarcsecond and not too different. In \CARS,
the seeing values for a pointing typically do not differ by more than
$0\myarcsec 2$-$0\myarcsec 3$ (see \tabref{tab:CARSquality}).

\subsection{Limiting magnitudes}
\label{sec:mag_lim}
The images filtered in that way are then analysed for their
sky-background properties. For the accurate estimation of photometric
redshifts it is important to have a reasonable estimate for the 
limiting magnitude at a given object position. 
Therefore, we create limiting magnitude
maps from the RMS fluctuations of the sky-background in small parts of
the image. Here we use 1$\sigma$ limiting magnitudes calculated in a
circular aperture of 2$\times$ stellar FWHM diameter. This procedure ensures
that varying depths over the field are properly taken into account in
the colour estimation. It may well be that an object would be detected
in one part of the image whereas it is undetectable in a different
part due to the dither pattern or stray-light leading to inhomogeneous
depth. By assigning position-dependent limiting magnitudes to each
object in all bands we can later decide which flux measurements are
significant and which are not.

\subsection{Object detection}
\label{sec:obj_det}
The object detection is performed with \SExtractor
\citep[see][]{bea96} in dual-image mode and we consider all objects
\refcomm{having at least 5 connected pixels exceeding $2\sigma$ of the
sky-background variation}. \refcomm{We will base our
primary science analyses (galaxy cluster searches and weak lensing 
applications) on the $i'$-band data. Hence, we generate our object catalogues
based on this colour rather than on a combination of all
available
colours such as a $\chi^2$ image \citep[see e.g.][]{mrb03}.}   
We use the unconvolved
$i'$-band image as the detection image and measure fluxes for the
colour estimation and the photometric redshifts on the convolved
frames. Colour indices are estimated from the differences of isophotal
magnitudes taking into account local limiting magnitudes, i.e. if
a magnitude is measured to be fainter than the local limiting
magnitude, then this limit is used instead of the measured magnitude
to estimate an upper/lower bound for the colour index.

Additionally, we also measure the total $i'$-band magnitudes on the
unconvolved image so that total magnitudes in the other bands can in
principle be calculated from those and from the colour
indices. However, it should be noted that our approach to run
\SExtractor in dual-image-mode with the unconvolved $i'$-band image for
detection will never lead to accurate total magnitudes in the
$u^*g'r'z'$-bands. While adding/subtracting the appropriate colour index
to/from the total $i'$-band magnitude yields accurate total magnitudes
in one of the other bands for bright objects without a colour
gradient, it can yield strongly biased results in other cases. Only
catalogues created in single-image-mode on the different bands assure
a reliable estimation of total magnitudes. Since our emphasis here is
on estimating colours as accurately as possible, we do not pursue this
issue further.

\subsection{Creation of image masks}
\label{sec:masking}
\begin{figure}[ht]
  \centering
  \fbox{\includegraphics[width=0.9\columnwidth,height=0.75\columnwidth]{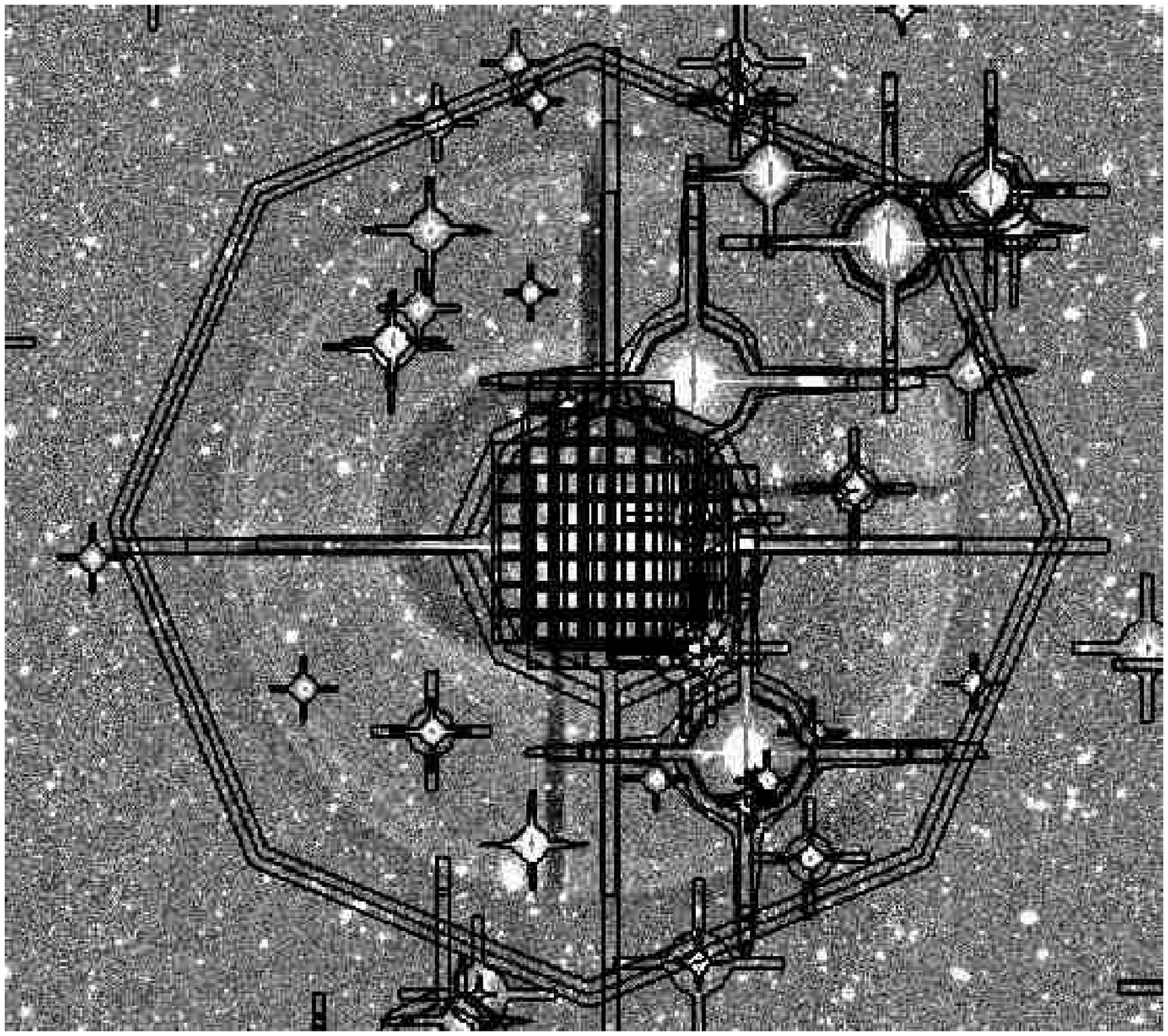}}
  \fbox{\includegraphics[width=0.9\columnwidth,height=0.75\columnwidth]{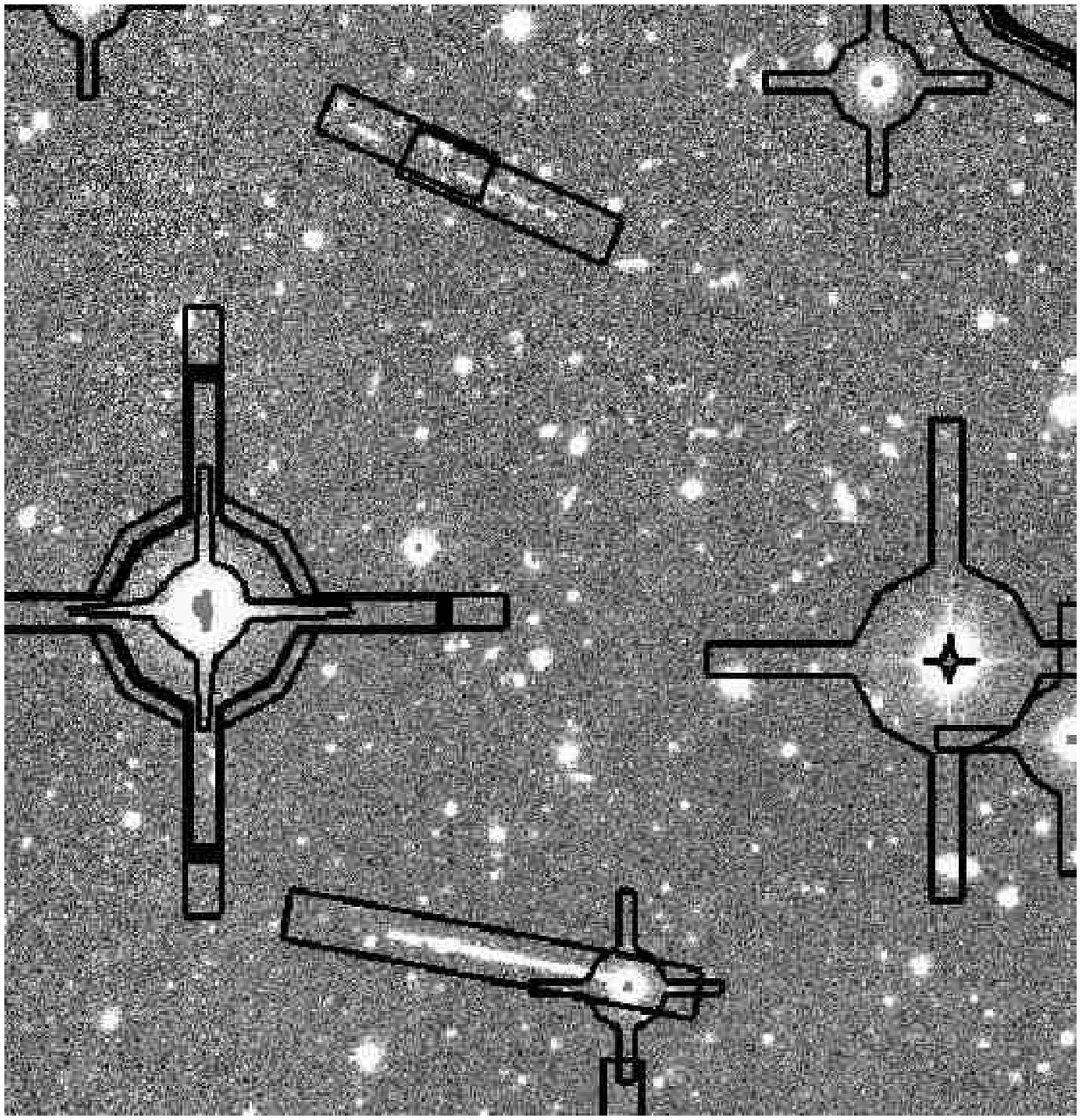}}
  \caption{\label{fig:imagemasks} Semi-automatic image masking: Shown
    is the result of our semi-automatic image masking for areas of the
    field \texttt{W4m0m0}. The polygon squares result from our object
    density analysis and the stars cover sources identified in the
    \texttt{GSC-1}, \texttt{GSC-2.3.2} and \texttt{USNO-A2} Standard
    Star Catalogues (upper panel; multiple masks around stars appear
    for sources identified in various catalogues). 
    The lower panel shows results from our asteroid masking
    procedure.}
\end{figure}

All \CARS pointings suffer from bright stars and other large-
and small-scale astronomical features that we would like to exclude
from the following analysis. At least we want to know the location and
shape of those areas so that catalogues can be cleaned from objects in
problematic areas. Of course the regions which need to be masked
heavily depend on the science project for which our data are used.
Our main scientific drivers for the \CARS data are the photometric
identification of galaxy-clusters and their subsequent investigation
with photometric redshift and weak gravitational lensing techniques.
These applications require the accurate determination of galaxy
surface brightness moments to at least fourth order. Hence, we want to
exclude all image areas in which the light distribution of faint
objects (often confined to a very small number of image pixels) is
probably altered by other sources. Amongst such defects are:
\begin{itemize}
\item Extended haloes of very bright stars
\item Diffraction spikes of stars
\item Areas around very large galaxies
\item Various kinds of image reflections
\item Tracks of asteroids
\end{itemize}
\refcomm{A complete manual masking process for the large amount of \CARS
data would be a prohibitively long and man-power intensive task.  
We developed a software package which generates template masks for
most image features which we want to reject. If necessary, these automatically
generated masks are manually optimised later.}
Our tools are based on the following ideas:
\begin{enumerate}
\item Object detection algorithms such as \SExtractor identify
  astronomical sources by connected areas which exceed the sky
  background noise by a certain amount.  
  The pixel distribution of \THELI produced images of \emph{empty}
  fields has mode zero after sky subtraction. Large-scale artefacts
  like stellar reflection rings lead to local deviations of the
  background. By running \SExtractor with a fixed background
  value of zero and a very low detection threshold of $0.6\sigma$, this
  local variation of the background leads to a significant increase in the
  detection of spurious objects. We examine the \SExtractor catalogue
  for areas of significant over-densities and strong gradients in the object
  density distribution. Corresponding image regions are flagged as
  problematic.  The output of the procedure is an 8-bit FITS
  FLAG image with the size of the original image (masked areas are '1'
  and unmasked areas are '0' in this image) or/and a \saoimage/\dsnine
  polygon region file of masked areas.  See \citet{del07} for further
  details on the algorithm and its implementation.
\item Astronomical Standard Star Catalogues such as \texttt{USNO-B1},
  \texttt{GSC-2} or \texttt{SDSS-R5} list the positions and
  magnitudes of known astronomical sources up to a magnitude of about
  18. In the \CARS data, the large majority of these objects with
  $m\leq 16$ are bright or moderately bright stars whose
  surroundings should be excluded from object catalogues (faint haloes,
  diffraction spikes).  Moreover, stellar sources have well defined
  shapes over the complete \MegaPrime field-of-view. The extent
  of the central light concentration and the width and the height of
  stellar diffraction spikes can be modeled as function of
  apparent magnitude.  
  On the basis of these observations we automatically
  create object masks for stellar objects:
   \begin{itemize}
   \item We retrieve object positions and magnitudes from the Standard
     Star Catalogues \texttt{GSC-1}, \texttt{GSC-2.3.2} and
     \texttt{USNO-A2}. We found that our selection criteria in these
     catalogues (magnitude limits, catalogue flags) result in slightly
     different source lists and hence the three samples complement
     each other. Our masking is performed independently on all three
     catalogues.
   \item At each catalogue position we lay down template masks for the
     central light halo and the diffraction spikes. The templates are
     scaled with (red photographic) magnitude to conservatively
     encompass the stellar areas. In addition, for very bright stars
     with $m<10.35$ we mask extended stellar diffraction haloes. For
     \MegaPrime these haloes have an extend of about $4\myarcmin 0$
     depending only weakly on magnitude. Moreover, these haloes occur
     with a radial offset towards the \MegaPrime centre. The halo
     displacement from the stellar centre as function of \MegaPrime position
     can well be described by $-0.022$ times the relative
     position of the star with respect to the camera centre.
   \item Finally, the masks are converted to \saoimage/\dsnine polygon
     region files which can further be processed by the \ww programme
     \citep[see][]{bem07} to construct a FLAG\_IMAGE file.
   \end{itemize}
 \item Tracks of fast moving asteroids typically show up as a series
   of high S/N, lined up, short dashed and highly elliptical objects
   in co-added \CARS images. They are present in the data because our
   strictly linear co-addition process does not include any pixel
   rejection/clipping procedure. We try to detect and mask them in our
   multi-colour object catalogues.  We identify an asteroid candidate
   if a minimum number of $N$ objects are located within $0.4$ pixels
   from a line connecting any two objects within overlapping boxes of
   \mbox{$M \times M$} pixels$^2$. We run this algorithm for the two
   parameter sets $N=4$; $M=100$ and $N=5$; $M=175$ and merge the
   resulting candidate lists. This combination was found empirically
   to give good results on the \CARS data set.  For real asteroids the
   ellipticities of contributing objects are usually highly aligned.
   As in weak lensing theory \citep[see e.g.][]{bas01} we compute the
   two-component ellipticity
   \begin{equation}
   \left( \epsilon_1, \epsilon_2 \right) = \frac{1-r}{1+r} \left( \cos 2\theta, \sin 2\theta\right)\, , 
   \end{equation}
   which depends on the object axis ratio $r$ and position angle $\theta$
   as determined by \SExtractor.  The expectation value of both
   ellipticity components is zero if the ellipticities of different
   objects are not aligned.  We then compute the alignment estimator
   \begin{equation}
    A=\sqrt{\left[\mathrm{median}(\epsilon_1)\right]^2+\left[\mathrm{median}(\epsilon_2)\right]^2 }
   \end{equation}
   from the ellipticities of all objects belonging to a candidate.  We
   only keep asteroid candidates with \mbox{$A>0.20$}; \mbox{$A>0.24$}
   (first and second parameter set)
   in order to minimise the false flagging of galaxies in areas with
   increased object number density, such as galaxy clusters.  These
   parameters were optimised for typical \CARS seeing conditions and image
   depths. For the \texttt{W4m0m0} field the algorithm automatically
   masks 30 out of 32 visually identified asteroid tracks, with one false
   positive, for an object number density of
   \mbox{$35/\mathrm{arcmin}^2$}.
\end{enumerate}
\refcomm{We note that the different algorithms are
complementary to each other.  While large-scale features such as
very large galaxies or image borders influence the object density,
small scale defects from medium bright stars (diffraction spikes;
outer extended haloes) are caught by masking known catalogue
sources.}
We independently run the object density analysis on all five colours
of a \CARS pointing. However, the stellar and asteroid track masks are
calculated for the $i'$-band only. 
The latter ones need some manual revision which is done on the basis
of the $i'$-band image only.
Hence, asteroid tracks in the $u^*g'r'z'$ bands are not included in
our object masks. \refcomm{Other problems which require
manual optimisation of the image masks are: (1) The
object density distribution analysis also masks rich galaxy
clusters. (2) Some objects labelled as stellar source in the
Standard Star Catalogues are galaxies. (3) For images
with an exceptional good seeing of $0\myarcsec 6$ or better the
high density of objects leads to a significant number of false
positives in the asteroid masking.} 
The final masks from the individual colours are merged
and collected in one \saoimage/\dsnine polygon region file. The
masking information is also transfered to our multi-colour catalogues as 
a \texttt{MASK} key which allows an easy  filtering of problematic
sources later. \figref{fig:imagemasks} shows examples of our masking 
procedure.
\section{Photometric redshifts}
\label{sec:photoz}
From the multi-colour catalogues described in the preceding section we
estimate photometric redshifts for all objects in two steps. In a
first pass we use available spectroscopic information from the
VVDS\footnote{spectroscopic data were obtained from
  \burl{http://cencosw.oamp.fr/VVDS/}} to \emph{correct/recalibrate}
our photometric zeropoints on a patch-wide basis. Afterwards we obtain
photo-$z$ estimates for our objects \citep[see][]{hwb08}.  In the
following we set the minimal photometric error to 0.1mag in order to
avoid very small purely statistical errors for high-S/N objects and to
take into account our estimated internal and external photometric
accuracies (see \sectionref{sec:phot-calibr-summ}). Throughout this
work we use \MegaPrime filter response curves which were computed by Mathias
Schultheis and Nicolas Regnault. They are available at 
\burl{http://terapix.iap.fr/forum/showthread.php?tid=136}\footnote{Note
that there are at least two more sets of \MegaPrime filter
curves available on the WWW: On the CFHT web pages 
(\burl{http://www.cfht.hawaii.edu/Instruments/Filters/megaprime.html})
and on Stephen Gwyn \MegaPipe pages 
(\burl{http://www1.cadc-ccda.hia-iha.nrc-cnrc.gc.ca/megapipe/docs/filters.html})}. 

The following analysis only includes secure VVDS objects
(marked by flags 3, 4, 23 and 24; \refcomm{in total these are 4463
objects for \texttt{W1} [up to a limiting magnitude of 
$i'_{\rm AB}\approx 24$] and 9617 for \texttt{W4} 
[up to $i'_{\rm AB}\approx 22.5$]}). \refcomm{Here and in the
following we match objects from our source lists with those from 
external catalogues if their position agrees to better than $1\myarcsec
0$.} 
First we run the new version of \texttt{Hyperz}
\citep{bmp00}\footnote{publicly available at
  \burl{http://www.ast.obs-mip.fr/users/roser/hyperz/}} on 13 fields
with overlap to the VVDS\footnote{spectroscopic data were obtained
  from \burl{http://cencosw.oamp.fr/}.}, four of which are in \texttt{W1} and nine in
\texttt{W4}. We use the CWW
template set \citep{cww80} supplied by \texttt{Hyperz} and add two
starburst templates from \cite{kcb96}.  Additionally, we fix the
redshift to the spectroscopic redshift for every object. In this way
we find the best fitting template at the spectroscopic redshift for
every object.  \texttt{Hyperz} puts out the magnitudes of the best-fit
templates and enables us to compare these to our original estimates.
We average the differences between the observed and the best-fit
template's magnitudes over all objects. In this way we derive
corrections for the zeropoints in the five bands. 
\refcomm{We only use spectra of galaxies with $i'_{\rm AB}\leq 21.5$ 
which have a high S/N photometric measurement in all filter bands; these
were 654 sources in \texttt{W1} and 2158 objects in \texttt{W4}.}
The mean and the
scatter of the corrections in the four \texttt{W1} fields are $\Delta
u^*=-0.064\pm0.015$, $\Delta g'=0.069\pm0.005$, $\Delta r'=0.027\pm0.019$,
$\Delta i'=-0.004\pm0.018$, and $\Delta z'=0.007\pm0.007$. In the nine
\texttt{W4} fields we find $\Delta u^*=-0.088\pm0.011$, $\Delta
g'=0.136\pm0.029$, $\Delta r'=0.019\pm0.03$, $\Delta i'=0.008\pm0.023$,
and $\Delta z'=-0.010\pm0.014$. Note that the photo-$z$ code is only
sensitive to colours so that the absolute values of the corrections in
the different bands should not be misunderstood as pure calibration
errors. \refcomm{Prior to the calibration step, we did not modify the
\texttt{W3} and \texttt{W4} $u^*$ zeropoints for identified
systematic calibration problems (see \appendixref{sec:phot-calibr}).
As all \texttt{W3} and \texttt{W4} fields are equally affected by it
we expect that it is taken into account properly by our correction
procedure.} We also did not apply any galactic extinction
corrections to our catalogues.  

For the \texttt{W1} fields that do not overlap with the VVDS we use
the zeropoint corrections from the field \texttt{W1p2p3}, the one with
the highest density of spectroscopic redshifts in the \texttt{W1}
region.  Since the regions \texttt{W3} and \texttt{W4} show different
$u^*$-band calibration systematics than \texttt{W1} (see
\sectionref{sec:sloancomp}), we correct all \texttt{W3} fields and the
two \texttt{W4} fields without VVDS overlap with the values from
\texttt{W4p1m1}, again the most densly covered field in this region.

Then we run \texttt{Bayesian Photometric Redshifts} \citep[\BPZ; see
][]{ben00}\footnote{publicly available at
  \burl{http://acs.pha.jhu.edu/~txitxo/}} on the catalogues with the
corrected photometry using the same template set as before.  The
Bayesian approach of \BPZ combines spectral template $\chi^2$
minimisation with a redshift/magnitude prior. \refcomm{The prior was calibrated
from HDF-N observations and the Canada-France Redshift Survey
\citep[see][]{llc95}.} It contains the probability of a galaxy
having redshift $z$ and spectral type $T$ given its apparent magnitude
$m$. A detailed description of the code and the prior can be found in
\cite{ben00}. We restrict the fitting of the photo-$z$'s to $z\leq
3.9$ due to the limited depth of the Wide data. The Bayesian redshift
estimates are added to our multi-colour catalogues. \refcomm{Note that
not all objects in our catalogues have well determined photometric
measurements in the full $u^*$ to $z'$ wavelength coverage. This can
have physical reasons (e.g.  high-redshift dropout galaxies which
are fainter than the magnitude limit in blue passpands) or it can be
connected to problems in the data itself (e.g. pixels without
information in one of the filter bands).  Our current catalogues
miss information to cleanly distinguish between these cases but only
allow us to identify problematic photometry by either large
photometric errors or a flux measurement below the formal detection
limit. In all cases with a magnitude estimate below the limiting
magnitude, or a magnitude error larger than 1mag, we configured \BPZ
to treat the object as non-detected with a flux error equal to the
$1\sigma$ limiting magnitude. This leads to unreliable results if
the large photometric error results e.g. from image defects and not
from intrinsic source properties.  To allow an easy rejection of
such problematic sources each object in our catalogues obtains
photometry quality flags for all filter bands.}

The \refcomm{internal} accuracy of the \BPZ photo-$z$'s is described by
the ODDS parameter \citep[see e.g.][]{mib04} assigning a probability
to the Bayesian redshift estimate by integrating the posterior
probability distribution in an interval that corresponds to the 95\%
confidence interval for a single-peaked Gaussian. By rejecting the
most unsecure objects with a low ODDS value one can obtain much
cleaner subsamples; see also \cite{hwb08}.

\refcomm{If not stated otherwise we use in quality assessments of
our \BPZ photo-$z$'s the following subsample of our catalogue data:
\begin{enumerate}
\item We reject all objects falling within an object mask 
(see \sectionref{sec:masking}).
\item We select galaxies by means of the \SExtractor star-galaxy
  classifier \texttt{CLASS\_STAR} and reject all sources with
  $\mbox{\texttt{CLASS\_STAR}}>0.95$.
\item We include only objects with reliable photometry in all five
  filter bands (see above). 
\item Finally we reject all sources with $\mbox{ODDS}<0.9$.
\end{enumerate}
Our catalogues contain in total 3.9 million galaxies outside an object
mask (rejection steps 1 and 2) and finally 1.45 million sources (about 
13 galaxies per sq. arcmin) with
reliable \BPZ photo-$z$ estimates (object sample after all
rejections).} 

\begin{figure}
   \centering \includegraphics[width=\columnwidth]{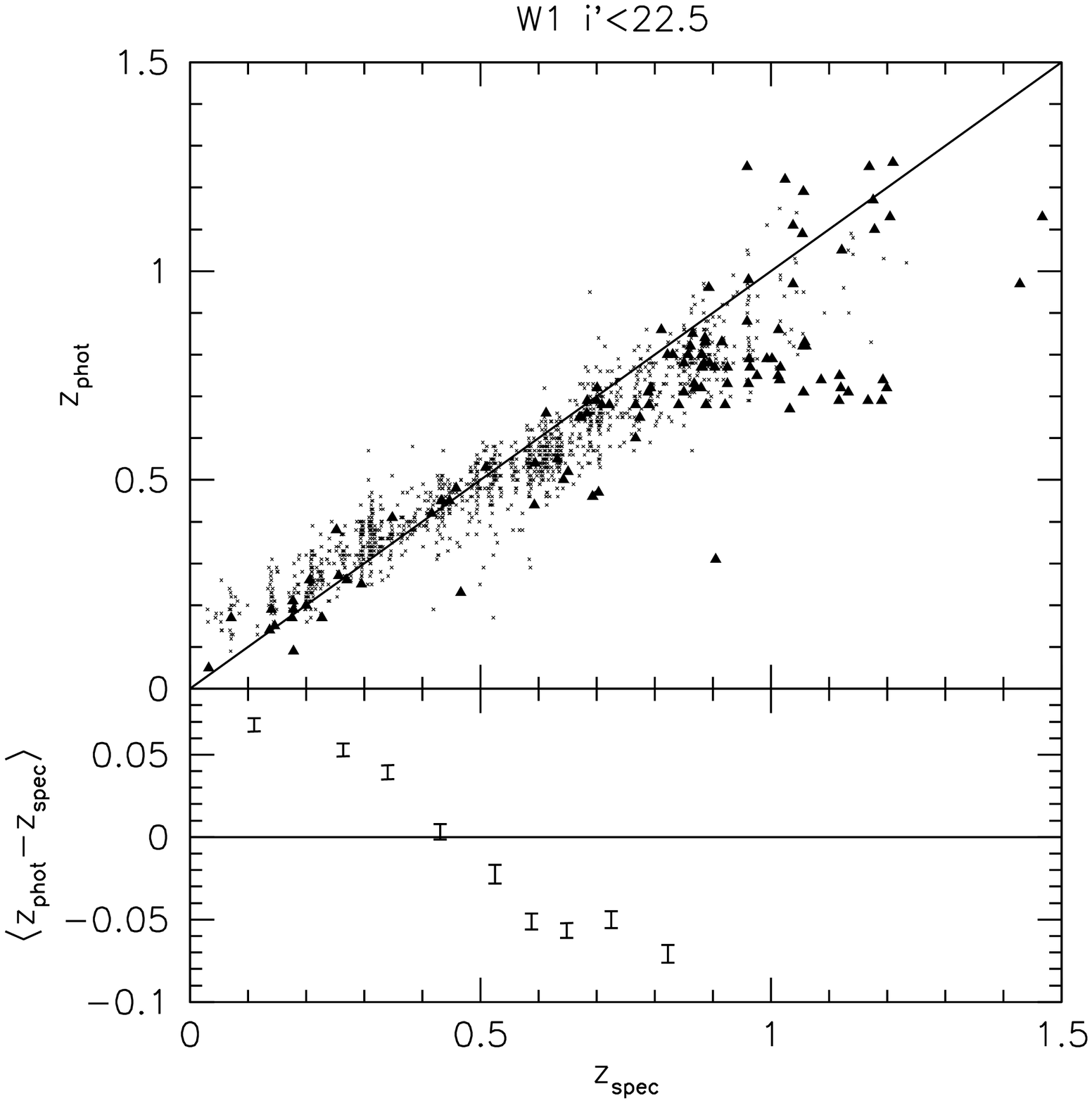}
   \centering \includegraphics[width=\columnwidth]{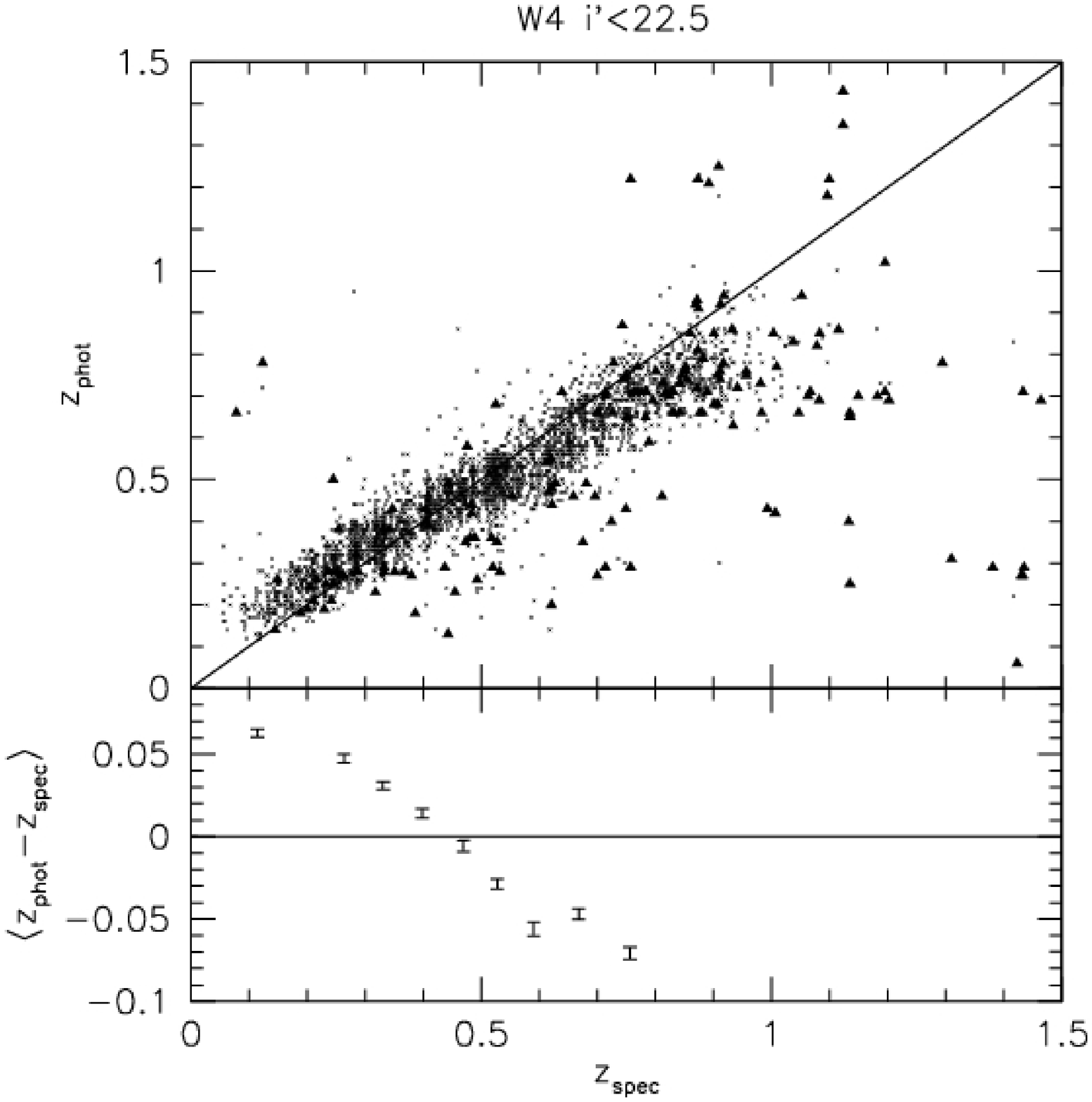}
   \caption{\label{fig:zz}Photometric vs. spectroscopic redshifts in
     the \texttt{W1} and \texttt{W4} regions: We show in the upper
     panels 1349 (\texttt{W1}) and 3312 (\texttt{W4}) galaxies with
     $i'_{\rm AB}<22.5$, reliable VVDS flags, good photometry in all
     five filter bands and $\mbox{ODDS}>0.9$ (points). Triangles
     represent galaxies with $0<\mbox{ODDS}<0.9$ (117 sources in
     \texttt{W1} and 170 objects in \texttt{W4}). Lower panels show a
     binned distribution of $\langle z_{\rm phot}-z_{\rm spec}\rangle$
     from the $\mbox{ODDS}>0.9$ samples of the upper panels.}
\end{figure}

\refcomm{We first compare our photo-$z$'s from \texttt{W1} and
\texttt{W4} to spectroscopic redshifts from the VVDS in a similar
way as presented in \cite{hwb08}.  Note that these spectra were
previously used to calibrate the data!  \tabref{tab:zz} summarises
the results indicating a homogeneous dispersion $\sigma_{\Delta
  z/(1+z)}\approx 0.04-0.05$ and an outlier rate (defined as the
percentage of galaxies with
$(z_{\mathrm{phot}}-z_{\mathrm{spec}})/(1+z_{\mathrm{spec}})>0.15$)
of 1-2\% up to $i'_{\rm AB}=24$. The $\sigma_{\Delta
  z/(1+z)}$ statistics is estimated after outliers have been rejected. 
If we perform the spectro-$z$ vs.
photo-$z$ comparisons with the $\mbox{ODDS}>0.0$ sample (but all
other filters as described above) the dispersion is nearly unchanged
while the outlier rate rises by a factor 3 to 8. This confirms that
the ODDS parameter is a good selection criterion to reject outliers
and to obtain samples of homogeneous photo-$z$ quality up to about
$i'_{\rm AB}\approx 24$. A plot of the photo-$z$ vs. VVDS
spectro-$z$ results in the regions \texttt{W1} and \texttt{W4} is
shown in \figref{fig:zz}.}
\begin{table}
\begin{minipage}[t]{\columnwidth}
  \caption{\label{tab:zz} Statistics of the comparison between
    photometric and spectroscopic VVDS redshifts:
    The third column
    gives the number of uniquely matched sources between our
    catalogues and high-confidence VVDS objects (see text).
    The completeness value (fourth column) lists from those objects
    the percentage with a high-confidence \BPZ photo-$z$ estimate
    ($\mbox{ODDS}>0.9$).}
  \centering
  \renewcommand{\footnoterule}{} 
\begin{tabular}{lrrrrr}
\hline
\hline
\multicolumn{1}{c}{Field} & \multicolumn{1}{c}{$m_{\rm lim}$}
      & \multicolumn{1}{c}{$N$} & \multicolumn{1}{c}{compl.}
      & \multicolumn{1}{c}{outl. rate $\eta$\footnote{defined as the
          percentage of galaxies with $(z_{\mathrm{phot}}-z_{\mathrm{spec}})/(1+z_{\mathrm{spec}})>0.15$}}
      & \multicolumn{1}{c}{$\Delta z/(1+z)$\footnote{bias and scatter of $(z_{\mathrm{phot}}-z_{\mathrm{spec}})/(1+z_{\mathrm{spec}})$ after outlier rejection}} \\
 & \multicolumn{1}{c}{[AB]} & \multicolumn{1}{c}{[\%]} & & \multicolumn{1}{c}{[\%]} & \\
\hline
\texttt{W1p2p2} & 22.5 & 212  & 91.51 & 1.03 & $0.000\pm0.052$\\
                & 24.0 & 517  & 73.11 & 1.85 & $-0.002\pm0.050$\\
\texttt{W1p2p3} & 22.5 & 1136 & 92.52 & 0.86 & $-0.008\pm0.051$\\
                & 24.0 & 2456 & 77.69 & 1.62 & $-0.011\pm0.049$\\
\texttt{W1p3p2} & 22.5 & 14   & 92.86 & 0.00 & $0.009\pm0.040$\\
                & 24.0 & 24   & 70.83 & 0.00 & $0.011\pm0.047$\\
\texttt{W1p3p3} & 22.5 & 104  & 87.50 & 2.20 & $0.004\pm0.046$\\
                & 24.0 & 257  & 65.37 & 1.79 & $0.009\pm0.050$\\
\hline                                                 
\texttt{W4m0m0} & 22.5 & 223  & 93.27 & 1.44 & $-0.015\pm0.045$\\
\texttt{W4m0m1} & 22.5 & 354  & 94.63 & 2.69 & $0.004\pm0.049$\\
\texttt{W4m0m2} & 22.5 & 132  & 98.48 & 0.77 & $-0.001\pm0.048$\\
\texttt{W4p1m0} & 22.5 & 395  & 92.91 & 0.82 & $0.006\pm0.045$\\
\texttt{W4p1m1} & 22.5 & 908  & 95.70 & 1.96 & $-0.010\pm0.051$\\
\texttt{W4p1m2} & 22.5 & 416  & 95.19 & 1.26 & $0.001\pm0.051$\\
\texttt{W4p2m0} & 22.5 & 274  & 94.89 & 0.77 & $-0.013\pm0.045$\\
\texttt{W4p2m1} & 22.5 & 517  & 96.52 & 1.00 & $-0.006\pm0.051$\\
\texttt{W4p2m2} & 22.5 & 263  & 94.30 & 0.40 & $-0.007\pm0.050$\\
\end{tabular}
\end{minipage}
\end{table}
While the figure shows an overall good performance of our photo-$z$
estimation it reveals residual systematics. A significant tilt is
present in the $z_{\mathrm{phot}}$ vs. $z_{\mathrm{spec}}$ comparison
leading to a systematic overestimation of up to $0.1-0.2$ of the
redshift at low $z_{\mathrm{spec}}$ and to a underestimation at high
$z_{\mathrm{spec}}$.  \refcomm{The tilt crosses the zero axis at
$z=0.5$ and hence it cancels negative and positive contributions to
statistics involving $\Delta z$ (see \figref{fig:zz}). The $\Delta
z/(1+z)$ statistics for the complete \Wone ($N=1466$) and \Wfour
($N=3488$) samples are: $\Delta z/(1+z) = -0.006\pm 0.051$ (\Wone)
and $\Delta z/(1+z) = -0.005 +/- 0.050$ (\Wfour). If we split the
sample at $z=0.5$ we obtain for $z<0.5$: $\Delta z/(1+z) = 0.03\pm
0.043$ (\Wone: $N=595$) and $\Delta z/(1+z) = 0.025\pm 0.037$
(\Wfour: $N=1690$). Accordingly for $0.5<z<1.0$: $\Delta z/(1+z) =
-0.032\pm 0.035$ (\Wone: $N=809$) and $\Delta z/(1+z) = -0.034\pm
0.038$ (\Wone: $N=1728$).}

\refcomm{We do not try to remedy these systematics in this article but
we will investigate it in a companion paper (Hildebrandt et al. in
prep.). The overestimation at the low-$z$ is mainly caused by the
redshift prior in \texttt{BPZ}. It seems to give too little
probability to the low-$z$ population in the \CARS data. A
modification of the original prior in this sense removes the
observed bias for $0<z<0.5$.  The high-$z$ underestimation of our
redshifts can be corrected by a recalibration of the original
\citet{cww80} and \citet{kcb96} template sets; see also
\citet{fcp06}.}  In the following we further check the consistency
and quality of our current photo-$z$ sample.

\subsection{Internal and external quality checks on our photo-$z$ sample}
\label{sec:intern-extern-qual}

\begin{figure}
   \centering \includegraphics[width=\columnwidth]{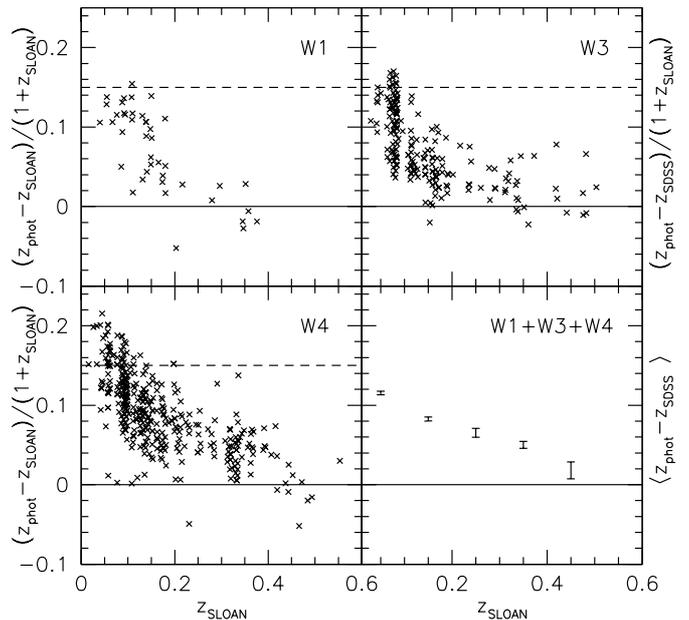}
   \caption{\label{fig:sloanzcomp} Comparison of \CARS BPZ photo-$z$
     against SDSS spectra ($0<z<0.6$): The plot shows $N=44$ objects 
     for \texttt{W1},
     $N=208$ for \texttt{W3} and $N=400$ for \texttt{W4}.
     Note the different ordinate in the lower right
     panel!}
\end{figure}

\begin{figure}
   \centering \includegraphics[width=0.85\columnwidth]{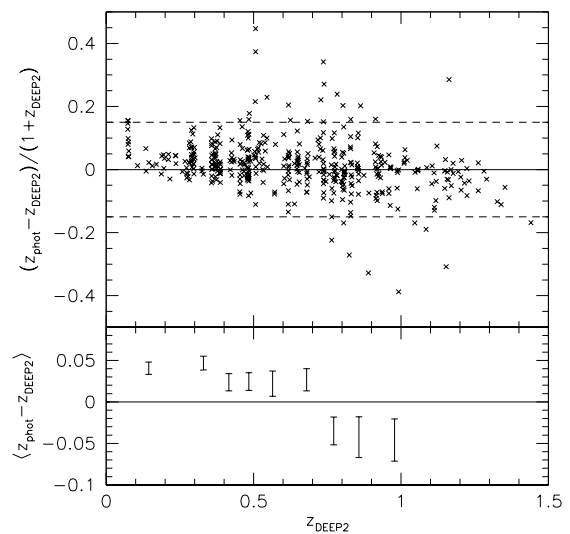}
   \caption{\label{fig:DEEPz} Comparison of \CARS \BPZ photo-$z$
     against DEEP2 spectra for $0<z<1.5$: shown are $N=448$ common
     objects between the DEEP2 redshift survey and the \CARS fields 
     \texttt{W3m1m2} and \texttt{W3m1m3}.}
\end{figure}

\begin{figure}
   \centering \includegraphics[width=\columnwidth]{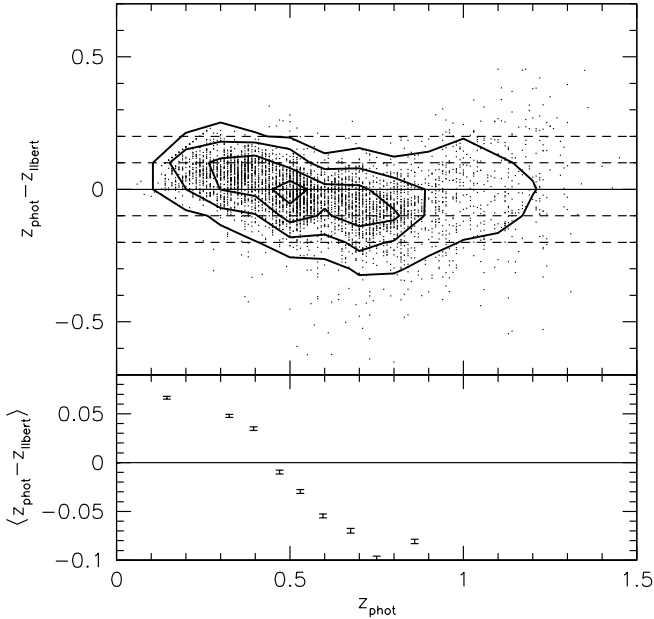}
   \caption{\label{fig:ilbertz} Comparison of our \BPZ photo-$z$
   estimates on \texttt{W1p2p3} with those from \citet{iam06} on the
   CFHTLS-Deep \texttt{D1} field. The comparison includes 24558
   common objects with a pre-filtering for high-confidence sources in
   both catalogues and $i'_{\rm AB}\leq 24$ (see text for details).
   $853$ objects ($3.89\%$) lie outside the plotting region of 
   $-0.7<z_{\rm phot}-z_{\rm Ilbert} < 0.7$.
   We show one point out of five for clarity of
   the plot; contours indicate areas of 0.8, 0.5, 0.25 and 0.05 times
   the peak-value of the point-density distribution.}
\end{figure}

Besides with the VVDS the \CARS patches overlap with public spectra from the
SDSS\footnote{spectroscopic data were obtained from
  \burl{http://cas.sdss.org.astro/en/tools/search/SQS.asp}} 
(\texttt{W1-W4}) and the DEEP2\footnote{spectroscopic data were obtained from
  \burl{http://deep.berkley.edu/DR3}} redshift survey
\citep[\texttt{W3}; ][]{dgk07}. Hence, we can test our photo-$z$s which
were partly calibrated and verified against the VVDS with an
independent set of spectroscopic data. The comparisons are shown in 
\figref{fig:sloanzcomp} and \figref{fig:DEEPz}; formal quality
  parameters are listed in \tabref{tab:sdssdeep2}.  
\begin{table}
  \caption{Quality parameters for the comparison of \CARS \BPZ
    photo-$z$ against SDSS and DEEP2 spectra. The third column gives
    the redshift range probed by the spectroscopic samples; see
    \tabref{tab:zz} for the meaning of the other columns.}
\label{tab:sdssdeep2}      
\centering          
\begin{tabular}{l|lllll}    
\hline\hline       
\multicolumn{1}{c|}{Field} & \multicolumn{1}{c}{Survey} &
\multicolumn{1}{c}{$z$-coverage} &
\multicolumn{1}{c}{$N$} & \multicolumn{1}{c}{$\eta$ [\%]} &
\multicolumn{1}{c}{$\Delta z/(1+z)$}\\ 
\hline
\texttt{W1} & SDSS  & $0 < z < 0.6$ & 44   & 2.4  & $0.069\pm 0.050$\\ 
\texttt{W3} & SDSS  & $0 < z < 0.6$ & 208  & 4.8  & $0.068\pm 0.042$\\ 
\texttt{W4} & SDSS  & $0 < z < 0.6$ & 400  & 12.8 & $0.080\pm 0.039$\\
\hline
\texttt{W3} & DEEP2 & $0 < z < 1.5$ & 448 & 7.3  & $0.010\pm 0.050$\\  
\hline                  
\end{tabular}
\end{table}
\refcomm{We observe exactly the same systematics identified in the 
comparisons with the VVDS spectra: A systematic tilt with
overestimates of about $0.05-0.15$
in the low-$z$ regime and a reverse trend for $z>0.5$.}
Our tests indicate that the photo-$z$ quality and remaining
systematics for the current \CARS data set are comparable in the mean
for all fields; regardless whether the galaxies profited directly from
a previous calibration with spectra or whether we transfered
corrections obtained with a galaxy subset to other fields.  The
trends we see with our \BPZ photometric redshifts also show up in
a comparison with previously obtained photo-$z$ estimates with the
photometric redshift code \texttt{LePhare}\footnote{see
  \burl{http://www.oamp.fr/people/arnouts/LE_PHARE.html}} on the
CFHTLS-Deep field \texttt{D1} \citep[see][]{iam06}. We show a direct
comparison of the \texttt{D1} and our \texttt{W1p2p3} photometric
redshift estimates in \figref{fig:ilbertz}. \refcomm{The figure confirms an
overall very good agreement and a systematic tilt for $0<z<1$ in our 
estimates; our high-confidence \BPZ photo-$z$ sample with 
$i'_{\rm AB}\leq 24$ has 24558 common
objects with the \citet{iam06} catalogue. The latter was filtered
for $(z_{\rm up 1\sigma}-z_{\rm inf 1\sigma})/(1+z_{\rm Ilbert}) <
0.25$. For the complete common sample with $i'_{\rm AB}\leq 24$
we measure $\Delta z=(z_{\rm phot}-z_{\rm Ilbert})=-0.02\pm
0.11$.}

\begin{figure}
   \centering \includegraphics[width=\columnwidth]{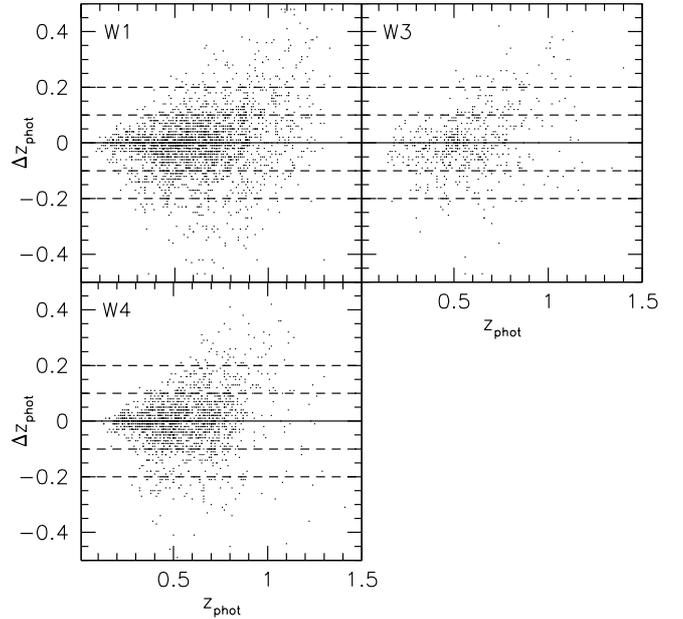}
   \caption{\label{fig:photozoverlap} Consistency of our photometric
     redshift estimates: We compare independent photo-$z$ measurements
     from objects appearing in multiple \CARS pointings in each patch
     (\Wone: $N=24329$; \Wthree: $N=6211$; \Wfour: $N=16849$).
     Dashed lines mark regions with
     $\Delta z_{\rm phot}=0.1, 0.2$. Only 1 out of 10 points is shown
     for clarity of the plots. See the text for further details.}
\end{figure}

\begin{figure}
   \centering \includegraphics[width=\columnwidth]{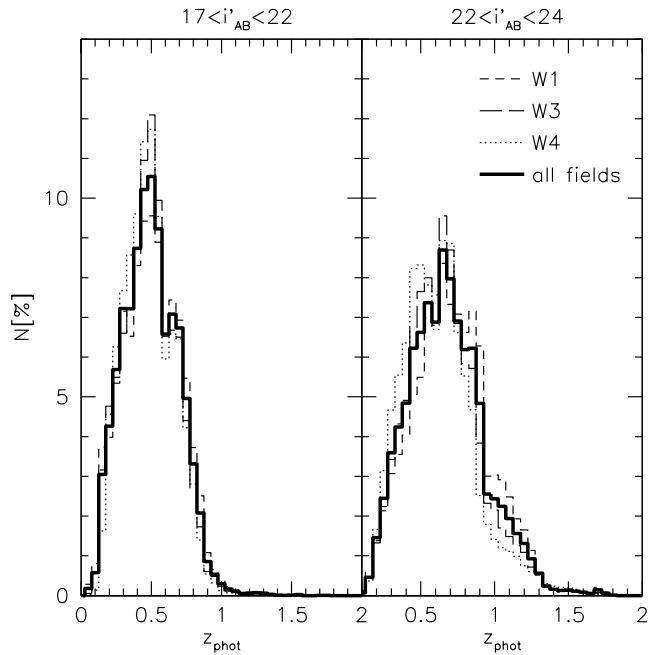}
   \caption{\label{fig:zdist} Normalised distributionw of our
     high-confidence photometric redshift estimates for all \CARS
     patches (\Wone: $N=205956$ for $17<i'_{\rm AB}<22$ and 
     $N=487593$ for $22<i'_{\rm AB}<24$; 
     \Wthree: $N=52295$ for $17<i'_{\rm AB}<22$ and 
     $N=119589$ for $22<i'_{\rm AB}<24$; 
     \Wfour: $N=104417$ for $17<i'_{\rm AB}<22$ and 
     $N=252969$ for $22<i'_{\rm AB}<24$); 
     all distributions have only very few objects beyond 
     redshift 2 (not shown).}
\end{figure}

Finally we perform two internal consistency checks on our estimates.
The first one is a comparison of independent estimates from overlap 
objects in different \CARS pointings (see
\figref{fig:fieldlayout}). \refcomm{In \figref{fig:photozoverlap} we show
on a patch basis $\Delta z_{\rm phot}=(z_{\rm phot 1}-z_{\rm phot 2})$ 
for all overlap sources. The means and scatters of this quantity for 
individual patches are: $\Delta z_{\rm phot}=0.0002\pm 0.0772$ (\texttt{W1};
24329 objects),  $\Delta z_{\rm phot}=-0.00005\pm 0.0767$
(\texttt{W3}; 6211 objects) and $\Delta z_{\rm phot}=-0.0006\pm
0.0723$ (\texttt{W4}; 16849 objects).
The plot and the numbers
demonstrate a homogeneous and robust redshift estimation over the
whole \CARS area.} Note
that this comparison includes areas with and without spectroscopic 
calibration overlap.
The second internal consistency check are the redshift distributions
for the three \CARS patches. \figref{fig:zdist} shows that they are
very comparable for all three \CARS patches.

\subsubsection{Angular cross-correlation of galaxy populations in
  different photo-$z$ redshift bins}
\label{sec:photozcorrel}
In the following we apply a correlation function analysis 
to further quantify the
reliability and suitability of photometric redshift estimates. 
The theoretical background will be detailed in Benjamin et al. (in prep.).
Judging photo-$z$ quality by calibration with external spectroscopy
provides us with an overall picture of the dispersion of our estimates
and the total rate of \emph{catastrophic outliers}. First, this
quality control is limited to the magnitude and/or redshift range of
our external comparison sample and second we often need
a more detailed picture on the photo-$z$ (re)distribution of galaxies in
redshift space. For instance, weak lensing tomography studies of the
cosmic shear effect do not require a precise redshift estimate for
each individual galaxy. However, we need to reliably separate galaxies
into redshift bins and we require a precise understanding of our errors
concerning redshift misidentifications and inter-bin contamination
factors \citep[see e.g.][]{htb06}.

To quantify the crucial error contribution due to large systematic redshift
misidentifications we investigate the angular cross-correlation
function of galaxies in different photo-$z$ slices. The
cross-correlation technique has already been advocated as a way to
reconstruct the source redshift distribution \citep[][]{skz06}, or, in
combination with spectroscopic redshift, as a way to improve the
photo-$z$ calibration \citep[see][]{msh07}.

\begin{figure*}
\centering
\includegraphics[width=18.25cm]{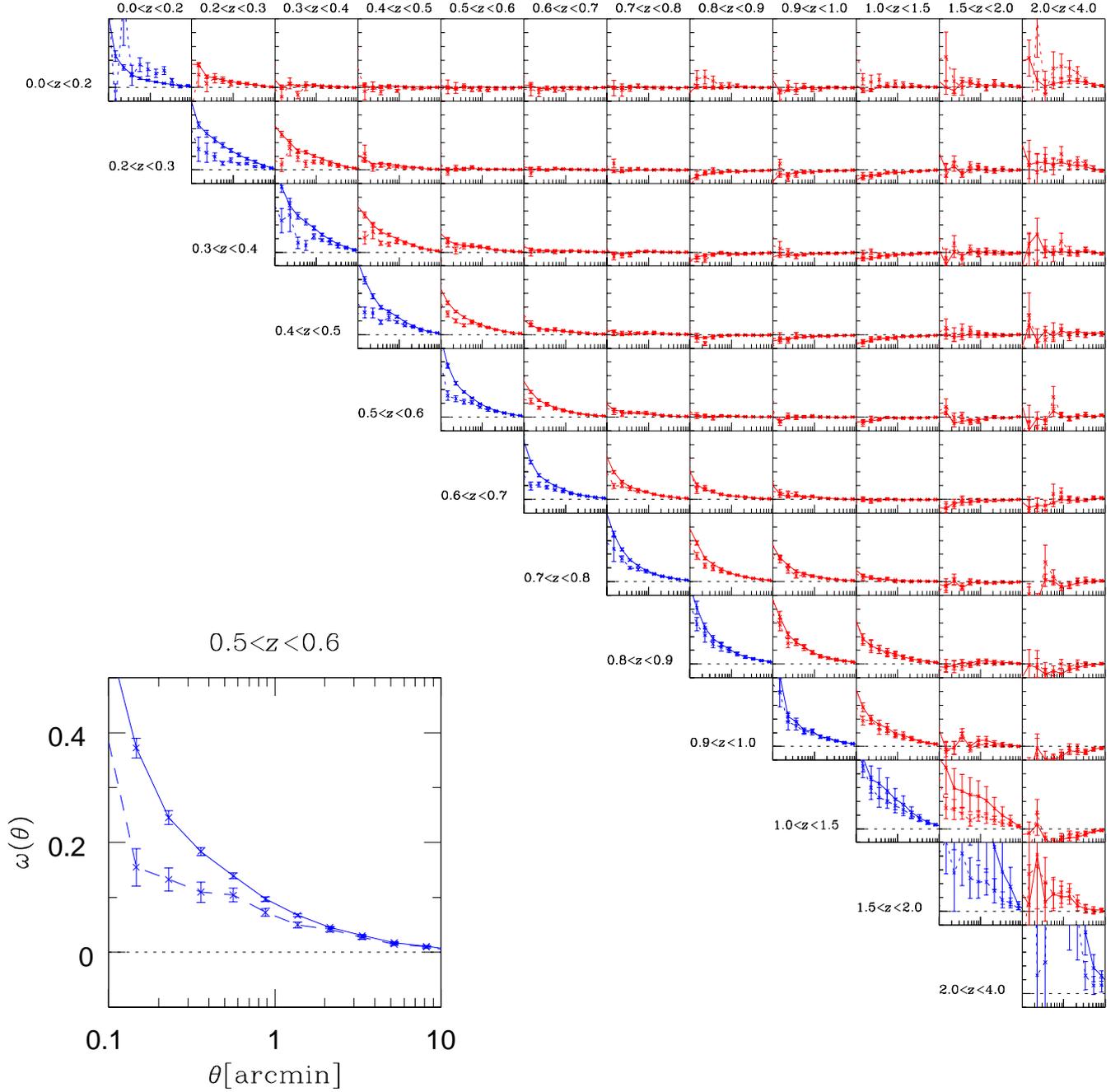}
\caption{\label{fig:cross}The angular cross-correlation of objects in
  different photo-$z$ slices (\emph{solid} lines represent objects
  with $i'_{\rm AB}<24$ and \emph{dashed} lines represent objects with
  $23<i'_{\rm AB}<24$): The matrix plot represents the complete \CARS area and
  each panel shows the qualitative behaviour of auto- (diagonal) and
  crosscorrelation (off-diagonal) measurement for different redshift
  slices. The rows represent bins from $z=0$ to $z=3.9$ (steps
  $z=0.0;0.2;0.3;0.4;0.5;0.6;0.7;0.8;0.9;1.0;1.5;2.0;3.9$). 
  The lower left panel is a zoom-in to the corresponding panel of the
  matrix plot. Note that in the cross-correlation analysis we include
  all objects with $\mbox{ODDS}>0.8$. See text for further details.}
\end{figure*}

For accurate redshift estimates we expect to see a strong
auto-correlation in individual redshift bins and, due to the expected
photo-$z$ scattering ($\sigma_{\Delta z}\approx 0.05-0.1$), a
lower-level cross-correlation signal in neighbouring slices.  But one
would not expect to see a cross-correlation signal for slices that are
physically far apart. Contamination by catastrophic photo-$z$ failures
would lead to significant amplitudes of the angular cross-correlation
function of photo-$z$ slices that are well separated in redshift. In
\figref{fig:cross} we show a correlation slice-analysis of the \CARS
photo-$z$ estimates for objects with $i'_{\rm AB}<24$. It represents a
matrix plot including the angular cross-correlation functions of all
pairs of photo-$z$ slices.  The figure illustrates the expected
behaviour: We observe a very significant autocorrelation and a
decreasing cross-correlation in neighbouring bins. Our analysis also
shows a decent signal for the highest redshift bins ($z\geq 1.5$) with
low-$z$ slices.  This shows that high-$z$ tails, which are often
observed in redshift distributions derived from photo-$z$s, are
populated with true low-$z$ galaxies.  We note that the
$\mbox{ODDS}>0.9$ filtering applied hitherto rejects large parts of
dubious sources with a very broad or doubly peaked photometric
redshift probability distribution. To visualise the effect of low-$z$
high-$z$ contamination we relaxed our filtering to $\mbox{ODDS}>0.8$
in the cross-correlation analysis.

In \figref{fig:cross} we show separately the correlation functions of
the faintest objects with $23<i'_{\rm AB}<24$. \refcomm{We see that the faint
population of our galaxies behaves exactly in the same way as the
complete sample.}
This indicates that the photo-$z$ accuracy does
not degrade in the lower S/N regime.

While \figref{fig:cross} already allows us to draw important
qualitative conclusions we can derive formulae for the mutual
contamination of redshift bins. A complete matrix description of the
formalism, which will allow a consistent analysis of the results in
\figref{fig:cross}, will be presented elsewhere (Benjamin et al., 
in prep.). In this paper we limit the
discussion to a strict pairwise cross-correlation analysis, i.~e.  we
present quantitative results only for cases where the whole
redshift sample is split in exactly two bins.

The basic idea is to infer the degree of contamination between
two redshift slices from the measurement of the cross-correlation
$w_{12}$ between the bins. 

The fraction of galaxies from bin $1$ present in bin $2$, as a fraction
of the true number of galaxies in bin $1$ ($\NiT$), is $\fij$.
Likewise the fraction of galaxies in bin $1$ from bin $2$ is given by
$f_{21}$, which is defined such that the number of galaxies present in
bin $1$ from bin $2$ is given by $\NjT \fji$. Hence the observed number
of galaxies in each bin ($N^{\rm o}$) can be expressed as,
\begin{eqnarray}
 \Nio&=&\NiT(1-\fij) + \NjT\fji,\nonumber\\
 \Njo&=&\NjT(1-\fji) + \NiT\fij. \label{Nos}
\end{eqnarray}
\noindent The first term of each equation accounts for those
galaxies that do not leave the given bin, the second term accounts
for those interloping galaxies from the other bin. Inverting these
equations, the true number of galaxies can be expressed as a
function of the observed numbers and the fractions $\fij$ and
$\fji$,
\begin{eqnarray}
 \NiT&=&\frac{\Nio-\fji(\Nio+\Njo)}{1-\fij-\fji},\nonumber\\
 \NjT&=&\frac{\Njo-\fij(\Nio+\Njo)}{1-\fij-\fji} \label{eq:NTs}.
\end{eqnarray}
\noindent Note that $\Nio + \Njo = \NiT + \NjT$, thus the total number
of galaxies is preserved, as should be the case. It is also obvious
that for cases where $\fij + \fji$ is unity there is a zero in the
denominator. What is less clear, is that in these cases the numerator
is also zero, which can be seen by plugging Eq.~\ref{Nos} into the
numerator. In this case the system of equations is degenerate, and
will not admit a unique solution. This should not pose a practical
limitation since it is expected that the fractional contamination
between bins is small, and specifically less than 0.5. 

In order to calculate how the cross-correlation function is changed
for non-vanishing coefficients $f_{12}, f_{21}$, it is sufficient to
consider the natural estimator of the angular correlation function,
as opposed to that presented by \citet{las93}. The natural
estimator works well at small and intermediate scales where edge
effects are not an issue, provided that there is a sufficient
density of points \citep[see][for a comparison of
the estimators]{kss00}. The observed angular cross correlation functions are
given by,
\begin{eqnarray}
1 + \wiio&=&\frac{\Dio}{\RR}, \\
1 + \wijo&=&\frac{\Dijo}{\RR}, \label{angcorr}
\end{eqnarray}
\noindent where $\Dio$ is the observed number of pairs separated by
angle $\theta$ within bin $1$, similarly $\Dijo$ is the number of
pairs between bins $1$ and $2$, and $\RR$ is the number of pairs between
objects from random fields of identical geometry.

Considering how galaxy pairs are split between the two bins $1$ and
$2$, one can show that the observed number of pairs depends on a
combination of the true number of pairs and the contamination
fractions:

\begin{eqnarray}
\Dijo&=&\DijT((1-\fij)(1-\fji) + \fji\fij) \nonumber\\
&&+\DiT(1-\fij)\fij\nonumber\\
&&+\DjT\fji(1-\fji).
\end{eqnarray}
\noindent Plugging this relation into \eqref{angcorr}, and noting
that the term $\Dijo/\RR$ must be normalised by 
$N^{\rm R}_1N^{\rm R}_2/\Nio\Njo$,
where $N^{\rm R}_{1,2}$ is the number of objects in the random samples, the
following equation can be derived for the observed angular
cross-correlation function,
\begin{eqnarray}
1+\wijo&=&(1+\wiiT)\frac{(\NiT)^2}{\Nio\Njo}\fij(1-\fij)\nonumber \\
&&+(1+\wjjT)\frac{(\NjT)^2}{\Nio\Njo}\fji(1-\fji)\nonumber \\
&&+(1+\wijT)\frac{\NiT\NjT}{\Nio\Njo}(1-\fij-\fji+2\fij\fji).\label{eq:wijo}
\end{eqnarray}
\noindent Note that the observed cross-correlation function depends on
the unknown true number of galaxies in the bins and the unknown true
auto-correlation function. The true galaxy number can be expressed in
terms of the observed number of galaxies and the contamination
fractions via \eqref{eq:NTs}.  It is possible to express the true
auto-correlation as functions of contamination fractions, the
number of observed galaxies and the observed auto-correlation
functions (Benjamin et al., in prep),
\begin{eqnarray}
\wooT&=&\wooo\left(\frac{\Noo}{\NoT}\right)^2\frac{(1-\fwo)^2}{(1-\fow)^2(1-\fwo)^2-\fow^2\fwo^2}\nonumber\\
&&-\wwwo\left(\frac{\Nwo}{\NoT}\right)^2\frac{\fwo^2}{(1-\fow)^2(1-\fwo)^2-\fow^2\fwo^2}\nonumber\\
&&-\wowT\left(\frac{\NwT}{\NoT}\right)\frac{2\fwo(1-\fwo)}{(1-\fow)(1-\fwo)+\fow\fwo}.\label{eq:wiiT}
\end{eqnarray}
By exchanging $1$ and $2$ in \eqref{eq:wiiT} an equivalent expression 
for the auto-correlation of bin $2$ is obtained. To finally use
eqs. (\ref{eq:wijo}) and (\ref{eq:wiiT})
we make the explicit assumption that the true
cross-correlation between the two redshift bins is zero ($\wijT=0$),
i.e. all the observed cross-correlation is due to contamination.  This
prescription allows us to use the observed correlation functions and
number of galaxies to determine the contamination fractions, $\fij$
and $\fji$, for a pair of redshift bins.


We note that the outlined formalism cannot be trivially extended
to a multi-bin setup, since it assumes a pair of bins and
ignores possible contamination from other redshifts. However, it
already allows us to recover the fraction of objects that cross a given
redshift $z_{\rm cut}$ due to photometric redshift errors, and an analysis can
be done as a function of $z_{\rm cut}$.

\begin{figure*}
\centering
\includegraphics[width=18.25cm]{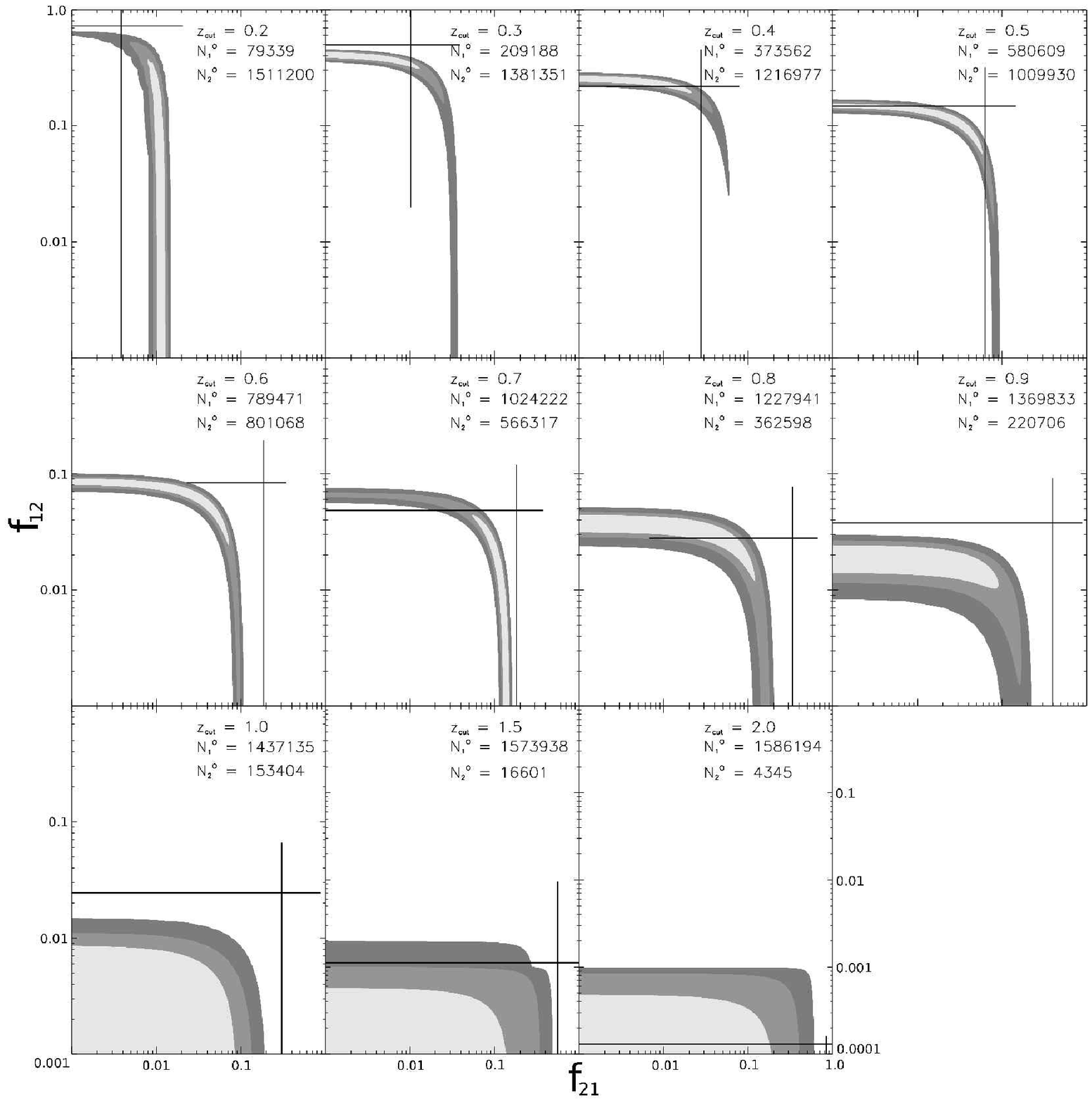}
\caption{\label{fig:pairwise}The estimated contamination fractions,
  $\fij$ and $\fji$ resulting from a strict pairwise cross-correlation
  analysis. The two redshift bins are defined to be $0.0 < z_{\rm 1} <
  z_{\rm cut}$ and $z_{\rm cut} < z_{\rm 2} < 4.0$. Contours indicate
  the 1, 2 and 3-sigma confidence regions, having progressively darker
  shades of grey.  Legends in the figure give the observed number of
  galaxies in each bin. The data points
  with error bars denote the measured contamination fraction found for
  those galaxies in the VVDS \texttt{W1} spectroscopic sample (see
  text for details).  Note that for $z_{\rm cut}=1.5$ and
  $z_{\rm cut}=2.0$ an extended vertical scale is used in order to
  show the measured contamination.}
\end{figure*}

We apply the pairwise analysis on our data by cutting it at $z_{\rm
  cut}=0.2;0.3;0.4;0.5;0.6;0.7;0.8;0.9;1.0;1.5;2.0$ yielding a low
redshift bin $0.0 < z_{1} < z_{\rm cut}$ and a high redshift bin
$z_{\rm cut} < z_{2} < 4.0$.  The angular auto and cross-correlation
functions from the pair of bins are used to estimate the contamination
fractions $\fij$ and $\fji$ by fitting the observed cross-correlation
with \eqref{eq:wijo}.  \refcomm{The analysis was performed with eleven
equally spaced cross-correlation bins ranging from $0\myarcmin 9$ to
$10\myarcmin 0$. We checked with an analysis of three and five bins
that our results do not depend significantly on this choice.}  This
step is followed by a minimum chi-square analysis and the likelihood 
contours in the contamination fraction parameter space are presented in
\figref{fig:pairwise}. The degeneracy between the two contamination
fractions is clearly evident, and lower and upper limits can be
estimated.

For the lowest redshift cuts there is a strong degeneracy, with $0.0 <
\fij < 0.6$, and $\fji \sim 0.01$, hence a potentially large fraction
of low redshift galaxies are expected to be at higher redshift.
Likewise for the highest redshift cuts a potentially large fraction of
high redshift galaxies are expected to be at lower redshifts.  It is
important to reiterate that the fraction $\fij$ is the
number of galaxies that move from bin $1$ to bin $2$ as a fraction of
the true number of galaxies in bin $1$. Hence, the observed large
degeneracies in the low and high redshift cuts are a consequence of
the very different occupation numbers in the two bins. For the
intermediate redshift cut range $0.4 < z_{\rm cut}< 0.9$ the 
contamination factors are around $10\%-20\%$.

\refcomm{The spectroscopic data in \Wone allow us to directly calculate
contamination fractions $\fij$ and $\fji$ for the field
\texttt{W1p2p3} and to check whether our estimates obtained via
correlation functions are reasonable. The VVDS in this field has the
same depth as the limiting magnitude of our correlation analysis
sample ($i'_{\rm AB}\leq 24.0$). We subdivide galaxies according to
spectroscopic (true) redshifts and determine directly contamination 
fractions with the photometric redshifts. Error contributions on 
this quantity are the Poisson noise and the redshift sample variance
in a field of 1 sq. degree. The latter was estimated in \citet{vwh06}
to be 15 times the Poisson contribution. We plot our results in
\figref{fig:pairwise}. We see that the contamination fractions
determined with spectroscopic redshifts in one field are very
comparable to the correlation function estimates for the whole 
\CARS surveys. This directly shows the validity of our analysis.}
%
\section{Available data products}
\label{sec:releaseproducts}
We make available on request our multi-colour catalogues including
photo-$z$ estimates of 37 \CARS fields (corresponding to $\sim
30\mathrm{deg}^2$ effective survey area after image masking).  The
data package includes object catalogues, the derived image masks and
\texttt{JPEG} images to inspect colour data and extracted sources.
The catalogues are available as \texttt{FITS} binary tables and a
subset of the most important entries as \texttt{UNIX-ASCII} text
files.

The \texttt{FITS} catalogue version includes most of the original
\SExtractor keywords and for their meaning we refer to the \SExtractor
manual \citep{ber03}. All these basic keys are measured in a
\SExtractor run in dual-image mode where we use the unconvolved
$i$-band image for detection as well as for photometric measurements
(see Sect.~\ref{sec:obj_det}).

Additional keywords created in subsequent \SExtractor runs with the
PSF-matched images in the five bands for photometric colour
measurements are indicated by an additional $\_\mathrm{x}$ where
$\mathrm{x} \in \left[ u^*,g',r',i',z'\right]$. In particular, these are the
different kinds of fluxes, magnitudes and corresponding error
estimates (e.g. \texttt{FLUX\_AUTO\_x}, \texttt{FLUXERR\_AUTO\_x},
\texttt{MAG\_ISO\_x}, \texttt{MAGERR\_ISO\_x}, etc.); \refcomm{note
  that magnitude error estimates in the catalogues do not take into
  account systematic zeropoint offsets but only statistical errors due
  to photon noise.}  We estimate 24 different aperture fluxes and
magnitudes with diameters ranging from 4 to 55 pixels ($\hat=
0\farcs744$ to $10\farcs23$). We add the $1\sigma$ limiting magnitudes
\texttt{MAG\_LIM\_x} as described in \sectionref{sec:mag_lim}. All
magnitudes are provided in \MegaPrime instrumental AB magnitudes. We
note that we did not apply any magnitude correction to the catalogue
entries also if our tests performed in \sectionref{sec:phot-calibr}
might justify them. This especially applies for discrepancies present
in the $u^*$-band calibration of the \texttt{W3} and \texttt{W4}
pointings (see \sectionref{sec:phot-calibr}). \refcomm{To allow an
easy identification of objects with problematic photometry we add
the flags \texttt{NBPZ\_GOODFILT} indicating the number of filters
with reliable photometry, \texttt{NBPZ\_BADFILT} giving the number
of filters with $\mbox{\texttt{MAGERR\_ISO\_x}} \geq 1.0$ and
\texttt{NBPZ\_LIMFILT} listing the number of filters with
\texttt{MAG\_ISO\_x} fainter than our formal magnitude limit (see
\sectionref{sec:CARSdata}). Which of these three properties applies
to which filters is encoded in additional keys.}

Furthermore, we provide a global mask key \texttt{MASK} which is 0 for
objects that do not lie inside one of our object masks and 1
otherwise. This key takes into account all masks from our object
density, bright star and asteroid track analyses as described in
\sectionref{sec:masking}.

Finally, the catalogues contain photo-$z$ relevant quantities from the
output of \BPZ. Besides the Bayesian redshift estimate, \texttt{Z\_B},
we include the \texttt{ODDS} probability, the SED corresponding to the
Bayesian redshift (\texttt{T\_B}), the corresponding $\chi^2$, the
95\% confidence interval (\texttt{Z\_B\_MIN} and \texttt{Z\_B\_MAX})
as well as the maximum-likelihood redshift and type estimate
(\texttt{Z\_ML} and \texttt{T\_ML}), which are put out by \BPZ before
the prior is applied.

The most important catalogue entries are summarised in \tabref{tab:keys}.

\begin{table*}
\begin{minipage}[t]{\textwidth}
  \caption{Description of the most important \texttt{FITS} keys in the
    \CARS multi-colour catalogues. The \texttt{ASCII} catalogue
    version contains one aperture magnitude at a diameter of
    $1\myarcsec 86$. The \texttt{FITS} version lists 24 aperture
    magnitudes for diameters from $0\myarcsec 744$ to $10\myarcsec
    23$.}
\label{tab:keys}
\centering
\renewcommand{\footnoterule}{}  
\begin{tabular}{lllc}
  \hline
  \hline
  key name & description & measured on & \texttt{ASCII} catalogue\\
  \hline
  \texttt{SeqNr} & Running object number & $-$ & \checkmark \\
  \texttt{ALPHA\_J2000} & Right ascension & unconvolved $i'$-band image
  & \checkmark \\
  \texttt{DELTA\_J2000} & Declination & unconvolved $i'$-band image &
  \checkmark \\
  \texttt{Xpos} & x pixel position & unconvolved $i'$-band image &
  \checkmark \\
  \texttt{Ypos} & y pixel position & unconvolved $i'$-band image &
  \checkmark \\
  \texttt{MAG\_AUTO} & total $i'$-band magnitude & unconvolved $i'$-band
  image & \checkmark\\
  \texttt{MAGERR\_AUTO} & total $i'$-band magnitude error & unconvolved
  $i'$-band image & \checkmark \\
  \texttt{MAG\_ISO\_x\footnote{$\mathrm{x} \in \left[
        u,g,r,i,z\right]$}} & isophotal magnitude in x-band &
  PSF-equalised x-band image & \checkmark\\
  \texttt{MAGERR\_ISO\_x} & isophotal magnitude error in x-band &
  PSF-equalised x-band image & \checkmark\\
  \texttt{MAG\_APER\_x} & aperture magnitude vector in x-band &
  PSF-equalised x-band image & \checkmark\\
  \texttt{MAGERR\_APER\_x} & aperture magnitude error vector in x-band
  & PSF-equalised x-band image & \checkmark\\
  \texttt{FWHM\_WORLD} & FWHM assuming a Gaussian core & unconvolved
  $i'$-band image & \checkmark\\
  \texttt{FLUX\_RADIUS} & half-light-radius & unconvolved $i'$-band
  image & \checkmark\\
  \texttt{A\_WORLD} & profile RMS along major axis & unconvolved
  $i'$-band image & \checkmark\\
  \texttt{B\_WORLD} & profile RMS along minor axis & unconvolved
  $i'$-band image & \checkmark\\
  \texttt{THETA\_J2000} & position angle & unconvolved $i'$-band image
  & \checkmark\\
  \texttt{CLASS\_STAR} & star-galaxy classifier & unconvolved $i'$-band
  image & \checkmark\\
  \texttt{Flag} & \SExtractor extraction flags & unconvolved $i'$-band
  image & \checkmark\\
  \texttt{FLUX\_ISO\_x} & isophotal flux in x-band & PSF-equalised
  x-band image & $-$\\
  \texttt{FLUXERR\_ISO\_x} & isophotal flux error in x-band &
  PSF-equalised x-band image & $-$\\
  \texttt{FLUX\_APER\_x} & aperture flux vector in x-band &
  PSF-equalised x-band image & $-$\\
  \texttt{FLUXERR\_APER\_x} & aperture flux error vector in x-band &
  PSF-equalised x-band image & $-$\\
  \hline
  \texttt{MAG\_LIM\_x} & limiting magnitude in x-band & unconvolved
  x-band image & \checkmark\\
  \hline
  \texttt{Z\_B} & Bayesian photo-$z$ estimate & $-$ & \checkmark \\
  \texttt{Z\_B\_MIN} & lower bound of the 95\% confidence interval & $-$
  & $-$\\
  \texttt{Z\_B\_MAX} & upper bound of the 95\% confidence interval & $-$
  & $-$\\
  \texttt{T\_B} & best-fit spectral type\footnote{Ell$=1$, Sbc$=2$,
    Scd$=3$, Im$=4$, SB3$=5$, SB2$=6$, plus 
    two interpolated types in colour-redshift space between each pair
    of these basis templates. Intermediate best-fit
    templates are indicated by a floating point number for
    \texttt{T\_B}.} 
  & $-$ & \checkmark\\
  \texttt{ODDS} & empirical odds\footnote{integrated probability
    inside an interval which is such that it contains 95\%
    probability for a single Gaussian} & $-$ & \checkmark \\
  \hline
  \texttt{NBPZ\_GOODFILT} & filters with reliable photometry & $-$ & \checkmark\\
  \texttt{NBPZ\_BADFILT} & filters with
  $\mbox{\texttt{MAGERR\_ISO}}\geq 1.0$ & $-$ & \checkmark\\
  \texttt{NBPZ\_LIMFILT} & filters with
  $\mbox{\texttt{MAG\_ISO\_x}}\geq \mbox{\texttt{MAG\_LIM\_x}}$ & $-$ & \checkmark\\
  \hline
  \texttt{MASK} & global mask key\footnote{unification of the
    different masks described in \sectionref{sec:masking}} & $-$ & \checkmark\\
\end{tabular}
\end{minipage}
\end{table*}

\section{Summary and conclusions}
\label{sec:summary}
We have presented high-quality five-band multi-colour data from 37 sq.
deg. of the \CARS survey. We gave a detailed description of our
data-handling procedures ranging from data selection to the final
catalogues including a first set of photometric redshift estimates.
Our algorithms provide an accurate astrometric alignment on the
sub-pixel level to extract precise object colour information. For the
large majority of our data the \Elixir photometric information allows
us to derive an unbiased absolute photometric calibration with a
scatter of $\sigma\approx 0.02-0.05$ on a pointing basis for
$g'r'i'z'$; tests against the official TERAPIX \Tthree CFHTLS-data
release show very significant zeropoint offsets for four out of 93
common fields.  In $u^*$ direct comparisons with SDSS suggest that our
zeropoints are systematically about 0.1 mag too faint.

We showed that our colour catalogues allow, with the help of
spectroscopic information, the estimation of reliable photometric
redshift estimates with the method of \citet{ben00}.  In our 37 sq.
deg. survey (about 30 sq. deg. in unmasked areas) we detect about 3.9
million objects classified as galaxies (\SExtractor
\texttt{CLASS\_STAR} $<0.95$). From those about 1.45 million ($10-15$
galaxies per sq. arcmin) have a formally reliable photo-$z$ estimate with
$\mbox{ODDS}>0.9$ (completeness 37.2\%). \refcomm{Comparing our photo-$z$
estimates with external spectroscopic data we find an
overall performance of $\sigma(\Delta z/(1+z))\approx 0.04-0.05$ up to
$i_{\rm AB}\approx 24$ with an outlier rate of $\eta \approx 1\%-3\%$.}
We applied a cross-correlation analysis to qualitatively investigate
redshift slice contamination between samples in different redshift
bins.  It indicates significant contamination of neighbouring redshift
slices with a width of $\Delta z\approx 0.1$ and a dying correlation
signal for bins more than $\Delta z\approx 0.3$ apart. Catastrophic
outliers occur between low-$z$ bins and galaxies with an estimate of
$z_{\rm phot}\geq 1.5$.  We performed a more quantitative analysis
only for the case when our whole redshift sample is divided in exactly two
redshift bins. With the help of spectroscopic redshifts from the deep
part of the VVDS it reconfirms the homogeneity of our photo-$z$ sample
over the entire \CARS area.  A more complete, in-depth analysis with
the correlation function technique will be presented in Benjamin et
al., (in prep.).

We note that the catalogues and the creation of photo-$z$s was
optimised for studies in the regime $0<z<1.4$ and objects with a
larger estimate should be filtered. The current catalogues are not
suited for studies of the high-$z$ regime such as $u^*$-band drop-out
searches.  While the photo-$z$ performance according to formal
parameters is very good our estimates show a systematic tilt for
$0<z<1$ (higher redshift ranges cannot be verified due to the lack of
spectroscopic information). \refcomm{Our estimates are too high by
$\Delta z\approx 0.1$ for low $z$ and the bias decreases linearly to
reach about $\Delta z\approx -0.1$ for $z\approx 1$. The
zero-crossing of the tilt is at $z\approx 0.5$. The mean bias is
about $\langle\Delta z\rangle \approx 0.03$ for $z<0.5$ and about
$\langle\Delta z\rangle \approx -0.03$ for $0.5<z<1$.  Improved and
bias-free \BPZ photo-$z$ estimates will be presented in Hildebrandt
et al. (in preparation). Additionally, photo-$z$ estimates with the
method of \citet{bab01} will be analysed and compared to our current
work in Brimioulle et al. (in preparation).}

The presented catalogues mark the first step for the primary science
goal of \CARS in the CFHTLS-Wide area: The assembling of a galaxy
cluster sample from low to high redshift and its subsequent
exploitation for cosmological studies. For the second step in this
effort, our multi-colour data are currently being used on several
cluster detection algorithms: The Voronoi tessellation technique from
\citet{rbf01}, the Postman matched filter algorithm
\citep[see][]{plg96} and the Red-Cluster Sequence technique
\citep[see][]{gly00}.

To trigger a larger variety of follow-up studies we make available our
catalogues on request.

\begin{acknowledgements}
  We thank Brice M{\'e}nard for help with the redshift
  cross-correlation analysis and Stephen Gwyn for clarifications on
  his \MegaPipe processing pipeline. We thank the anonymous referee 
  who helped us to improve the manuscript significantly.
  We acknowledge use of the
  Canadian Astronomy Data Centre, which is operated by the Dominion
  Astrophysical Observatory for the National Research Council of
  Canada's Herzberg Institute of Astrophysics.  This work was
  supported by the DFG Sonderforschungsbereich 375
  "Astro-Teilchenphysik", the DFG priority program SPP-1177 "Witnesses
  of Cosmic History: Formation and evolution of black holes, galaxies
  and their environment" (project IDs ER327/2-2, SCHN 342/7--1,
  Se1038/1), the German Ministry for Science and Education (BMBF)
  through DESY under the project 05AV5PDA/3 and the TR33 "The Dark
  Universe". H.~H. and M.~L. thank the European Community for the Marie Curie
  research training network "DUEL" doctoral fellowship
  MRTN-CT-2006-036133. T.~S. acknowledges financial support from the
  Netherlands Organization for Scientific Research (NWO).  
  L.~v.~W. and J.~B.
  are supported by NSERC and CIfAR. Part of the data reduction
  described in this work was performed on CFI funded equipment under
  project grant \#10052.  M.~L. thanks the University of Bonn and the
  University of British Columbia for hospitality.
\end{acknowledgements}
\bibliographystyle{aa}
\bibliography{CARS}
\appendix
\section{Details on the data handling of \CARS observations}
\label{app:processing}
In this appendix we give a more detailed description on our data
handling procedures of the \CARS survey. The excellent \Elixir
preprocessing of the \CARS data and the available meta-data
information (see below) allowed us to build up a complete survey
pipeline starting from data retrieval up to final co-added science
images. The construction of our data-processing and many choices for
our data-handling as described below were driven by the following two 
requirements:  
\begin{enumerate}
\item The system allows a 100\% automatic processing of the data with
  the need for manual intervention only at the final verification
  stage of co-added science images. 
  This particularly forbids manual passes through all
  individual \Elixir images either to visually grade data or to
  remove/mask artefacts. We needed to automatically reject
  problematic exposures or parts of them from the whole analysis 
  or we need to deal with remaining defects at the level of the final 
  science images. For instance, our automatic satellite track removal
  module reliably detects and removes about 95\% of all bright
  satellite trails. If necessary, the remaining 5\% need to be 
  masked manually in the final science images.
\item We want to independently and incrementally process individual
  pointings as soon as the full five band coverage of a particular area
  is becoming publicly available.
\end{enumerate}
We completely achieved the first goal and the complete data processing
of the \CARS data is done by one of the authors with two computers 
(a double processor Athlon2800+ with 4 GB Virtual memory and a quad
processor/dual core AMD Opteron 885 with 11 GB of RAM) and
a total disk storage capacity of about 10 Terabytes.
The second goal could not be met for the photometric calibration of
several fields and we needed to use information from adjacent
pointings to obtain an absolute flux calibration (see below).

Most of our algorithms to process optical data from multi-chip cameras
were described in \citet{esd05} and \citet{hed06} in the context of
\GaBoDS data \citep[see e. g.][]{ses03} from the 8-chip instrument
WFI@MPG/ESO2.2m. We therefore limit the discussion to peculiarities of
the \CARS data, necessary pipeline upgrades due to the four times
larger field-of-view of \MegaPrime and quality assessments of our
final science images.
\subsection{Data preselection and retrieval}
\label{sec:data-retr-select}
As described in \sectionref{sec:CARSdata}, the starting point of the
current \CARS data set are the \Elixir preprocessed images from the
CFHTLS-Wide Survey. Besides the images, comprehensive information on
the current status of the Survey and the observed data is available in
the form of a \texttt{CFHTLS exposure catalogue} (see
\burl{http://www.cfht.hawaii.edu/Science/CFHTLS-DATA/exposurescatalogs.html}).
From the CFHTLS-Wide we preselect all pointings which are publicly
available at 18/01/2008 and which have observations in the complete
$u^*g'r'i'z'$ filter set.

For each survey-image a \texttt{Service Observer quality flag} ranging
from 1-5 is available. A '1' states that the exposure was obtained
within survey specifications and has no obvious defects. A '2' means
that one of the predefined specifications for that exposure (seeing,
sky transparency or moon phase) was out of bounds. A flag of three or
higher indicates poor observing conditions or other severe defects
such as tracking problems during the exposure.  Only images with flags
'1' or '2' enter our processing. We visually inspected in total 300
\Elixir preprocessed \MegaPrime exposures (60 in each filter) to
verify the suitability of this quality assessment for a blind and
automatic preselection of \CARS data. On the other hand we did not
check whether a subset of the images with higher flags still could be
included in our survey.

We use the aforementioned information and the possibility to request
CADC files and data products directly within programs or
shell scripts\footnote{see
\burl{http://www1.cadc-ccda.hia-iha.nrc-cnrc.gc.ca/getData/doc}} to
automatically retrieve the images of interest. For the current work we
transfered in total 1246 \MegaPrime CFHTLS-Wide images from CADC.
\subsection{Data preprocessing and weight image creation}
\label{sec:data-preprocessing}
The \Elixir preprocessing includes all necessary operations to remove
the instrumental signature from raw data. The data on which we start
our analysis are bias-corrected and flat-fielded. Moreover, fringes
are removed in $i'$ and $z'$ observations, permanent bad CCD pixels
are marked and all images are corrected for photometric
non-uniformities across the \MegaPrime field-of-view; see the WWW pages
\burl{http://www.cfht.hawaii.edu/Science/CFHTLS-DATA/dataprocessing.html}
and
\burl{http://www.cfht.hawaii.edu/Science/CFHTLS-DATA/megaprimecalibration.html}
for a more detailed description of the \Elixir processing. 

The visual appearance of the \Elixir processed data is very good. Only
fringe residuals are observed for parts of the $z'$ data. As
discussed within \figref{fig:finalimages} individual chips of certain
exposures might not contain useful data. The first step of our own
processing is therefore to identify problematic chips by considering
pixel statistics and to mark them as unusable. More precisely, we
exclude chips with the following defects from any further analysis:
(1) The pixel value at the lower quartile of the chip
pixelvalue-distribution is 10 or lower. This means that large
fractions of the chip contain zeros; (2) More than 3\% of the pixels
in a chip are saturated. This means that a considerable chip-area is
\emph{contaminated} by a very bright star which would most probably
lead to problems in the later astrometric calibration.  Furthermore,
at the level of science analysis such areas would be excluded anyway;
(3) To astrometrically calibrate our data we first tie the $i'$
observations to the \texttt{USNO-B1} catalogue and then extract from
the stacked $i'$ image a deeper astrometric reference system for the
other colours. This is discussed in more detail in
\sectionref{sec:astr-calibr} below. For this reason we exclude from
$u^*g'r'z'$ all chips which have been identified as bad in the
corresponding $i'$-band data. In the following we updated all our
\THELI pipeline modules to smoothly handle an arbitrary geometry of
usable chips within a CCD-array.

To prepare the extraction of object catalogues for astrometric
calibration we create for each chip a corresponding weight
map which contains information on bad pixels and the relative noise
properties of the image pixels. The steps of our weight image
creation are described in detail in Sect. 6 of \citet{esd05}. For the
\CARS data we updated and expanded our procedures as follows:
\begin{enumerate}
\item For the WFI@MPG/ESO2.2m camera, the starting point of our weight
  maps was a normalised flat-field image. A flat-field maps the
  relative sensitivity of image pixels within a CCD array and allows
  us to take into account associated pixel noise variations during
  object extraction. For the \CARS data we neglect this effect and
  start the weight creation with a flat image with a pixel value of
  '1' on the whole array. We verified that this simplification has no
  significant effect on our object catalogues of single frames and of
  the final co-added science images later-on. It allows us to store
  weight images very efficiently and with significantly reduced
  hard-disk space.
\item Permanent bad pixels of the CCDs are marked in the \Elixir
  processed science frames by a pixel value of '0'. This information
  is transfered to our weight maps. For WFI@MPG/ESO2.2m data defect
  pixels had to be identified with dark frames and/or flat-field
  images.
\item We used to visually identify and to mask bright satellite tracks
  which must be excluded from the object extraction and
  co-addition process. To process the \CARS data we developed an
  automatic track detection and masking tool based on Hough transform
  techniques \citep[see e.~g.][]{dup72,van01}. To reliably find real
  tracks and to reject spurious detections due to bright stars and extended
  objects we use that a satellite typically contaminates several chips
  on the \MegaPrime mosaic. When a candidate is found on a particular chip we
  check for detections on expected positions in other
  detectors. Tracks which are found in two or more chips are masked on 
  all CCDs crossing their path including an additional margin of one
  CCD on both sides of the track. In this way also detectors
  on which a track cannot be detected individually are appropriately
  covered by an image mask. From manual inspection of 60 \MegaPrime
  exposures it is found that our implementation correctly detects and
  masks more than 95\% of all bright satellite tracks. The failures
  can mostly be traced to either dashed, non continuous satellite tracks or
  to short ones at the edges of the mosaic. We observed only a handful
  false-positive detections in very special configurations such
  as extended and bright object chains over chip boundaries.  

  Pixels attached to identified tracks are set to zero in the
  corresponding weight maps.
\end{enumerate}
\subsection{Astrometric calibration}
\label{sec:astr-calibr}
\begin{figure}[ht]
  \centering
  \includegraphics[width=0.95\columnwidth]{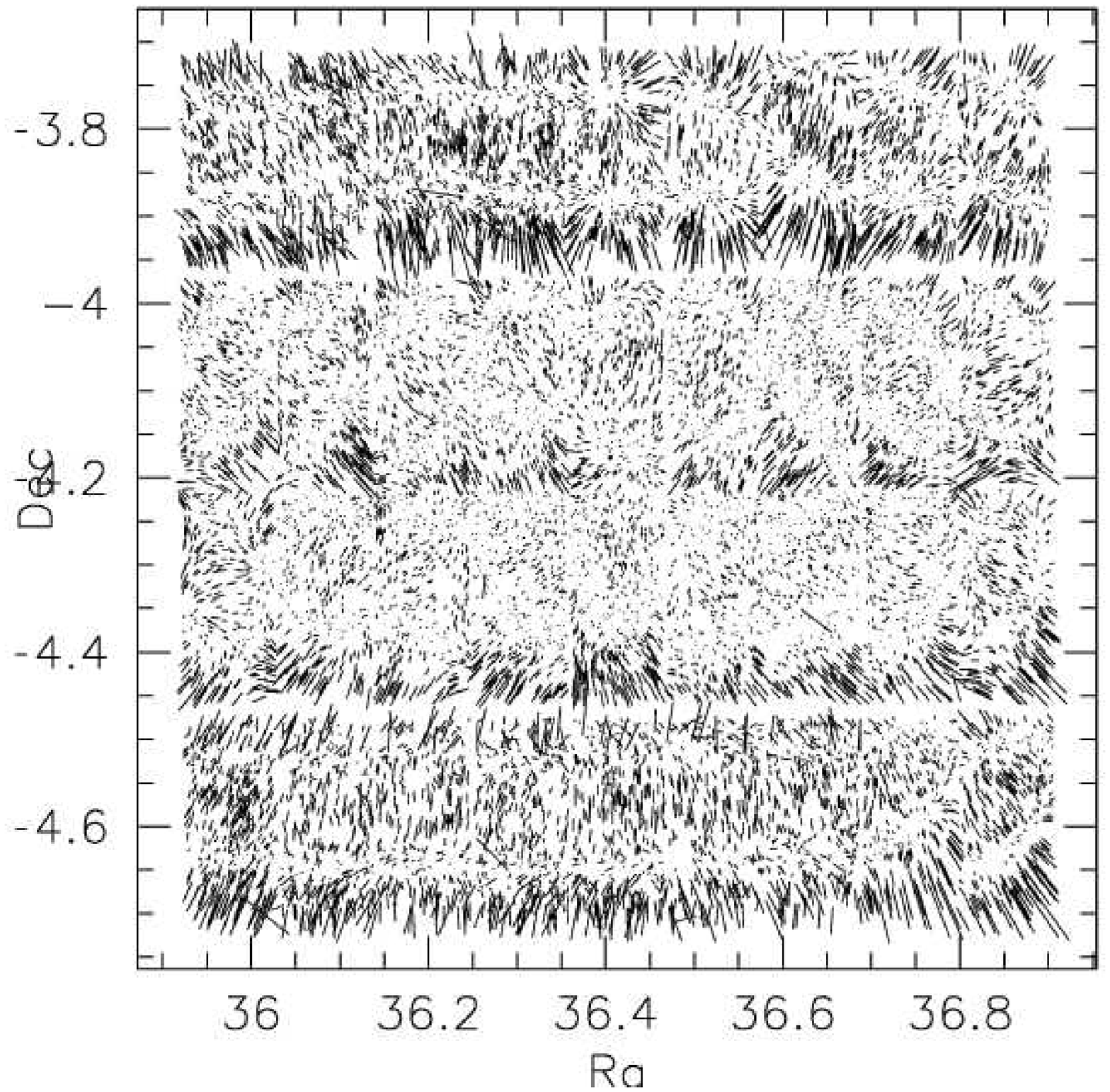}
  \includegraphics[width=0.95\columnwidth]{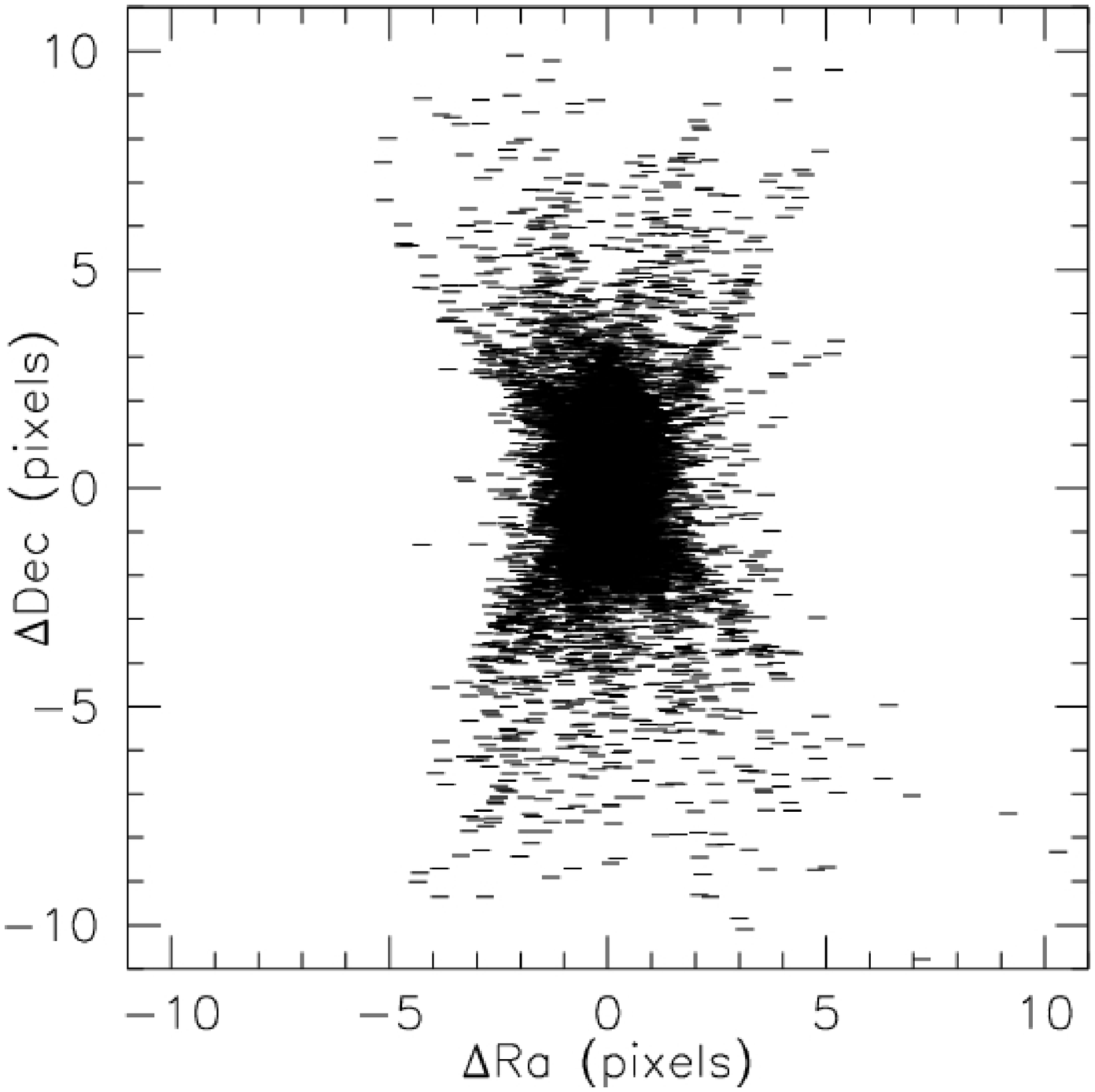}
  \caption{\label{fig:astromdistort} Higher order \MegaPrime
    distortions: The plots show the difference in object position
    after a first-order astrometric alignment, i.~e. corrections for
    linear shifts and rotations and a the full third-order astrometric
    solution estimated by \Astrometrix. The sticks in the upper plot
    indicate the positional displacement vector between the two
    solutions and the lower plot gives the absolute displacement
    numbers. We note that the second- and third-order terms contribute
    very significantly to the solution (see text for details).}
\end{figure}

\begin{figure}[ht]
  \centering
  \includegraphics[width=0.95\columnwidth]{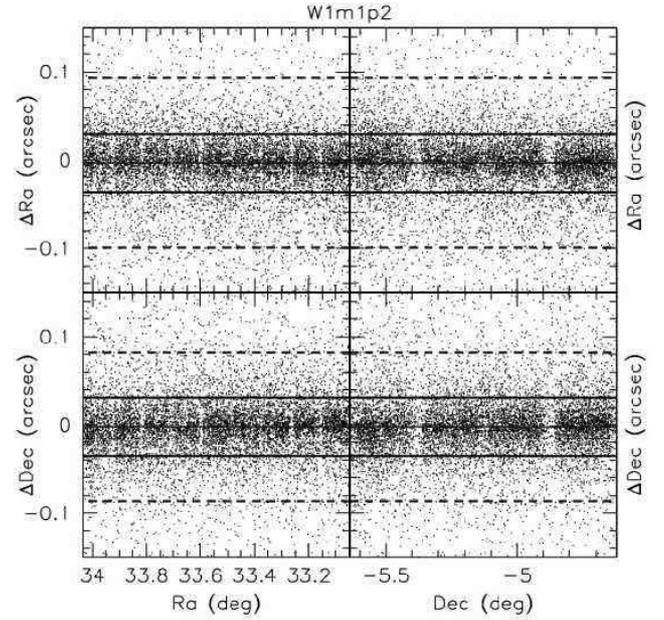}
  \caption{\label{fig:internalastrom} Internal Astrometric alignment
    of $i'$-band data from \texttt{W1m1p2}: Shown are the differences
    of sources in 7 dithered $i'$-band observations after astrometric
    calibration with the counterparts in the co-added image. 
    The 7 individual exposures were
    obtained with a dither pattern spanning about $45\myarcsec 0$ in
    Ra and $180\myarcsec 0$ in Dec to cover the \MegaPrime chip gaps.
    The plot covers the complete \MegaPrime area of about 1 sq. degree.
    The thick solid lines mark the region containing $68\%$ of all
    points and are at $\Delta {\rm Ra}=-0.003^{+0.029}_{-0.035}$ arcsec and 
    $\Delta {\rm Dec}=-0.002^{+0.031}_{-0.035}$ arcsec. Dashed lines show the
    corresponding area for $90\%$ of all points 
    ($\Delta {\rm Ra}=-0.003^{+0.092}_{-0.099}$ arcsec and 
    $\Delta {\rm Dec}=-0.002^{+0.081}_{-0.086}$) arcsec. Only 1 out of 10
    points is shown for clarity of the plot.}
\end{figure}

\begin{figure}[ht]
  \centering
  \includegraphics[width=0.95\columnwidth]{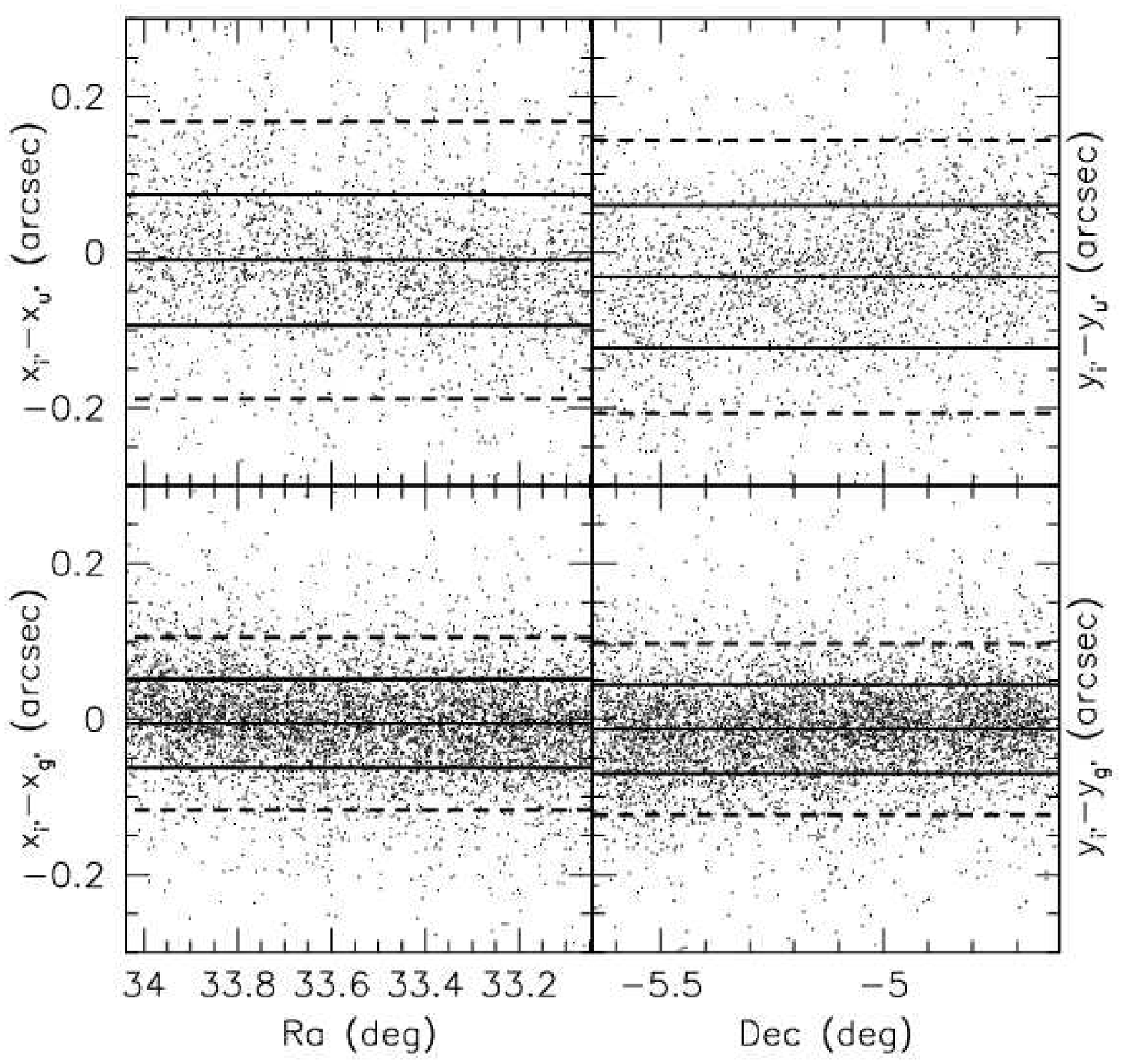}
  \caption{\label{fig:intercolourastrom} Astrometric alignment of
    different colours of \CARS field \texttt{W1m1p2}: We show
    differences in object positions from different colours $u^*g'i'$
    of the \CARS pointing \texttt{W1m1p2}. 
    The plot shows that the different
    bands are well aligned with sub-pixel accuracy although slight
    trends in the residuals with $u^*$ are visible. Solid (dashed)
    lines enclose areas containing 68\% (90\%) of all points. They are
    at $x_{i'}-x_{g'}=-0.006^{+0.05}_{-0.06}(^{+0.10}_{-0.11})$ arcsec,
    $y_{i'}-y_{g'}=-0.012^{+0.04}_{-0.07}(^{+0.09}_{-0.12})$ arcsec and
    $x_{i'}-x_{u^*}=-0.009^{+0.07}_{-0.09}(^{+0.16}_{-0.18})$ arcsec,
    $y_{i'}-y_{u^*}=-0.03^{+0.06}_{-0.12}(^{+0.14}_{-0.20})$ arcsec.
    See text for further details.}
\end{figure}

With the preprocessed \Elixir images and the weight maps, the
astrometric calibration of \CARS data sets and the associated quality
assessment follows very closely the procedures outlined in Sect. 5 of
\citet{esd05}:
\begin{enumerate}
\item \SExtractor is used on all images to extract sources
  with at least 5 pixels having $5\sigma$ above the sky-background.
\item The \Astrometrix programme \citep[see][]{rad02,mrb03} is run on
  the $i'$-band to determine a third order astrometric solution for
  each individual chip. We use the \texttt{USNO-B1} standard-star
  catalogue \citep[see][]{mlc03} as our astrometric reference frame.
  We co-add the $i'$-band data (see below) and extract a high $S/N$
  object catalogue (sources with at least 20 pixels of $20\sigma$
  above the sky-background noise) from it.  This catalogue is used as
  reference for the astrometric calibration of the $u^*g'r'z'$ data.
  It allows us to use a dense reference catalogue with high positional
  precision. This turned out to be essential to map robustly
  significant higher-order astrometric distortions of \MegaPrime. 
  \refcomm{For this instrument, the contributions
  of second- and third-order terms to the astrometric correction 
  are up to $2\myarcsec 0$. As a comparison, for the four times smaller 
  WFI@MPG/ESO2.2m camera these values are on the order of $0\myarcsec
  5$}; see also \figref{fig:astromdistort}. 
  We use the $i'$-band data as reference
  because (1) the individual exposures are already reasonably deep and
  hence can be well calibrated with the \texttt{USNO-B1} sources and
  (2) with seven dithered exposures the co-added $i'$-band typically
  has the best coverage and filling-factor of the \MegaPrime area with
  its chip gaps in individual exposures.
\item For our scientific objectives, weak gravitational lensing, and
  multi-band studies with photometric redshifts it is essential to
  obtain a very high internal astrometric accuracy in the lensing band
  ($i'$ in our case). As discussed in \citet{esd05} higher order
  object brightness-moments (everything above the zeroth order moment,
  i.~e.  the object flux) are significantly changed if individual
  frames of the WFI@MPG/ESO2.2m camera are aligned with an accuracy
  $>0.5$pixel (pixel scale $0\myarcsec 238$). For our \CARS data we
  reach an accuracy of about $0\myarcsec 03$-$0\myarcsec 04$, i.~e.
  about $1/5$th of a \MegaPrime pixel. \refcomm{This was tested by comparing
  object positions (with $i_{\rm AB}<20$) 
  from astrometrically corrected individual frames
  with their cousins in the final co-added images. Results for the
  field \texttt{W1m1p2} in the $i'$ filter are shown in 
  \figref{fig:internalastrom}.
  In a similar way the inter-colour alignment of different filters per
  pointing is tested. From the final co-added images we compare object 
  positions of high $S/N$ sources ($i_{\rm AB}<20$). For all pointings
  we reach an alignment between the colours below one pixel. We note
  however that the inter-colour alignment between $g'r'i'$ is 
  better (typically 0.5 pixels) than between $i'u^*z'$ (between 0.5 and
  1 pixel). This is expected because individual frames from $u^*$ and
  $z'$ have fewer high $S/N$ sources for astrometric calibration than
  the intermediate filter bands. We show results from the $i'-g'$ and
  $i'-u^*$ comparisons of \texttt{W1m1p2} in \figref{fig:intercolourastrom}.} 
\end{enumerate}
\subsection{Photometric calibration}
\label{sec:phot-calibr}
The \Elixir preprocessed images come with all necessary meta-data 
to translate pixel counts to instrumental AB magnitudes; see
\burl{http://www.cfht.hawaii.edu/Science/CFHTLS-DATA/megaprimecalibration.html#P2}
for a full description of the \Elixir procedures to derive photometric
parameters. In addition, the \texttt{CFHTLS exposure catalogue}
contains a flag whether an image was taken under photometric sky
conditions or not. With this information we try to derive a photometric
zeropoint for each colour of each pointing (we call this a 
\texttt{set} in the following) with the two-stage process
described in \citet{hed06}:
\begin{enumerate}
\item We use \Photometrix to bring all individual images to the same
  flux scale by estimating the magnitude differences of overlap
  sources. This gives us for each image $i$ a relative zeropoint
  $ZP_{{\rm rel},i}$ which tells us the magnitude offset of that image
  w.r.t. the mean relative zeropoint of all images, i.e. we demand
  $\sum_i ZP_{{\rm rel},i} = 0$. Note that this procedure relatively
  calibrates data obtained under photometric and non-photometric
  conditions. An absolute flux scaling can now be obtained from the
  photometric subset.
\item For images being observed under photometric conditions we
  calculate a \texttt{corrected zeropoint} $ZP_{{\rm corr},j}$
  according to $ZP_{{\rm corr},j} = ZP + am\cdot ext + ZP_{{\rm
      rel},j}$, where $am$ is the airmass during observation, $ZP$ the
  instrumental AB zeropoint and $ext$ the colour dependent extinction
  coefficient. The last two quantities are part of the provided
  \Elixir meta-data. For images obtained under photometric conditions,
  the relative zeropoints compensate for atmospheric extinction and
  the corrected zeropoints agree within measurement errors in the
  ideal case. We later use the
  mean $\langle ZP_{{\rm corr},j}\rangle$ as zeropoint for our co-added images.
\end{enumerate}
Note that we can determine an absolute zeropoint only if at least one
exposure of a given \texttt{set} was obtained under photometric
conditions. For the \CARS data this is not the case for 12 \texttt{sets} and we
estimated a zeropoint for those after image co-addition by flux
comparison with objects of adjacent, calibrated pointings.
\begin{figure}[ht]
  \centering
  \includegraphics[width=0.95\columnwidth]{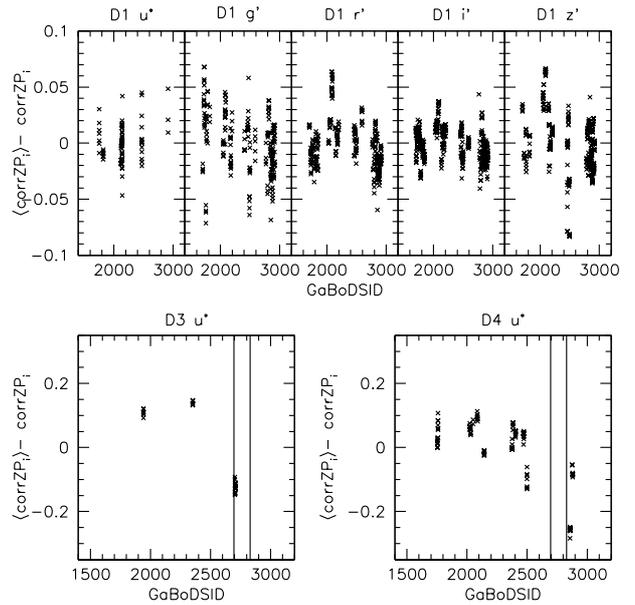}
  \caption{\label{fig:corrzps} Long-term studies of corrected
    zeropoints in CFHTLS-Deep data: We analysed all publicly available
    photometric data from the CFHTLS-Deep fields \texttt{D1} (Ra:
    02:26:00; Dec: $-$04:30:00; all colours), \texttt{D3} (Ra:
    14:19:28; Dec: +52:40:41; $u^*$-band) and \texttt{D4} (Ra:
    22:15:31; Dec: $-$17:44:06; $u^*$-band).  The panels show the
    long-term evolution of \texttt{corrected photometric zeropoints}
    in these fields from 25/06/2003 - 29/12/2006. On the measured
    zeropoint distributions we performed an iterative $3\sigma$
    clipping to exclude obvious outliers. The quoted GaBoDSID is a
    running number counting the nights from 31/12/1998. 
    The vertical lines in the \texttt{D3} and \texttt{D4}
    panels show the period of \texttt{W3} and \texttt{W4} $u^*$
    observations for which we observe larger discrepancies in
    comparisons with SDSS magnitudes; see text for further details.}
\end{figure}
As discussed in \citet{hed06} the \texttt{corrected zeropoints} offer a good
opportunity to verify the quality of absolute photometric calibration. If a
field is observed over an extended period, the comparison of zeropoints
estimated from different nights gives a robust indication on the
long-term stability of photometric instrument properties and on the
calibration process itself.
 
However, the \CARS fields from the CFHTLS-Wide Survey were observed
mostly in a compact period during a single night and hence do not
allow for this test directly. To perform this important quality
control we consider observations of the CFHTLS-Deep Survey. This part
of the CFHTLS continously observes four one square degree fields with
the goal to detect Supernovae and to measure their light curves; see
\citet{agr06} for more details on this survey. Because we noticed severe
problems with our $u^*$ flux calibration for \Wthree and \Wfour later-on (see
\sectionref{sec:sloancomp} below) we investigated CFHTLS-Deep images
which were publicly available at 01/01/2008 and which had the
photometric flag in the \texttt{CFHTLS exposure catalogue}.  \refcomm{We
studied the long-term evolution of the corrected zeropoints from June
2003 to December 2006. Results for the fields \texttt{D1}, \texttt{D3}
and \texttt{D4} are shown in \figref{fig:corrzps}.  For the
\texttt{D1} area we studied in detail photometric stability in all
five filter bands. \figref{fig:corrzps} shows that the photometric
calibration of this field over time is very consistent with formal
standard deviations of only up to about 0.03 mag. Note however that
the peak differences of magnitude zeropoints span more than 0.1 mag!
For the $u^*$-band observations of \texttt{D3} and \texttt{D4} we
observe a considerably larger scatter with extremely low values for
the corrected zeropoints (too high \CARS magnitude zeropoints), from around
April 2006 to November 2006. This hints to a calibration problem of
$u^*$-band CFHTLS data in that period and the 
data suggest a necessary correction of about $-0.2$ mag for 
$u^*$-band \CARS observations from spring to fall 2006. Due to its visibility
\texttt{D1} has no observations in that period. We verified that the
\texttt{D4} results for the colours $g'r'i'z'$ are similar to those in
\texttt{D1} and hence the problem seems to be confined to $u^*$.
These discrepancies in the $u^*$ calibration are also documented in
\citet{gwy08} and on TERAPIX WWW pages describing the official
\texttt{T0004} CFHTLS data release.
(\texttt{http://terapix.iap.fr/article.php?id\_article=713}).}

The Deep data allow us also to check homogeneity and reproducibility
of our photometric calibration over the \MegaPrime field-of-view. In
each colour we created three independent co-added images from
\texttt{D1}.  Each stack contains five images obtained under
photometric conditions in November 2003, 2004 and 2005.
\refcomm{From the different stacks in each colour we match 
bright ($17<m<20$) sources whose positions agree to 
$0\myarcsec 5$ or better. Magnitudes are compared and mean offsets
and standard deviations are estimated. \tabref{tab:deepmagcomp} 
summarises the results.} 
Except for the
$z'$-band the total magnitudes agree to better than $0.04$ for
these stacks. The scatter around these absolute offsets is around
$0.02$ mag for $u^*g'r'i'$ and about $0.03-0.04$ mag for $z'$; we
attribute the higher value in $z'$ to fringe residuals in this band.
These values give us an estimate on the internal photometric accuracy
of our data, i.~e. on the error propagation of inaccuracies in our
photometric calibration procedures to final magnitude estimates. We
consider them as upper limits because possible errors in the
determination of the photometric superflat within \Elixir contribute
to the quoted numbers.  We note that we obtain consistent errors when
comparing the Deep stacks with the \CARS pointing \texttt{W1p2p3}.
Their distance on the sky is $5\myarcmin 0$ in Ra, $18\myarcmin 0$ in
Dec and we can compare fluxes from objects which fall in different
areas of the \MegaPrime mosaic. The test between the Deep stack from
2003 and \texttt{W1p2p3} yields: $\Delta u^* = 0.003\pm 0.022$;
$\Delta g' = 0.014\pm 0.014$; $\Delta r' = 0.053\pm 0.012$; $\Delta i'
= -0.019\pm 0.014$; $\Delta z' = -0.034\pm 0.034$.
\small
\begin{table}
\caption{\label{tab:deepmagcomp} Comparison of magnitudes from three
  independent stacks of CFHTLS-Deep D1 data in each colour: The upper
  row shows magnitude differences between stacks of 2003 and 2004, the
  lower one between data from 2003 and 2005. See text for details.}
\begin{tabular}{rrrrr}
\hline\hline       
\multicolumn{1}{c}{$\Delta u^* \times 100$} & \multicolumn{1}{c}{$\Delta g' \times 100$} & 
\multicolumn{1}{c}{$\Delta r' \times 100$} & \multicolumn{1}{c}{$\Delta i' \times 100$} & 
\multicolumn{1}{c}{$\Delta z' \times 100$} \\ 
\hline
$-3.3\pm 1.8$ & $1.3\pm 1.5$ & $-0.5\pm 1.3$ & $-1.4 \pm 2.0$ & $4.6\pm 3.1$ \\
$-2.6\pm 2.0$ & $3.3\pm 1.6$ & $0.6\pm 1.6$ & $-0.0 \pm 2.0$ & $-7.3\pm 3.8$ \\
\hline
\end{tabular}
\end{table}
\normalsize
\subsection{Image co-addition}
\label{sec:image-co-addition}
After the photometric calibration we check whether the individual
exposures of a given \texttt{set} were obtained under varying photometric
conditions which is indicated by a large range of relative
zeropoints. Low-value outliers point to images which were observed
under unfavourable sky-conditions w.r.t. the rest of the \texttt{set}.
We estimate the median $med(ZP_{{\rm rel},i})$ of the
relative zeropoint distribution and sort out exposures with a relative
zeropoint of $med(ZP_{{\rm rel},i})-0.1$ or smaller.  Affected are the
five \CARS \texttt{sets} \texttt{W1p1p2-i'}, \texttt{W1p3m0-i'},
\texttt{W1p4m0-i'}, \texttt{W4m1m1-z'} and \texttt{W4m1m2-z'}.  
At this stage we also reject short calibration exposures ($t_{\rm exp} < 100$s)
from further processing.

Finally the exposures belonging to a \texttt{set} are sky-subtracted
with \SExtractor and co-added with the \Swarp programme
\citep[see][]{ber08}.  We use the LANCZOS3 kernel to remap original
image pixels according to our astrometric solutions.  The subsequent
co-addition is done with a statistically optimal weighted mean which
takes into account sky-background noise, weight maps and the relative
photometric zeropoints as described in Sect. 7 of \cite{esd05}. As sky
projection we use the TAN projection \citep[see][]{grc02} and all
colours from a specific pointing are mapped on the same pixel grid.
The origins of the TAN projection for each pointing are those defined
for the CFHTLS-Wide survey (see
\burl{http://terapix.iap.fr/cplt/oldSite/Descart/summarycfhtlswide.html}).
An example for a final co-added image is shown in
\figref{fig:finalimages}.

After co-addition we cut all images to a common size of $21k\times 21k$ 
which covers areas with useful data for all \CARS pointings. We 
visually check each product for obvious
defects. Because no pixel rejection takes place in our weighted mean 
stacking the co-added images show some remaining artefacts. We
observe faint satellite tracks which were not detected and
removed by our automatic satellite track masking tool (see
\sectionref{sec:data-preprocessing}), short asteroid trails and warm
pixels. Within the subsequent catalogue creation process we tried to 
mask these defects on the basis of the $i'$-band of each pointing 
(see \sectionref{sec:catalogues}).

\begin{figure}
   \centering \includegraphics[width=\columnwidth]{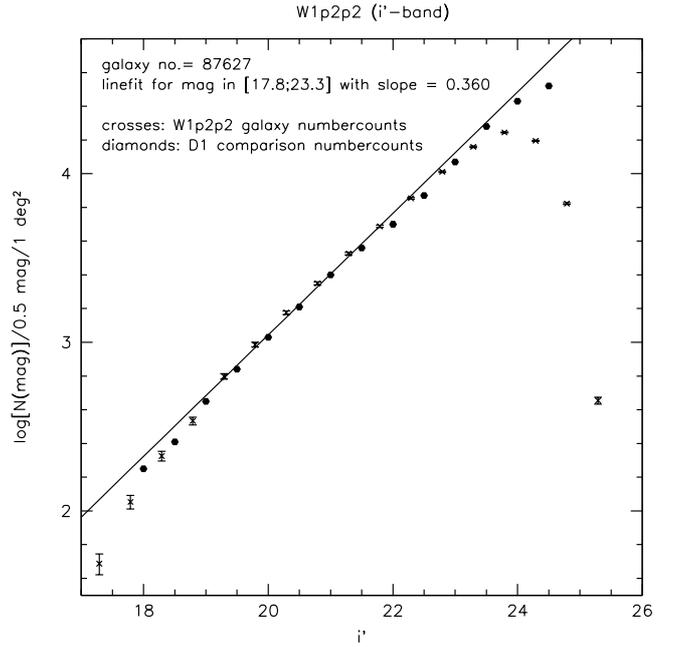}
      \caption{\label{fig:gal_counts} $i'$-band numbercounts for the
        field \texttt{W1p2p2}: galaxies are selected with
        \texttt{CLASS\_STAR} $<0.95$ to allow for a first crude check of
        our magnitude zeropoints with galaxy-number counts.}
\end{figure}

\begin{figure}
   \centering \includegraphics[width=\columnwidth]{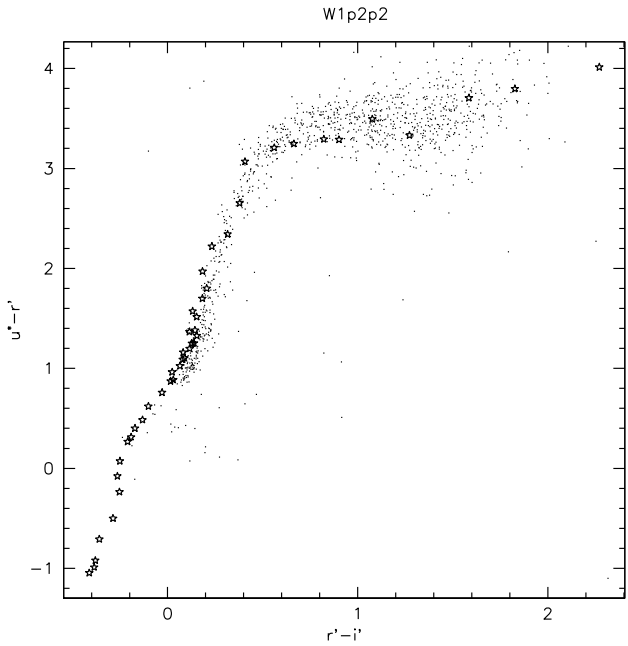}
      \caption{\label{fig:col_col} $u^*-r'$ vs. $r'-i'$ colour-colour
        diagram in the field \texttt{W1p2m2}. The dots represent the measured
        colours of stars (\SExtractor \texttt{CLASS\_STAR} $>0.95$ and
        $i_{\rm AB}<20$; $N=1089$) and the star symbols are colours of
        stars from the library of \cite{pic98}.}
\end{figure}

\refcomm{A first rough check for the photometric calibration of each
co-added image is done with the help of galaxy number counts. In all
colours we estimated reference counts from the CFHTLS \texttt{D1}
field and compare those to our \CARS Wide data. The reference counts
are about two magnitudes deeper than those from our co-added images.
Each of the one square degree fields yields a robust estimation of
these counts and allows us to quickly spot photometric calibrations
with obvious problems ($\Delta m \approx 0.2$). Galaxies are selected
by the \SExtractor \texttt{CLASS\_STAR} parameter 
($\mbox{\texttt{CLASS\_STAR}} < 0.95$); see
\figref{fig:gal_counts} for an example of the field \texttt{W1p2p2}.
With multi-colour observation for all fields we also can compare
colours from stellar sources with predictions from the \cite{pic98}
library.  We select bright, unsaturated stars by magnitude
($17<i'_{\rm AB}<22$) and by $\mbox{\texttt{CLASS\_STAR}} > 0.95$. 
With data in $u^*g'r'i'z'$ we plot ten possible colour-colour combinations
against the Pickles stellar library \citep{pic98} which allows us,
similar to the galaxy number counts, to identify grossly inaccurate
zeropoints with $\Delta m \approx0.1- 0.2$. See \figref{fig:col_col}
for an example track of \texttt{W1p2p2}.}

\refcomm{A much more rigorous and accurate test for the photometric quality of
our data is given by direct comparison with external and well
calibrated data sets. This is the topic of the following sections.}
\subsection{Comparison of \CARS data with SDSS photometry}
\label{sec:sloancomp}
The overlap of all three \CARS patches with the SDSS 
(see \sectionref{sec:CARSdata}) allows us a
direct comparison of object fluxes with Sloan photometry. To convert the
instrumental \MegaPrime AB magnitudes from stellar objects to the SDSS
system we use the following transformation formulae: 
\begin{eqnarray}
\label{eq:sloantransform}
u^*_{\rm AB} & = & u_{\rm SDSS} - 0.241 \cdot (u_{\rm SDSS} - g_{\rm SDSS})
\nonumber \\
g'_{\rm AB}  & = & g_{\rm SDSS} - 0.153 \cdot (g_{\rm SDSS} - r_{\rm SDSS})
\nonumber \\
r'_{\rm AB}  & = & r_{\rm SDSS} - 0.024 \cdot (g_{\rm SDSS} - r_{\rm SDSS}) \\
i'_{\rm AB}  & = & i_{\rm SDSS} - 0.085 \cdot (r_{\rm SDSS} - i_{\rm SDSS})
\nonumber \\
z'_{\rm AB}  & = & z_{\rm SDSS} + 0.074 \cdot (i_{\rm SDSS} - z_{\rm SDSS}) 
\nonumber
\end{eqnarray}
The relations for $g'r'i'z'$ were determined within the CFHTLS-Deep Supernova
project (see
\burl{http://www.astro.uvic.ca/~pritchet/SN/Calib/ColourTerms-2006Jun19/index.html#Sec04});
the $u^*$ transformation comes from the CFHT instrument page (see
\burl{http://cfht.hawaii.edu/Instruments/Imaging/MegaPrime/generalinformation.html}).

\begin{figure}[ht]
  \centering
  \includegraphics[width=0.95\columnwidth]{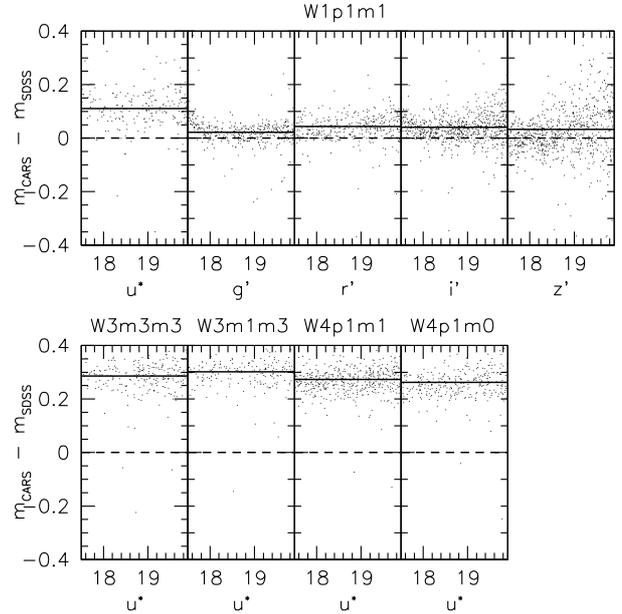}
  \caption{\label{fig:sdss_comp} Comparison of \CARS fluxes and SDSS
    magnitudes: We show magnitude offsets of \CARS data with SDSS
    overlap. The upper row shows the comparison for all five
    bands of the field \texttt{W1p1m1}. The dashed line in the plots
    indicates the zero offset and the solid line the observed mean
    difference; see text for a discussion of the results.}
\end{figure}

For all the following photometric comparison studies we extracted
\emph{single frame} photometric catalogues from all \CARS images and
we use the \SExtractor \texttt{MAG\_AUTO} estimate throughout,
i.~e. here we do not use the multi-colour catalogues described in 
\sectionref{sec:catalogues}. We compare magnitude estimates from
sources classified as stars in the SDSS and having a \CARS \texttt{MAG\_AUTO}
estimate of $17<m<20$. 
Representative results of our SDSS comparisons are shown in
\figref{fig:sdss_comp}. A complete listing of the measured magnitude
offsets and dispersions can be found in \tabref{tab:CARSquality}.  We
note a stable calibration in $g'r'i'$. For nearly all fields the mean
offset in these filters is well below 0.05 mag and the transformation
relations from \eqref{eq:sloantransform} are valid with $\sigma_{{
    g',r',i'}}\approx 0.02-0.04$ in the magnitude range $17<m<20$. For
$z'$ the mean offset reaches up to 0.08 mag and also the dispersion
broadens to $\sigma_{{z'}}\approx 0.04-0.07$. Larger disagreements
are observed for the $u^*$ filter.  For the three \texttt{W1} fields
with SDSS overlap we measure a consistent offset of $u'-u_{\rm SDSS} +
0.241 \cdot (u_{\rm SDSS} - g_{\rm SDSS}) \approx 0.1$. Two of the
$u^*$ fields (\texttt{W1p3m0} and \texttt{W1p4m0}) were observed on
01/01/2006 and 02/01/2006 and the third one, \texttt{W1p1m1}, on
13/12/2006 and hence we obtain this result for different calibration
periods. Our long-term zeropoint analysis of \texttt{D1} for which
$u^*$-band data have been obtained during December 2006 does not
indicate larger systematic calibration offsets.  Hence, at the current
stage, we have no explanation for this high, consistent offset between
our \texttt{W1} $u^*$ fluxes and the SDSS magnitudes. Even
considerably larger absolute offsets (up to 0.3 mag) are observed for
all \CARS $u^*$ pointings of \texttt{W3} and \texttt{W4}.  However, as
was discussed in
\sectionref{sec:phot-calibr} systematic zeropoint offsets for the
$u^*$ calibration are observed from April 2006 to November 2006 and all
\texttt{W3} and \texttt{W4} $u^*$ band observations were obtained just
in that period. The data presented in \figref{fig:corrzps} suggest a
necessary correction for the $u^*$-band zeropoint of about 0.2 mag
which would make the observed offsets consistent with the \texttt{W1}
results. For $u^*$, the dispersion with the transformation in
\eqref{eq:sloantransform} is about $\sigma_{u^*}\approx 0.03-0.06$
($17<u^*<20$).
\subsection{Photometric comparison of \CARS images with publicly
  available CFHTLS data}
\label{sec:comparison-cars-data}
We further check the quality of our photometric calibration by
comparison with publicly available CFHTLS data. We use the TERAPIX
\Tthree data release\footnote{At the time of writing the most recent
  TERAPIX \texttt{T0004} release (\refcomm{with a 35 sq. degree
    five-colour \CARS overlap}) was not publicly available.} (see
\burl{http://www1.cadc-ccda.hia-iha.nrc-cnrc.gc.ca/cadcbin/cfht/wdbi.cgi/cfht/tpx_fields/form}
and \burl{http://terapix.iap.fr/rubrique.php?id_rubrique=208}) which
consists of all available CFHTLS-Deep and CFHTLS-Wide observations
until 12/09/2005. \refcomm{It overlaps with our data on 24 sq. degrees with the
full five-colour coverage on nine sq. degrees.} 
Furthermore, Stephen Gwyn considers public CFHTLS
data within his \MegaPipe project which aims at providing calibrated
and co-added data from the complete \MegaPrime archive at CADC
\citep[see][]{gwy08} \refcomm{(\CARS overlap on 22 sq. degrees with five-colour
coverage on two sq. degrees)}. All three data sets start their processing from
the \Elixir images but each pipeline derives the photometric
calibration with different software modules and by including different
internal and external data sets.  Hence this comparison gives us
another check on the accuracy and limitations of our algorithms:
\begin{itemize}
\item The \Tthree data are processed on a patch basis, i.~e. to derive a
  photometric solution all available information from a CFHTLS-Wide
  patch are considered simultaneously. The global photometric analysis
  takes into account overlap sources from adjacent pointings and also
  allows modest variations of the derived \Elixir zeropoints to better
  ensure a consistent solution on the complete patch. In contrast, the
  \CARS data are treated on a strict pointing-by-pointing basis. Other
  differences in the processing which might influence direct flux
  comparisons between \Tthree and \CARS are:
  \begin{enumerate}
  \item To select suitable \Elixir images for further processing
    TERAPIX does not rely solely on the quality flag in the CFHTLS
    exposure catalogue but each image is regraded. Hence, the
    composition of image stacks might be different for some
    pointing/colour combinations.  We did not investigate this in
    detail.
  \item The \Tthree stacks are created with a median co-addition whereas
    \CARS uses a weighted mean statistics. Satellite tracks in individual 
    frames are not masked before co-addition in the \Tthree processing.
  \end{enumerate}
\item The \MegaPipe project directly uses the SDSS to photometrically
  calibrate \MegaPrime data on a pointing basis. For observations
  which overlap with Sloan, the relations from
  \eqref{eq:sloantransform} are used to derive zeropoints for the
  images. For observations outside the Sloan area, the procedure is
  similar to ours. If data are obtained under photometric conditions,
  the \Elixir calibration is used. Otherwise, a calibration with
  adjacent pointings having photometric information is tried.

  Similar to the \Tthree processing all \MegaPipe images are
  rechecked manually for their suitability to be processed further
  and the final stacking is done with a median co-addition.
\end{itemize}
The results of our flux comparisons with 93 \Tthree and 62 \MegaPipe
fields are detailed in \tabref{tab:CARSquality}. 

We note in general a very good agreement between our calibration and
that from TERAPIX \Tthree. For all but four pointings the discrepancy
is less than 0.04 mag. Notable differences occur for the stacks
\texttt{W1p2p3-$r'$} ($m_{\rm CARS}-m_{\rm T0003}=-0.063$ mag),
\texttt{W1m1p3-$u^*$} ($m_{\rm CARS}-m_{\rm T0003}=-0.18$ mag),
\texttt{W1m1p3-$z'$} ($m_{\rm CARS}-m_{\rm T0003}=-0.12$ mag) and
\texttt{W1m1p2-$u^*$} ($m_{\rm CARS}-m_{\rm T0003}=-0.12$ mag).  As
discussed in
\sectionref{sec:phot-calibr} the field \texttt{W1p2p3} overlaps with
CFHTLS-Deep 1 and we can confirm an offset of about 0.05 mag between
the \CARS \texttt{W1p2p3-$r'$} stack and corresponding Deep data. All
four individual images contributing to the \CARS image have been
obtained under photometric conditions which is confirmed by a very
narrow distribution (about 0.01 mag) of relative zeropoints. At the
current stage we do not have a conclusive explanation for the observed
discrepancy in this field. We note that \texttt{W1p4p3-$r'$} for which
we observe no discrepancy ($m_{\rm CARS}-m_{\rm T0003}=-0.007$) has
been observed in the same night (23/08/2003) as \texttt{W1p2p3-$r'$}.
Furthermore, these two \texttt{sets} share the same photometric
calibration data.

For the other three cases with a fairly large magnitude shift of more
than 0.1 mag all the science frames were obtained under
non-photometric conditions with a flux absorption of about 0.2 mag
which probably leads to larger errors in the estimation of fluxes and
relative zeropoints.  The images were absolutely calibrated with one
short exposed image obtained under photometric conditions.

The direct comparison of \CARS and the \MegaPipe images shows
considerably larger scatters. We investigated in more detail the case
of \MegaPipe \texttt{W3m1m2-$r'$} which shows a magnitude offset of
nearly 0.1 mag w.r.t. \CARS, \Tthree and the SDSS. It turned out that
an image obtained under unfavourable photometric conditions was
included in the calibration and stacking process although it should
have been rejected. The median-stacking of heterogeneous data
(\MegaPipe does, by default, not reject very short calibration
exposures as \CARS and \Tthree) and problematic images that slipped
through the grading process probably account for the observed scatters
in other cases (S. Gwyn, private communication).
%

\subsection{Photometric accuracy of \CARS data - Summary}
\label{sec:phot-calibr-summ}
In the preceding sections we evaluated the internal and external
photometric quality of our data. The results can be summarised as
follows:
\begin{itemize}
\item We evaluated the internal photometric accuracy of our co-added
  data with observations from the CFHTLS-Deep survey. From
  \texttt{D1} we constructed in each colour three independent stacks
  which contain data from 2003, 2004 and 2005 and compared fluxes from
  overlap sources. The \CARS \texttt{set} \texttt{W1p2p3} which was
  obtained in 2003 and is offset to \texttt{D1} is included in these
  tests. Around some absolute offsets, the magnitude comparisons show 
  internal scatters with 
  $\sigma_{int,  u^*g'r'i'} \approx 0.01-0.02$ in $u^*g'r'i'$ and about 
  $\sigma_{int,  z'} \approx 0.03-0.04$ in $z'$ and we quote these
  values as internal magnitude uncertainties over the \MegaPrime 
  field-of-view.
\item The accuracy of the absolute photometric calibration is
  primarily tested with a comparison to the Sloan Digital Sky Survey.
  The available \Elixir pre-calibration allows us to obtain an absolute
  photometric accuracy
  of about $\sigma_{abs, g'r'i'} \approx 0.01-0.04$ mag in the $g'r'i'$
  bands. Unbiased results are also obtained for $z'$ with an accuracy
  of $\sigma_{abs, z'} \approx 0.03-0.05$ mag. At the current stage we
  obtain a systematic bias of about 0.1 mag for the $u^*$-band. This
  holds directly for our data in \texttt{W1}. For \texttt{W3}
  and \texttt{W4} we arrive at the same conclusion if we take into 
  account systematics revealed by our zeropoint study of CFHTLS-Deep
  data. Given this result we quote the zeropoint uncertainty in the
  $u^*$-band with  $\sigma_{abs, u^*} \approx 0.15$ mag.
\item Because we process our data on a pointing basis we also need to
  calibrate our images \texttt{set} by \texttt{set} relying on \Elixir
  meta-data only. We generally do not take into account information
  from adjacent pointings. TERAPIX \Tthree data are treated with a
  more sophisticated procedure using all available information to
  simultaneously calibrate data on a patch-wide basis.  Our direct
  comparison shows that both pipelines lead to very comparable results
  with a small average magnitude scatter of about 0.02 mag. However, we
  observe four significant outliers (out of 93 common \CARS-\Tthree
  sets) with magnitude offsets of $0.05-0.18$ mag. Unfortunately no
  other external comparison is available for these fields. If we take the
  conservative approach to attribute these offsets to inaccuracies in
  our calibration and if the current
  \CARS data set is representative less than 5\% of our images severly
  suffer from a non-optimal photometric calibration procedure.
\end{itemize}


\subsection{Detailed \CARS data quality information}
\label{sec:CARSquality}
In \tabref{tab:CARSquality} we provide detailed information on the
characteristics of each \CARS data set. It contains the effective area
of each field after image masking (see \sectionref{sec:masking}), the
number of individual images contributing to each stack, the total
exposure time, the limiting magnitude as defined in
\sectionref{sec:CARSdata}, magnitude comparisons with Sloan, the
TERAPIX \Tthree and the \MegaPipe releases as described in
\sectionref{sec:sloancomp} and \sectionref{sec:comparison-cars-data},
the measured image seeing and special comments. The comments field
lists notable defects originating from the data itself or from our
reduction process. We do not list defects of astronomical origin (e.g.
very bright stars, external reflections) or problems which are present
in a large number of images (e.g. faint satellite tracks, asteroid
tracks, residual warm pixels, low-level fringe residuals which are
visible in most of the $z'$ images).  We use the following
abbreviations:
\begin{itemize}
\item \textbf{no ch. XX:} The stack contains no data around chip
  position(s) XX. We number the \MegaPrime mosaic chip from left to right and
  from bottom to top. The lower left (east-south) chip has number 1,
  the lower right (west-south) chip number 9 and the upper-right
  (west-north) chip number 36. Note that this labeling scheme differs
  from that used at CFHT.
\item \textbf{fr. res.:} The co-added image shows significant fringe residuals.
\item \textbf{m. ZP:} The zeropoint for this image was obtained
  manually by
  comparing object fluxes from the image with adjacent,
  photometrically calibrated pointings; see \sectionref{sec:phot-calibr}.
\item \textbf{sat. tr.:} The co-added image shows a bright satellite
  track which was not masked by our track detection module.
\end{itemize}
\begin{longtable}{llrrrrrrp{2.5cm}}
\caption{\label{tab:CARSquality}\CARS data quality overview: 
The first column lists: \CARS field-naming convention; CFHTLS
field-naming convention; effective field area after image masking. 
The seventh
column contains magnitude comparisons of \CARS fields with those
from TERAPIX \Tthree (indicated by \textbf{(T)}) and \MegaPipe
(indicated by \textbf{(M)}). Magnitude offsets are always given as
$m_{\rm CARS}-m_{\rm other}$. See the text for more details.} \\
\hline\hline
\multicolumn{1}{c}{Field/Area} & \multicolumn{1}{c}{Filter} & \multicolumn{1}{c}{N} & \multicolumn{1}{c}{expos. time} & \multicolumn{1}{c}{$m_{\rm lim}$} & \multicolumn{1}{c}{Sloan} & \multicolumn{1}{c}{T3/MegaP.} & \multicolumn{1}{c}{seeing} & \multicolumn{1}{c}{comments} \\
\multicolumn{1}{c}{[sq. deg.]} & & & \multicolumn{1}{c}{[s]} & \multicolumn{1}{c}{[AB mag]} & \multicolumn{1}{c}{$\Delta m \times 100$} & \multicolumn{1}{c}{$\Delta m \times 100$} & \multicolumn{1}{c}{[$''$]} & \\
\hline
\endfirsthead
\hline\hline
\multicolumn{1}{c}{Field/Area} & \multicolumn{1}{c}{Filter} & \multicolumn{1}{c}{N} & \multicolumn{1}{c}{expos. time} & \multicolumn{1}{c}{$m_{\rm lim}$} & \multicolumn{1}{c}{Sloan} & \multicolumn{1}{c}{T3/MegaP.} & \multicolumn{1}{c}{seeing} & \multicolumn{1}{c}{comments} \\
\multicolumn{1}{c}{[sq. deg.]} & & & \multicolumn{1}{c}{[s]} & \multicolumn{1}{c}{[AB mag]} & \multicolumn{1}{c}{$\Delta m \times 100$} & \multicolumn{1}{c}{$\Delta m \times 100$} & \multicolumn{1}{c}{[$''$]} & \\
\hline
\endhead
\hline
\endfoot
\texttt{W1m0p1} & u* & 5 & 3000.51 & 25.27 & - & $0.0 \pm 0.8$ \textbf{(T)}& 1.00 &  \\
\texttt{[w1.$-$0$+$1]} & & & & & & $-0.4 \pm 1.3$ \textbf{(M)} & & \\
(0.84) & g' & 5 & 2500.45 & 25.55 & - & $-0.4 \pm 0.5$ \textbf{(T)}& 0.90 &  \\
 & & & & & & $-7.1 \pm 0.9$ \textbf{(M)} & & \\
 & r' & 3 & 1500.28 & 24.72 & - & $-1.9 \pm 1.0$ \textbf{(T)}& 0.79 &  \\
 & i' & 7 & 4305.67 & 24.61 & - & $-0.1 \pm 0.8$ \textbf{(T)}& 0.85 &  \\
 & & & & & & $-0.8 \pm 0.7$ \textbf{(M)} & & \\
 & z' & 11 & 6601.19 & 23.88 & - & $0.2 \pm 1.5$ \textbf{(T)}& 0.79 &  \\
 & & & & & & $-7.7 \pm 1.1$ \textbf{(M)} & & \\
\hline
\texttt{W1m0p2} & u* & 5 & 3000.58 & 25.35 & - & $-0.6 \pm 1.1$ \textbf{(T)}& 1.00 &  \\
\texttt{[w1.$-$0$+$2]} & & & & & & $-2.2 \pm 1.5$ \textbf{(M)} & & \\
(0.76) & g' & 7 & 3500.46 & 25.72 & - & $-0.5 \pm 1.0$ \textbf{(T)}& 0.95 &  \\
 & & & & & & $-2.5 \pm 0.9$ \textbf{(M)} & & \\
 & r' & 2 & 1000.18 & 24.61 & - & $-0.5 \pm 0.5$ \textbf{(T)}& 0.82 & no ch. 21, 35 \\
 & i' & 7 & 4305.66 & 24.72 & - & $1.3 \pm 2.4$ \textbf{(T)}& 0.74 &  \\
 & & & & & & $0.2 \pm 1.8$ \textbf{(M)} & & \\
 & z' & 10 & 6000.83 & 23.64 & - & $-1.2 \pm 3.5$ \textbf{(T)}& 0.79 & fr. res. \\
 & & & & & & $-7.3 \pm 1.5$ \textbf{(M)} & & \\
\hline
\texttt{W1m0p3} & u* & 5 & 3000.58 & 25.27 & - & $-1.0 \pm 1.4$ \textbf{(T)}& 0.87 &  \\
\texttt{[w1.$-$0$+$3]} & & & & & & $5.3 \pm 1.2$ \textbf{(M)} & & \\
(0.75) & g' & 5 & 2500.49 & 25.56 & - & $-0.7 \pm 0.8$ \textbf{(T)}& 0.90 &  \\
 & & & & & & $-1.9 \pm 0.6$ \textbf{(M)} & & \\
 & r' & 2 & 1000.18 & 24.62 & - & $-1.2 \pm 0.5$ \textbf{(T)}& 0.85 & no ch. 21, 35 \\
 & i' & 7 & 4305.65 & 24.59 & - & $-0.5 \pm 0.7$ \textbf{(T)}& 0.74 &  \\
 & & & & & & $-0.8 \pm 0.5$ \textbf{(M)} & & \\
 & z' & 10 & 6000.84 & 23.59 & - & $-1.3 \pm 2.7$ \textbf{(T)}& 0.82 & fr. res. \\
 & & & & & & $-8.1 \pm 1.4$ \textbf{(M)} & & \\
\hline
\texttt{W1m1p1} & u* & 7 & 4200.60 & 25.50 & - & - & 1.00 &  \\
\texttt{[w1.$-$1$+$1]} & g' & 8 & 4000.76 & 25.73 & - & - & 0.66 &  \\
(0.84) & r' & 2 & 1000.20 & 24.51 & - & - & 0.59 &  \\
 & i' & 9 & 5535.73 & 24.53 & - & - & 0.79 &  \\
 & z' & 6 & 3600.40 & 23.30 & - & - & 0.75 &  \\
\hline
\texttt{W1m1p2} & u* & 5 & 3000.52 & 25.03 & - & $-12.0 \pm 0.6$ \textbf{(T)}& 0.92 &  \\
\texttt{[w1.$-$1$+$2]} & & & & & & $-9.2 \pm 1.5$ \textbf{(M)} & & \\
(0.82) & g' & 5 & 2500.48 & 25.44 & - & $3.3 \pm 0.7$ \textbf{(T)}& 0.87 &  \\
 & & & & & & $-4.1 \pm 1.0$ \textbf{(M)} & & \\
 & r' & 4 & 2000.38 & 24.80 & - & $1.0 \pm 2.2$ \textbf{(T)}& 0.87 &  \\
 & & & & & & $-2.1 \pm 2.0$ \textbf{(M)} & & \\
 & i' & 7 & 4305.66 & 24.52 & - & $-0.1 \pm 1.8$ \textbf{(T)}& 0.71 &  \\
 & & & & & & $1.6 \pm 2.1$ \textbf{(M)} & & \\
 & z' & 11 & 6601.18 & 23.87 & - & $-1.1 \pm 1.5$ \textbf{(T)}& 0.71 & fr. res. \\
 & & & & & & $0.3 \pm 0.8$ \textbf{(M)} & & \\
\hline
\texttt{W1m1p3} & u* & 5 & 3000.50 & 24.95 & - & $-18.3 \pm 0.5$ \textbf{(T)}& 0.79 &  \\
\texttt{[w1.$-$1$+$3]} & g' & 3 & 1500.30 & 25.22 & - & $1.0 \pm 0.8$ \textbf{(T)}& 0.77 &  \\
(0.76) & r' & 2 & 1000.23 & 24.58 & - & $1.4 \pm 0.8$ \textbf{(T)}& 0.79 & no ch. 21, 35 \\
 & i' & 7 & 4305.70 & 24.58 & - & $-0.3 \pm 1.0$ \textbf{(T)}& 0.74 &  \\
 & z' & 10 & 6001.00 & 23.43 & - & $-11.9 \pm 2.3$ \textbf{(T)}& 0.71 & fr. res. \\
\hline
\texttt{W1p1m1} & u* & 5 & 3000.42 & 25.26 & $12.2 \pm 5.5$ & - & 0.85 &  \\
\texttt{[w1.$+$1$-$1]} & g' & 6 & 3000.47 & 25.79 & $2.7 \pm 2.8$ & - & 0.87 &  \\
(0.85) & r' & 2 & 1000.16 & 24.50 & $4.2 \pm 3.4$ & - & 0.71 &  \\
 & i' & 7 & 4305.53 & 24.85 & $5.2 \pm 4.8$ & - & 0.71 &  \\
 & z' & 6 & 3600.40 & 23.52 & $1.3 \pm 5.7$ & - & 0.85 &  \\
\hline
\texttt{W1p1p1} & u* & 5 & 3000.47 & 25.28 & - & $0.6 \pm 0.9$ \textbf{(T)}& 0.92 &  \\
\texttt{[w1.$+$1$+$1]} & & & & & & $6.6 \pm 0.7$ \textbf{(M)} & & \\
(0.75) & g' & 10 & 5000.88 & 25.97 & - & $-0.3 \pm 0.7$ \textbf{(T)}& 0.95 &  \\
 & & & & & & $-2.0 \pm 0.5$ \textbf{(M)} & & \\
 & r' & 2 & 1000.18 & 24.57 & - & $-1.6 \pm 0.6$ \textbf{(T)}& 0.85 &  \\
 & i' & 7 & 4305.72 & 24.70 & - & $-0.1 \pm 0.8$ \textbf{(T)}& 0.82 &  \\
 & & & & & & $-0.5 \pm 0.5$ \textbf{(M)} & & \\
 & z' & 12 & 7201.21 & 23.94 & - & $0.9 \pm 1.5$ \textbf{(T)}& 0.74 & fr. res. \\
 & & & & & & $-9.3 \pm 1.3$ \textbf{(M)} & & \\
\hline
\texttt{W1p1p2} & u* & 5 & 3000.48 & 25.29 & - & $-3.9 \pm 1.0$ \textbf{(T)}& 1.00 &  \\
\texttt{[w1.$+$1$+$2]} & g' & 5 & 2500.42 & 25.60 & - & $-1.3 \pm 1.0$ \textbf{(T)}& 0.87 &  \\
(0.83) & r' & 2 & 1000.16 & 24.50 & - & $-0.1 \pm 0.6$ \textbf{(T)}& 0.74 &  \\
 & i' & 6 & 3690.57 & 24.45 & - & $-0.1 \pm 1.0$ \textbf{(T)}& 0.77 &  \\
 & z' & 10 & 6001.00 & 23.60 & - & $-0.4 \pm 1.8$ \textbf{(T)}& 0.63 & fr. res. \\
\hline
\texttt{W1p1p3} & u* & 5 & 3000.47 & 25.30 & - & $-2.7 \pm 1.1$ \textbf{(T)}& 0.95 & no ch. 21, 35 \\
\texttt{[w1.$+$1$+$3]} & & & & & & $-1.8 \pm 1.7$ \textbf{(M)} & & \\
(0.80) & g' & 5 & 2500.42 & 25.53 & - & $2.4 \pm 0.8$ \textbf{(T)}& 0.95 & no ch. 21, 35 \\
 & r' & 4 & 2000.22 & 24.72 & - & $-1.4 \pm 1.4$ \textbf{(T)}& 0.89 & no ch. 21, 35 \\
 & i' & 8 & 4920.84 & 24.63 & - & $-0.4 \pm 0.8$ \textbf{(T)}& 0.95 & no ch. 21, 35 \\
 & z' & 10 & 6001.03 & 23.63 & - & $-0.9 \pm 2.2$ \textbf{(T)}& 0.69 & no ch. 21, 35; fr. res. \\
 & & & & & & $-9.9 \pm 1.7$ \textbf{(M)} & & \\
\hline
\texttt{W1p2p1} & u* & 5 & 3000.53 & 25.32 & - & $-3.1 \pm 1.4$ \textbf{(T)}& 1.00 & no ch. 31 \\
\texttt{[w1.$+$2$+$1]} & g' & 5 & 2500.45 & 25.56 & - & $-1.9 \pm 1.1$ \textbf{(T)}& 1.00 & no ch. 31 \\
(0.84) & r' & 2 & 1000.18 & 24.42 & - & $-2.8 \pm 0.6$ \textbf{(T)}& 0.95 & no ch. 31 \\
 & i' & 8 & 4960.74 & 24.63 & - & $0.6 \pm 0.6$ \textbf{(T)}& 0.90 & no ch. 31 \\
 & z' & 10 & 6000.82 & 23.74 & - & $-0.9 \pm 2.2$ \textbf{(T)}& 0.71 & no ch. 31; fr. res. \\
\hline
\texttt{W1p2p2} & u* & 5 & 3000.56 & 25.21 & - & $-2.8 \pm 1.1$ \textbf{(T)}& 0.98 & no ch. 21, 35 \\
\texttt{[w1.$+$2$+$2]} & g' & 5 & 2500.45 & 25.63 & - & $1.4 \pm 1.2$ \textbf{(T)}& 0.95 &  \\
(0.81) & r' & 2 & 1000.18 & 24.53 & - & $-3.4 \pm 0.6$ \textbf{(T)}& 0.95 &  \\
 & i' & 8 & 4960.76 & 24.80 & - & $0.7 \pm 1.1$ \textbf{(T)}& 0.85 &  \\
 & z' & 10 & 6000.95 & 23.73 & - & $-1.1 \pm 1.3$ \textbf{(T)}& 0.74 &  \\
\hline
\texttt{W1p2p3} & u* & 7 & 5950.29 & 25.61 & - & $-0.9 \pm 0.5$ \textbf{(T)}& 1.00 & no ch. 31 \\
\texttt{[w1.$+$2$+$3]} & g' & 5 & 2500.35 & 25.62 & - & $-0.3 \pm 0.5$ \textbf{(T)}& 0.95 &  \\
(0.83) & r' & 4 & 2000.37 & 24.82 & - & $-6.3 \pm 0.9$ \textbf{(T)}& 0.71 &  \\
 & i' & 7 & 4340.55 & 24.59 & - & $-0.4 \pm 0.9$ \textbf{(T)}& 0.95 &  \\
 & z' & 9 & 7200.41 & 23.80 & - & $0.9 \pm 2.3$ \textbf{(T)}& 0.69 & no ch. 31; fr. res. \\
\hline
\texttt{W1p3m0} & u* & 5 & 3000.55 & 25.17 & $11.9 \pm 6.6$ & $6.4 \pm 2.2$ \textbf{(M)}& 0.63 & no ch. 31 \\
\texttt{[w1.$+$3$-$0]} & g' & 5 & 2500.43 & 25.54 & $-1.8 \pm 3.0$ & $-2.6 \pm 0.6$ \textbf{(T)}& 0.87 & no ch. 31; m. ZP \\
(0.81) & r' & 2 & 1000.17 & 24.33 & $2.4 \pm 3.0$ & $0.1 \pm 1.3$ \textbf{(T)}& 0.69 & no ch. 31 \\
 & i' & 5 & 3100.23 & 24.44 & $3.5 \pm 3.5$ & $3.1 \pm 1.0$ \textbf{(T)}& 0.71 & no ch. 31; sat. tr. \\
 & z' & 12 & 7201.49 & 23.59 & $-0.4 \pm 7.2$ & $-8.8 \pm 3.7$ \textbf{(M)}& 0.71 & no ch. 31; fr. res. \\
\hline
\texttt{W1p3p1} & u* & 5 & 3000.26 & 25.31 & - & - & 0.85 & no ch. 31 \\
\texttt{[w1.$+$3$+$1]} & g' & 6 & 3000.27 & 25.62 & - & $0.0 \pm 0.5$ \textbf{(T)}& 0.95 & no ch. 31 \\
(0.85) & r' & 2 & 1000.10 & 24.39 & - & $0.5 \pm 0.7$ \textbf{(T)}& 0.85 & no ch. 31 \\
 & i' & 7 & 4340.32 & 24.60 & - & $-0.8 \pm 1.1$ \textbf{(T)}& 0.95 & no ch. 31 \\
 & z' & 6 & 3600.47 & 23.55 & - & - & 0.69 & no ch. 31 \\
\hline
\texttt{W1p3p2} & u* & 7 & 4200.42 & 25.45 & - & $5.0 \pm 1.5$ \textbf{(M)}& 0.90 & no ch. 31 \\
\texttt{[w1.$+$3$+$2]} & g' & 5 & 2500.41 & 25.54 & - & $0.3 \pm 0.9$ \textbf{(T)}& 0.85 & no ch. 31; m. ZP \\
(0.82) & r' & 3 & 1500.22 & 24.51 & - & $-1.2 \pm 1.2$ \textbf{(T)}& 0.83 & no ch. 31 \\
 & i' & 6 & 3720.24 & 24.48 & - & $0.6 \pm 0.8$ \textbf{(T)}& 0.69 & no ch. 31 \\
 & z' & 6 & 3600.58 & 23.31 & - & $-7.3 \pm 0.9$ \textbf{(M)}& 0.55 & no ch. 31 \\
\hline
\texttt{W1p3p3} & u* & 4 & 2400.39 & 25.01 & - & $3.8 \pm 0.9$ \textbf{(M)}& 1.11 & no ch. 31 \\
\texttt{[w1.$+$3$+$3]} & g' & 5 & 2500.20 & 25.51 & - & $0.1 \pm 0.8$ \textbf{(T)}& 0.95 & no ch. 31 \\
(0.84) & r' & 2 & 1000.10 & 24.39 & - & $-2.9 \pm 0.6$ \textbf{(T)}& 0.87 & no ch. 31 \\
 & i' & 7 & 4340.33 & 24.51 & - & $-1.3 \pm 0.9$ \textbf{(T)}& 0.82 & no ch. 31 \\
 & z' & 6 & 3600.62 & 23.39 & - & $-7.2 \pm 0.9$ \textbf{(M)}& 0.55 & no ch. 31 \\
\hline
\texttt{W1p4m0} & u* & 5 & 3000.54 & 25.35 & $9.4 \pm 4.7$ & $6.4 \pm 0.7$ \textbf{(M)}& 0.72 & no ch. 31 \\
\texttt{[w1.$+$4$-$0]} & g' & 5 & 2500.41 & 25.50 & $-1.5 \pm 3.1$ & $-0.8 \pm 1.1$ \textbf{(T)}& 0.87 & no ch. 31; m. ZP \\
(0.82) & r' & 2 & 1000.19 & 24.33 & $2.7 \pm 2.5$ & $0.3 \pm 0.6$ \textbf{(T)}& 0.71 & no ch. 31 \\
 & i' & 4 & 2480.15 & 24.15 & $0.1 \pm 3.1$ & $0.9 \pm 1.1$ \textbf{(T)}& 0.95 & no ch. 31 \\
 & z' & 6 & 3600.72 & 23.48 & $0.4 \pm 5.9$ & $-7.1 \pm 0.8$ \textbf{(M)}& 0.69 & no ch. 31 \\
\hline
\texttt{W1p4p1} & u* & 5 & 3000.54 & 25.42 & - & - & 0.79 &  \\
\texttt{[w1.$+$4$+$1]} & g' & 5 & 2500.43 & 25.45 & - & $-1.5 \pm 0.7$ \textbf{(T)}& 0.85 & no ch. 31; m. ZP \\
(0.82) & r' & 2 & 1000.16 & 24.21 & - & $1.6 \pm 0.6$ \textbf{(T)}& 0.79 & m. ZP \\
 & i' & 7 & 4340.62 & 24.48 & - & $0.3 \pm 0.8$ \textbf{(T)}& 0.79 & m. ZP \\
 & z' & 6 & 3600.52 & 23.34 & - & - & 0.63 &  \\
\hline
\texttt{W1p4p2} & u* & 5 & 3000.60 & 25.29 & - & - & 0.77 &  \\
\texttt{[w1.$+$4$+$2]} & g' & 5 & 2500.44 & 25.49 & - & $-0.9 \pm 1.3$ \textbf{(T)}& 0.85 & no ch. 31; m. ZP \\
(0.87) & r' & 2 & 1000.16 & 24.22 & - & $-2.2 \pm 0.8$ \textbf{(T)}& 0.74 & m. ZP \\
 & i' & 7 & 4340.64 & 24.44 & - & $-0.9 \pm 1.1$ \textbf{(T)}& 0.87 & m. ZP \\
 & z' & 6 & 3600.50 & 23.45 & - & - & 0.55 &  \\
\hline
\texttt{W1p4p3} & u* & 5 & 3000.25 & 25.32 & - & $5.9 \pm 0.8$ \textbf{(M)}& 0.98 & no ch. 31 \\
\texttt{[w1.$+$4$+$3]} & g' & 5 & 2500.41 & 25.65 & - & $-0.9 \pm 2.3$ \textbf{(T)}& 0.95 & no ch. 31 \\
(0.83) & & & & & & $-1.4 \pm 1.1$ \textbf{(M)} & & \\
 & r' & 2 & 1000.17 & 24.46 & - & $-0.7 \pm 1.0$ \textbf{(T)}& 0.93 & no ch. 31 \\
 & i' & 7 & 4340.34 & 24.44 & - & $-1.0 \pm 0.7$ \textbf{(T)}& 0.95 & no ch. 31 \\
 & z' & 6 & 3600.38 & 23.29 & - & - & 0.77 & no ch. 31 \\
\hline
\hline
\texttt{W3m1m2} & u* & 5 & 3000.96 & 24.78 & $30.2 \pm 3.9$ & $25.9 \pm 2.5$ \textbf{(M)}& 0.87 &  \\
\texttt{[w3.$-$1$-$2]} & g' & 5 & 2501.02 & 25.48 & $0.2 \pm 1.9$ & $0.2 \pm 0.6$ \textbf{(T)}& 0.87 & m. ZP \\
(0.83) & & & & & & $-4.9 \pm 1.5$ \textbf{(M)} & & \\
 & r' & 2 & 1000.47 & 24.51 & $-0.1 \pm 2.7$ & $-1.3 \pm 0.7$ \textbf{(T)}& 0.63 &  \\
 & & & & & & $-9.9 \pm 3.7$ \textbf{(M)} & & \\
 & i' & 7 & 4306.47 & 24.36 & $0.6 \pm 3.8$ & $-0.6 \pm 1.3$ \textbf{(T)}& 0.66 &  \\
 & z' & 6 & 3601.20 & 23.41 & $-6.1 \pm 4.2$ & $-9.4 \pm 2.3$ \textbf{(M)}& 0.67 &  \\
\hline
\texttt{W3m1m3} & u* & 5 & 3000.98 & 24.87 & $29.9 \pm 3.8$ & $26.9 \pm 2.6$ \textbf{(M)}& 0.74 & no ch. 21 \\
\texttt{[w3.$-$1$-$3]} & g' & 5 & 2500.90 & 25.63 & $0.4 \pm 2.0$ & $0.1 \pm 0.5$ \textbf{(T)}& 0.87 & no ch. 21; m. ZP \\
(0.82) & & & & & & $-4.7 \pm 1.1$ \textbf{(M)} & & \\
 & r' & 2 & 1000.28 & 24.52 & $2.9 \pm 2.2$ & $-2.2 \pm 1.6$ \textbf{(M)}& 0.97 & no ch. 21 \\
 & i' & 7 & 4306.61 & 24.36 & $0.6 \pm 2.7$ & $0.3 \pm 0.9$ \textbf{(T)}& 0.66 & no ch. 21; m. ZP \\
 & & & & & & $-4.4 \pm 1.6$ \textbf{(M)} & & \\
 & z' & 6 & 3601.17 & 23.36 & $-3.8 \pm 5.6$ & $-9.1 \pm 2.2$ \textbf{(M)}& 0.59 & no ch. 21 \\
\hline
\texttt{W3m2m2} & u* & 5 & 3001.06 & 25.33 & $29.4 \pm 4.1$ & $25.3 \pm 2.2$ \textbf{(M)}& 0.77 &  \\
\texttt{[w3.$-$2$-$2]} & g' & 5 & 2500.89 & 25.59 & $0.7 \pm 2.2$ & $-0.3 \pm 0.5$ \textbf{(T)}& 0.90 &  \\
(0.85) & & & & & & $-3.8 \pm 1.0$ \textbf{(M)} & & \\
 & r' & 2 & 1000.42 & 24.56 & $0.7 \pm 2.7$ & $-1.1 \pm 1.0$ \textbf{(T)}& 0.57 &  \\
 & & & & & & $-4.8 \pm 2.2$ \textbf{(M)} & & \\
 & i' & 7 & 4306.23 & 24.46 & $0.3 \pm 2.7$ & $0.2 \pm 0.8$ \textbf{(T)}& 0.66 &  \\
 & z' & 6 & 3601.22 & 23.43 & $-3.3 \pm 4.9$ & $-8.5 \pm 1.7$ \textbf{(M)}& 0.64 &  \\
\hline
\texttt{W3m3m2} & u* & 5 & 3000.99 & 25.27 & $29.0 \pm 5.0$ & - & 0.68 &  \\
\texttt{[w3.$-$3$-$2]} & g' & 5 & 2500.87 & 25.40 & $0.4 \pm 2.2$ & $-0.2 \pm 1.1$ \textbf{(T)}& 0.66 &  \\
(0.84) & r' & 2 & 1000.40 & 24.20 & $4.5 \pm 2.4$ & $1.9 \pm 0.7$ \textbf{(T)}& 0.66 &  \\
 & i' & 7 & 4306.27 & 24.43 & $0.1 \pm 3.3$ & $0.0 \pm 0.9$ \textbf{(T)}& 0.53 &  \\
 & z' & 6 & 3601.20 & 23.30 & $-2.3 \pm 4.7$ & - & 0.55 &  \\
\hline
\texttt{W3m3m3} & u* & 5 & 3001.02 & 25.29 & $28.8 \pm 4.1$ & - & 0.74 &  \\
\texttt{[w3.$-$3$-$3]} & g' & 5 & 2500.93 & 25.52 & $0.2 \pm 2.3$ & $-0.5 \pm 0.6$ \textbf{(T)}& 0.82 &  \\
(0.83) & r' & 2 & 1000.42 & 24.25 & $4.4 \pm 2.6$ & $1.4 \pm 0.6$ \textbf{(T)}& 0.66 &  \\
 & i' & 7 & 4306.19 & 24.49 & $-0.8 \pm 3.0$ & $0.1 \pm 0.8$ \textbf{(T)}& 0.50 &  \\
 & z' & 6 & 3601.21 & 23.49 & $-3.1 \pm 5.5$ & - & 0.58 &  \\
\hline
\hline
\texttt{W4m0m0} & u* & 5 & 3000.26 & 25.38 & $26.4 \pm 3.8$ & - & 1.03 &  \\
\texttt{[w4.$+$0$+$0]} & g' & 5 & 2500.40 & 25.43 & $-0.1 \pm 2.0$ & $-1.9 \pm 0.6$ \textbf{(M)}& 0.79 &  \\
(0.82) & r' & 3 & 1500.22 & 24.23 & $-0.2 \pm 2.2$ & $1.0 \pm 0.6$ \textbf{(M)}& 0.61 &  \\
 & i' & 7 & 4305.65 & 24.62 & $5.3 \pm 3.3$ & - & 0.71 &  \\
 & z' & 12 & 7200.87 & 23.78 & $-2.3 \pm 4.6$ & - & 0.66 &  \\
\hline
\texttt{W4m0m1} & u* & 5 & 3000.29 & 25.30 & $23.2 \pm 4.0$ & - & 0.74 &  \\
\texttt{[w4.$+$0$-$1]} & g' & 10 & 5000.68 & 25.85 & $-0.2 \pm 2.5$ & $-7.7 \pm 1.6$ \textbf{(M)}& 0.82 &  \\
(0.83) & r' & 2 & 1000.10 & 24.30 & $1.7 \pm 2.3$ & $-5.4 \pm 1.7$ \textbf{(M)}& 0.67 &  \\
 & i' & 7 & 4305.36 & 24.62 & $-0.8 \pm 3.0$ & - & 0.56 &  \\
 & z' & 6 & 3600.27 & 23.29 & $-3.2 \pm 5.0$ & - & 0.50 &  \\
\hline
\texttt{W4m0m2} & u* & 5 & 3000.31 & 25.41 & $24.0 \pm 4.5$ & - & 0.71 &  \\
\texttt{[w4.$+$0$-$2]} & g' & 10 & 5000.58 & 25.83 & $-0.3 \pm 1.9$ & $-6.2 \pm 1.2$ \textbf{(M)}& 0.77 &  \\
(0.81) & r' & 2 & 1000.10 & 24.28 & $2.1 \pm 2.2$ & - & 0.58 &  \\
 & i' & 7 & 4305.58 & 24.77 & $0.7 \pm 2.6$ & - & 0.61 &  \\
 & z' & 6 & 3600.34 & 23.62 & $-4.4 \pm 4.1$ & - & 0.63 &  \\
\hline
\texttt{W4m1m1} & u* & 5 & 3000.24 & 25.26 & $24.2 \pm 3.9$ & - & 0.69 &  \\
\texttt{[w4.$-$1$-$1]} & g' & 13 & 6500.96 & 26.00 & $2.9 \pm 2.4$ & - & 0.79 &  \\
(0.79) & r' & 2 & 1000.16 & 24.32 & $1.4 \pm 2.5$ & - & 0.87 &  \\
 & i' & 6 & 3690.44 & 24.62 & $-0.2 \pm 3.2$ & - & 0.71 &  \\
 & z' & 5 & 3000.24 & 23.28 & $-1.2 \pm 4.8$ & - & 0.48 &  \\
\hline
\texttt{W4m1m2} & u* & 6 & 3600.41 & 25.45 & $33.2 \pm 8.1$ & - & 0.66 &  \\
\texttt{[w4.$-$1$-$2]} & g' & 5 & 2500.53 & 25.54 & $-1.8 \pm 1.8$ & $-7.6 \pm 1.3$ \textbf{(M)}& 0.79 &  \\
(0.82) & r' & 2 & 1000.15 & 24.34 & $2.9 \pm 2.4$ & - & 0.50 &  \\
 & i' & 7 & 4305.43 & 24.60 & $1.8 \pm 2.9$ & - & 0.72 &  \\
 & z' & 5 & 3000.38 & 23.37 & $-0.3 \pm 4.7$ & - & 0.51 &  \\
\hline
\texttt{W4p1m0} & u* & 5 & 3000.43 & 25.29 & $26.2 \pm 4.1$ & - & 0.90 &  \\
\texttt{[w4.$+$1$+$0]} & g' & 5 & 2500.30 & 25.40 & $1.6 \pm 2.1$ & $-2.3 \pm 0.6$ \textbf{(M)}& 0.67 &  \\
(0.75) & r' & 2 & 1000.16 & 24.35 & $3.0 \pm 2.2$ & - & 0.94 &  \\
 & i' & 7 & 4305.52 & 24.31 & $0.4 \pm 3.3$ & - & 0.53 &  \\
 & z' & 6 & 3600.36 & 23.33 & $-5.0 \pm 4.2$ & - & 0.55 &  \\
\hline
\texttt{W4p1m1} & u* & 5 & 3000.22 & 25.33 & $27.5 \pm 3.8$ & - & 0.85 &  \\
\texttt{[w4.$+$1$-$1]} & g' & 5 & 2500.34 & 25.33 & $1.2 \pm 2.0$ & - & 0.83 &  \\
(0.80) & r' & 2 & 1000.17 & 24.28 & $-0.8 \pm 2.5$ & - & 0.67 &  \\
 & i' & 14 & 8611.03 & 24.86 & $1.4 \pm 3.7$ & - & 0.66 &  \\
 & z' & 6 & 3600.33 & 23.35 & $-4.2 \pm 4.7$ & - & 0.63 &  \\
\hline
\texttt{W4p1m2} & u* & 5 & 3000.39 & 25.15 & $12.7 \pm 3.9$ & - & 0.87 &  \\
\texttt{[w4.$+$1$-$2]} & g' & 5 & 2500.31 & 25.37 & $0.9 \pm 1.9$ & $-4.9 \pm 1.3$ \textbf{(M)}& 0.85 &  \\
(0.82) & r' & 2 & 1000.10 & 24.34 & $1.7 \pm 2.1$ & - & 0.61 &  \\
 & i' & 7 & 4305.41 & 24.57 & $1.1 \pm 3.4$ & $-5.8 \pm 1.1$ \textbf{(M)}& 0.71 &  \\
 & z' & 5 & 3000.48 & 23.12 & $-2.1 \pm 5.0$ & - & 0.53 &  \\
\hline
\texttt{W4p2m0} & u* & 5 & 3000.31 & 25.24 & $28.2 \pm 4.4$ & - & 0.79 &  \\
\texttt{[w4.$+$2$-$0]} & g' & 5 & 2500.47 & 25.30 & $-1.6 \pm 2.4$ & $-4.4 \pm 1.6$ \textbf{(M)}& 0.74 &  \\
(0.77) & r' & 2 & 1000.14 & 24.36 & $2.0 \pm 2.3$ & - & 0.63 &  \\
 & i' & 7 & 4305.53 & 24.68 & $0.0 \pm 3.0$ & - & 0.57 &  \\
 & z' & 6 & 3600.39 & 23.20 & $-4.9 \pm 3.9$ & - & 0.79 &  \\
\hline
\texttt{W4p2m1} & u* & 5 & 3000.36 & 25.18 & $19.7 \pm 3.7$ & $14.5 \pm 2.0$ \textbf{(M)}& 0.98 &  \\
\texttt{[w4.$+$2$-$1]} & g' & 5 & 2500.39 & 25.45 & $-0.7 \pm 1.9$ & $-6.4 \pm 1.1$ \textbf{(M)}& 0.85 &  \\
(0.80) & r' & 2 & 1000.22 & 24.12 & $1.6 \pm 2.4$ & - & 0.85 &  \\
 & i' & 7 & 4305.69 & 24.53 & $-3.4 \pm 2.9$ & $-9.9 \pm 0.9$ \textbf{(M)}& 0.66 &  \\
 & z' & 12 & 7200.72 & 23.77 & $-8.0 \pm 4.5$ & $-15.7 \pm 2.1$ \textbf{(M)}& 0.74 &  \\
\hline
\texttt{W4p2m2} & u* & 4 & 2400.24 & 25.18 & $16.4 \pm 3.7$ & - & 1.00 &  \\
\texttt{[w4.$+$2$-$2]} & g' & 5 & 2500.37 & 25.36 & $-1.3 \pm 2.1$ & - & 0.77 &  \\
(0.83) & r' & 2 & 1000.22 & 24.05 & $2.5 \pm 2.5$ & - & 0.90 &  \\
 & i' & 13 & 7995.87 & 24.98 & $1.9 \pm 4.1$ & - & 0.63 &  \\
 & z' & 10 & 6000.55 & 23.66 & $-3.7 \pm 4.0$ & - & 0.72 &  \\
\hline
\hline
\end{longtable}
\end{document}